\documentclass[aps,prx,twocolumn,floatfix,superscriptaddress]{revtex4-2}
\usepackage{tabularx} 
\usepackage{amsmath}  
\usepackage{amssymb}
\usepackage{graphicx} 
\usepackage[margin=1in,letterpaper]{geometry} 
\usepackage[final]{hyperref} 
\usepackage{multirow}
\hypersetup{
	colorlinks=true,       
	linkcolor=blue,        
	citecolor=blue,        
	filecolor=magenta,     
	urlcolor=blue         
}
\usepackage{blindtext}
\usepackage{verbatim}
\usepackage{siunitx}
\usepackage[normalem]{ulem}

\newcommand{\Cl}[2]{#1^{#2}}
\newcommand{\Cls}[3]{#1^{#2}_{#3}}

\newcommand{\dl}{d}
\newcommand{\fl}{f}
\newcommand{\A}{{a}}
\newcommand{\B}{{b}}
\newcommand{\one}{\B}
\newcommand{\two}{\A}

\newcommand{\Fig}[1]{Fig.~\ref{#1}}
\newcommand{\Eq}[1]{Eq.~(\ref{#1})}
\newcommand{\Eqs}[2]{Eqs.~(\ref{#1}) and (\ref{#2})}
\newcommand{\Eqsd}[2]{Eqs.~(\ref{#1})-(\ref{#2})}

\newcommand{\bvec}[1]{\mathbf{#1}}
\newcommand{\xhat}{\bvec{\hat{x}}}
\newcommand{\yhat}{\bvec{\hat{y}}}
\newcommand{\zhat}{\bvec{\hat{z}}}
\newcommand{\mhat}{\bvec{\hat{m}}}

\newcommand{\Ehat}{\bvec{\hat{E}}}

\newcommand{\torque}{\boldsymbol{\tau}}
\newcommand{\df}{\dl/\fl}

\newcommand{\TauF}[2]{\Cls{\torque}{#1}{#2}}
\newcommand{\tauF}[2]{\Cls{\tau}{#1}{#2}}

\newcommand{\Taud}{\TauF{}{\dl}}
\newcommand{\Tauf}{\TauF{}{\fl}}

\newcommand{\Tauad}{\TauF{\A}{\dl}}
\newcommand{\Tauaf}{\TauF{\A}{\fl}}

\newcommand{\taud}{\tauF{}{\dl}}
\newcommand{\tauf}{\tauF{}{\fl}}

\newcommand{\tauad}{\tauF{\A}{\dl}}
\newcommand{\tauaf}{\tauF{\A}{\fl}}
\newcommand{\tauadf}{\tauF{\A}{\df}}

\newcommand{\tauonedf}{\tauF{\one}{\df}}

\newcommand{\tautwod}{\tauF{\two}{\dl}}
\newcommand{\tautwof}{\tauF{\two}{\fl}}
\newcommand{\tautwodf}{\tauF{\two}{\df}}

\newcommand{\FP}{\bvec{s}}
\newcommand{\FPLdl}[1]{\Cls{\FP}{#1}{\dl}}
\newcommand{\FPLfl}[1]{\Cls{\FP}{#1}{\fl}}

\newcommand{\FPdl}{\FPLdl{}}
\newcommand{\FPfl}{\FPLfl{}}

\newcommand{\FPdla}{\FPLdl{\A}}
\newcommand{\FPfla}{\FPLfl{\A}}

\newcommand{\FPdltwo}{\FPLdl{\two}}
\newcommand{\FPfltwo}{\FPLfl{\two}}

\newcommand{\gL}[1]{\Cl{g}{#1}}
\newcommand{\gone}{\gL{\one}}
\newcommand{\gtwo}{\gL{\two}}

\newcommand{\dsL}[1]{\Cls{d}{#1}{s}}
\newcommand{\dxL}[1]{\Cls{d}{#1}{\times}}
\newcommand{\dmL}[1]{\Cls{d}{#1}{m}}
\newcommand{\dDL}[1]{\Cls{d}{#1}{D}}
\newcommand{\dtwoL}[1]{\Cls{d}{#1}{2}}
\newcommand{\dzL}[1]{\Cls{d}{#1}{z}}

\newcommand{\fsL}[1]{\Cls{f}{#1}{s}}
\newcommand{\fxL}[1]{\Cls{f}{#1}{\times}}
\newcommand{\fmL}[1]{\Cls{f}{#1}{m}}
\newcommand{\fDL}[1]{\Cls{f}{#1}{D}}
\newcommand{\ftwoL}[1]{\Cls{f}{#1}{2}}
\newcommand{\fzL}[1]{\Cls{f}{#1}{z}}

\newcommand{\ds}{\dsL{}}
\newcommand{\dx}{\dxL{}}
\newcommand{\dm}{\dmL{}}
\newcommand{\dD}{\dDL{}}
\newcommand{\dtwo}{\dtwoL{}}
\newcommand{\dz}{\dzL{}}

\newcommand{\fs}{\fsL{}}
\newcommand{\fx}{\fxL{}}
\newcommand{\fm}{\fmL{}}
\newcommand{\fD}{\fDL{}}
\newcommand{\ftwo}{\ftwoL{}}
\newcommand{\fz}{\fzL{}}

\newcommand{\dsa}{\dsL{\A}}
\newcommand{\dxa}{\dxL{\A}}
\newcommand{\dma}{\dmL{\A}}
\newcommand{\dDa}{\dDL{\A}}
\newcommand{\dtwoa}{\dtwoL{\A}}
\newcommand{\dza}{\dzL{\A}}

\newcommand{\fsa}{\fsL{\A}}
\newcommand{\fxa}{\fxL{\A}}
\newcommand{\fma}{\fmL{\A}}
\newcommand{\fDa}{\fDL{\A}}
\newcommand{\ftwoa}{\ftwoL{\A}}
\newcommand{\fza}{\fzL{\A}}

\newcommand{\fsb}{\fsL{\B}}

\newcommand{\dsone}{\dsL{\one}}
\newcommand{\dxone}{\dxL{\one}}
\newcommand{\dmone}{\dmL{\one}}

\newcommand{\fsone}{\fsL{\one}}
\newcommand{\fxone}{\fxL{\one}}
\newcommand{\fmone}{\fmL{\one}}

\newcommand{\dstwo}{\dsL{\two}}
\newcommand{\dxtwo}{\dxL{\two}}
\newcommand{\dmtwo}{\dmL{\two}}

\newcommand{\fstwo}{\fsL{\two}}
\newcommand{\fxtwo}{\fxL{\two}}
\newcommand{\fmtwo}{\fmL{\two}}

\newcommand{\threevec}[1]{\mathbf{#1}}

\newcommand{\pdf}[2]{\frac{\partial #1}{\partial #2}}

\newcommand{\ReGm}{\text{Re}[G_{\uparrow\downarrow}]}
\newcommand{\ImGm}{\text{Im}[G_{\uparrow\downarrow}]}

\newcommand{\ReGma}{\text{Re}[G^\A_{\uparrow\downarrow}]}
\newcommand{\ImGma}{\text{Im}[G^\A_{\uparrow\downarrow}]}
\newcommand{\Gca}{G^\A_c}

\newcommand{\ReGmb}{\text{Re}[G^\B_{\uparrow\downarrow}]}
\newcommand{\ImGmb}{\text{Im}[G^\B_{\uparrow\downarrow}]}
\newcommand{\Gcb}{G^\B_c}

\begin{document}

\title{Direct and indirect spin current generation and spin-orbit torques in ferromagnet/nonmagnet/ferromagnet trilayers}
\author{V. P. Amin}
\affiliation{Department of Physics, Indiana University, Indianapolis, IN 46202}
\author{G. G. Baez Flores}
\author{A. A. Kovalev}
\author{K. D. Belashchenko}
\affiliation{Department of Physics and Astronomy and Nebraska Center for Materials and Nanoscience, University of Nebraska-Lincoln, Lincoln, Nebraska 68588, USA}
\date{\today}

\begin{abstract}
Spin-orbit torques in ferromagnet/nonmagnet/ferromagnet trilayers are studied using a combination of symmetry analysis, circuit theory, semiclassical simulations, and first-principles calculations using the non-equilibrium Green’s function method with supercell disorder averaging. We focus on unconventional processes involving the interplay between the two ferromagnetic layers, which are classified into direct and indirect mechanisms. The direct mechanism involves spin current generation by one ferromagnetic layer and its subsequent absorption by the other. In the indirect mechanism, the in-plane spin-polarized current from one ferromagnetic layer ``leaks'' into the other layer, where it is converted into an out-of-plane spin current and reabsorbed by the original layer. The direct mechanism results in a predominantly dampinglike torque, which damps the magnetization towards a certain direction $\FPdl$. The indirect mechanism results in a predominantly fieldlike torque with respect to 
a generally different direction $\FPfl$. Similar to the current-in-plane giant magnetoresistance, the indirect mechanism is only active if the thickness of the nonmagnetic spacer is smaller than or comparable to the mean-free path. Numerical calculations for a semiclassical model based on the Boltzmann equation confirm the presence of both direct and indirect mechanisms of spin current generation. First-principles calculations reveal sizeable unconventional spin-orbit torques in Co/Cu/Co, Py/Cu/Py, and Co/Pt/Co trilayers and provide strong evidence of indirect spin current generation.
\end{abstract} 

\maketitle

\section{Introduction}

The discoveries of the giant magnetoresistance~\cite{PhysRevB.39.4828,PhysRevLett.61.2472} and tunnel magnetoresistance~\cite{Maekawa1982,PhysRevLett.74.3273,PhysRevB.63.054416,Yuasa2004,Parkin2004} effects led to the development of many spintronic devices requiring the control of magnetization. For example, spin transfer torque magnetic random access memories (STT-MRAM) employ the tunnel magnetoresistance  to read out the state of an individual bit~\cite{Kent2015}, while writing is done using spin-transfer torque~\cite{Slonczewski1996,PhysRevB.54.9353}. The latter effect relies on the transfer of angular momentum between layers and requires a relatively large charge current flow perpendicular to the layer planes. Later, it was discovered that the magnetization of a ferromagnetic layer can also be controlled by running an in-plane charge current through an adjacent nonmagnetic layer with sufficient spin-orbit interaction---this process is called \emph{spin-orbit torque} \cite{ManchonRMP2019}. Like spin-transfer torques, spin-orbit torques allow all-electrical control of magnetization while offering additional advantages, such as higher endurance and the potential for better power efficiency \cite{Ramaswamy2018,Krizakova2022}.

Spin-orbit torques in magnetic nanostructures are attributed to several mechanisms, including the spin-Hall effect~\cite{RevModPhys.87.1213,PhysRevLett.106.036601,Liu2012,PhysRevLett.109.096602,Zhu2021}, Rashba-Edelstein effect~\cite{Chernyshov2009,MihaiMiron2010,Miron2011,Manchon2015}, orbital Hall effect~\cite{PhysRevResearch.2.033401,PhysRevResearch.2.013177}, planar Hall effect~\cite{Safranski2018,PhysRevLett.124.197204}, magnetic spin-Hall effect~\cite{SAHE,PhysRevResearch.2.023065}, and interfacial spin current generation~\cite{Amin2020,PhysRevB.101.020407}. Unfortunately, the multitude of mechanisms responsible for spin-orbit torques lead to difficulties in interpreting experimental results. On the other hand, understanding these mechanisms is crucial to optimize the switching efficiency of spin-orbit torque-based devices. Further, efficient field-free magnetization switching in layers with perpendicular magnetization requires additional modifications, such as employing systems with lower structural \cite{Yu2014} or crystallographic ~\cite{MacNeill2016,Liu2021} symmetry or ferromagnetic trilayers~\cite{Humphries2017,Baek2018}. Field-free switching has already been demonstrated in ferromagnetic trilayers~\cite{Baek2018}.

The spin-diffusion model~\cite{PhysRevB.53.6554,PhysRevB.66.224424,Barnas.Fert.eaPRB2005} is often sufficient for identifying various mechanisms of spin-orbit torque~\cite{PhysRevLett.121.136805,PhysRevB.94.104419,PhysRevB.94.104420}. However, as we explain below, the interpretation of certain experimental \cite{Humphries2017} and theoretical \cite{Freimuth2018} results for ferromagnetic trilayers within the spin-diffusion model requires an implausibly large imaginary part of the spin-mixing conductance.
On the other hand, it is well understood that the current-in-plane giant magnetoresistance effect involves interlayer scattering with a length scale on the order of a mean free path, a process beyond the spin-diffusion model~\cite{Tsymbal2001}. The interplay between such processes and spin-orbit coupling and their contributions to spin-orbit torques in ferromagnetic trilayers are the subject of the present paper.

In this work, we classify and calculate spin-orbit torques in ferromagnet/normal metal/ferromagnet trilayers that have no analogue in nonmagnet/ferromagnet bilayers, including those not describable by spin diffusion models. We refer to such torques as \emph{nonlocal} 
because they require coupling between the two ferromagnetic layers via spin currents. We identify two previously unrecognized phenomena due to nonlocal torques. First, we find a disorder-dependent torque that exceeds typical torques in heavy metal/ferromagnet bilayers when in the parallel magnetization configuration but is strongly reduced in the antiparallel configuration. We argue this behavior is the spin-orbit torque analogue of the current-in-plane giant magnetoresistance and call it \emph{giant magnetotorquance}. Second, when the magnetizations of the two ferromagnetic layers are orthogonal to each other, we identify unconventional fieldlike torques generated by the interplay of spin-orbit coupling and interlayer scattering. 

In general, the nonlocal torques may be categorized into \emph{direct} and \emph{indirect} mechanisms. In the direct mechanism, one ferromagnetic layer emits a spin current than is absorbed by the other ferromagnetic layer. In the indirect mechanism, the in-plane spin-polarized current from one ferromagnetic layer ``leaks'' into the other ferromagnetic layer, where it is converted into an out-of-plane spin current and reabsorbed by the original layer. 

The main results of this paper are the following:

\begin{enumerate}
\item   A symmetry-based characterization of spin-orbit torques in ferromagnetic trilayers.
\item   A classification of the physical mechanisms of nonlocal spin-orbit torques, including those beyond the spin-diffusion model.
\item   Semiclassical calculations qualitatively illustrating nonlocal spin-orbit torques.
\item   Ab-initio calculations showing that nonlocal spin-orbit torques in ferromagnetic trilayers have comparable strength to conventional spin-orbit torques in bilayers.
\end{enumerate}

The paper is organized as follows. Section \ref{sec:background} gives a brief background on spin-orbit torques in both nonmagnet/ferromagnet bilayers and ferromagnetic trilayers. Section \ref{SectOverview} characterizes the spin-orbit torques from the symmetry point of view. Section \ref{mechanisms} introduces the direct and indirect nonlocal mechanisms and uses magnetoelectronic circuit theory to deduce how these mechanisms contribute to spin-orbit torques. Section \ref{semiclassical} presents semiclassical calculations of nonlocal spin-orbit torques, and Section \ref{SectFP} reports on first-principles calculations of these torques. The paper is summarized in Section \ref{conclusions}.

\begin{figure*}
    \centering
    \includegraphics[width=\textwidth]{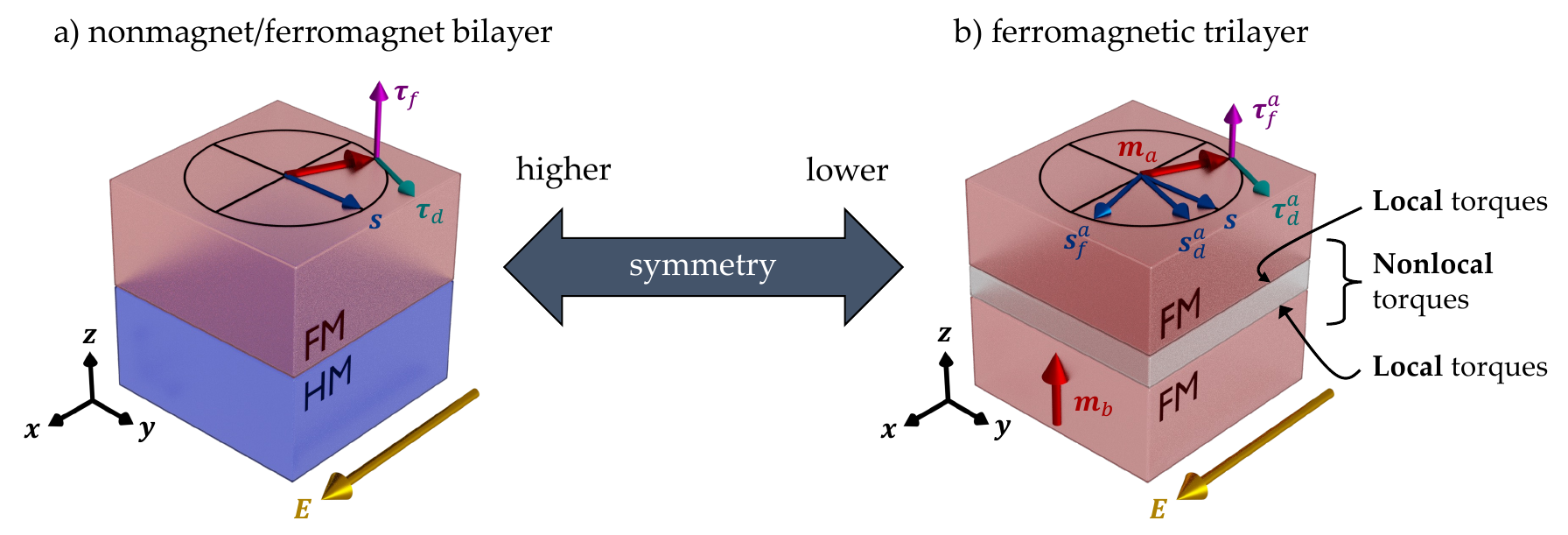}
    \caption{Spin-orbit torques in bilayers and trilayers. (a) In nonmagnet/ferromagnet bilayers, symmetry guarantees the dampinglike torque $\Taud \propto \mhat \times (\FP \times \mhat)$ and fieldlike torque $\Tauf \propto \mhat \times \FP$ vanish when $\mhat ~||~\FP$, where $\FP = \zhat \times \Ehat$. (b) For trilayers, additional symmetries are broken such that the dampinglike and fieldlike torques are defined relative to individual fields $\FPdla$ and $\FPfla$ respectively, such that $\Tauad \propto \mhat_\A \times (\FPdla \times \mhat_\A)$ and $\Tauaf \propto \mhat_\A \times \FPfla$ where $a \in [1,2]$ labels the layer. The fields $\FPdla$ and $\FPfla$ can point along any direction in general (including out-of-plane). Some of the physical mechanisms that determine $\FPdla$ and $\FPfla$ arise from spin transport between the ferromagnetic layers that cannot be captured by spin diffusion theory, as we show in sections \ref{mechanisms} and \ref{semiclassical}.}
    \label{FixedPoints}
\end{figure*}

\section{Background}
\label{sec:background}

\emph{Heavy Metal/Ferromagnet Bilayers}---The first experimental investigations of spin-orbit torques were conducted in heavy metal/ferromagnet bilayers to leverage the strong spin-orbit coupling in the heavy metal in close proximity to the magnetization of the ferromagnet. For these systems, spin-orbit torques are ideally suited to switch the magnetization between the in-plane directions $\pm\FP$, where $\FP = \zhat \times \Ehat$. This is because symmetry mandates that the torque vanishes when the magnetization direction ($\mhat$) is parallel to $\pm\FP$, where for $\mhat \parallel \FP$ there is a stable equilibrium and for $\mhat \parallel -\FP$ there is an unstable equilibrium. For this reason, we refer to $\FP$ as the \emph{fixed point}. This makes deterministic switching via spin-orbit torque ideal in systems with in-plane magnetic anisotropy but impossible in systems with perpendicular magnetic anisotropy, unless an effective magnetic field is used to lower the symmetry of the system and shift the fixed point.

Early theoretical descriptions attributed spin-orbit torques to two mechanisms: 1) the spin Hall effect in the heavy metal generates spin current that flows into the ferromagnet and exerts a spin transfer torque~\cite{RevModPhys.87.1213,PhysRevLett.106.036601,Liu2012,PhysRevLett.109.096602,Zhu2021}; 2) the Rashba-Edelstein effect \cite{ALG1989,Edelstein1990,Ganichev-handbook} generates spin accumulation at the heavy metal/ferromagnet interface that exerts an exchange torque on the magnetization~\cite{Chernyshov2009,MihaiMiron2010,Miron2011,Manchon2015}. However, over the last several years the number of known spin-orbit torque mechanisms has exploded. In addition to the spin Hall and Rashba-Edelstein mechanisms, spin-orbit torques are now attributed to interfacial effects (such as interface-generated spin currents)~\cite{PhysRevMaterials.3.011401,Amin2020,PhysRevB.101.020407}, bulk effects in the nonmagnet (such as the orbital Hall effect~\cite{PhysRevResearch.2.033401,PhysRevResearch.2.013177}), and bulk effects in the ferromagnet (orbital Hall, spin and anomalous Hall, planar Hall, and magnetic spin Hall effects~\cite{SAHE,PhysRevResearch.2.023065}). 

Experimentally distinguishing the various mechanisms is difficult, because clear signatures of each mechanism (such as their magnetization dependence) are not known. For example, given a fixed electric field direction, the sign of the spin-orbit torque that enables magnetization switching (i.e., the dampinglike torque) often directly corresponds to the sign of the heavy metal's spin Hall conductivity, suggesting that the magnetization switching is dominated by the spin Hall contribution. However, there is no reason why other mechanisms, such as orbital Hall effects or interface-generated spin currents, cannot also switch sign based on the choice of the heavy metal. More theoretical and experimental work is required to determine which of the wide variety of mechanisms dominate spin-orbit torques in bilayers. 

\emph{Ferromagnetic Trilayers}---Over time, investigations of spin-orbit torque turned to materials with lower symmetry than conventional heavy metal/ferromagnet bilayers, including two-dimensional materials like transition metal dichalcogenides \cite{MacNeill2016}, magnetic heterostructures with additional structural asymmetry ~\cite{Yu2014,MacNeill2016,Liu2021}, and ferromagnetic trilayers~\cite{Taniguchi2015,Humphries2017,Baek2018}. In trilayers, which consist of two ferromagnetic layers sandwiching a nonmagnetic spacer, the magnetizations of the two ferromagnetic layer need not be collinear; thus, these systems have lower symmetry than bilayers owing to the symmetries broken by the additional magnetization. This lower symmetry leads to less constrained, qualitatively different torques. In particular, the torque on one ferromagnetic layer with magnetization direction $\mhat$ need not vanish when $\mhat = \FP$.
In general, the net torque on a given ferromagnetic layer in trilayers need not vanish for \emph{any} magnetization direction, meaning that no fixed point exists.

Humphries \emph{et al.} \cite{Humphries2017} reported sizable but unconventional torques acting on the permalloy (Py) layer in a Py/Cu/PML trilayer, where PML is a perpendicularly magnetized multilayer. These torques were unconventional because they vanish for $\mhat_\text{Py} \parallel \mhat_\mathrm{PML}\times\FP$ instead of $\mhat_\text{Py} \parallel \FP$. Thus, the fixed point of the Py layer in the trilayer was ``rotated' compared with a bilayer and depended on the magnetization direction of the PML layer. Further, Baek \emph{et al.} \cite{Baek2018} demonstrated field-free switching of a perpendicularly magnetized top CoFeB layer by similarly-defined unconventional torques in a trilayer system. In both experiments, the dampinglike torque can be explained by the emission of a spin current by the PML layer that flows out-of-plane with a ``spin-rotated''  polarization given by $\mhat_\mathrm{PML}\times\FP$ \cite{PhysRevLett.121.136805}. Such unconventional spin current generation is allowed in ferromagnets but not in nonmagnets due to the lower symmetry of ferromagnets.

Of particular note for us here is that the spin-rotated fieldlike torque [i.e., the torque $\propto\mhat_\text{Py} \times (\mhat_\mathrm{PML}\times\FP)$] was even larger than the spin-rotated dampinglike torque in the system studied in Ref. \cite{Humphries2017}.
Similarly, a large spin-rotated fieldlike torque, much larger than the spin-rotated dampinglike torque, was found in first-principles calculations \cite{Freimuth2018} for Co/Cu/Co trilayers using the Kubo linear response technique with eigenvalue broadening incorporated in the Green's function. Within the spin-diffusion theory, the existence of a large spin-rotated fieldlike torque requires an unusually large imaginary part of the spin-mixing conductance \cite{Brataas2000,Brataas2006}.

Thus, there is both experimental \cite{Humphries2017} and theoretical \cite{Freimuth2018} evidence of a strong spin-rotated fieldlike torque which has no satisfactory explanation within the conventional theory. 
Understanding the origin of this torque is a key motivation for the present work.

\section{Symmetry-based characterization of spin-orbit torques}
\label{SectOverview}

Symmetry is a powerful tool to determine the allowed orientations of spin-orbit torques in any material system. In an axially symmetric nonmagnet/ferromagnet bilayer, symmetry constrains the spin-orbit torque $\torque$ as follows:
\begin{align}
\label{TorqueBilayer}
    \torque &= \taud\, \mhat \times (\FP\times\mhat) + \tauf\, \mhat \times \FP + \tilde\torque(\mhat)
\end{align}
where $\FP = \zhat \times \Ehat$, $\zhat$ points out-of-plane, and $\Ehat$ is a unit vector pointing along the electric field. The most important consequence of symmetry is that $\torque$ vanishes when $\mhat~||~\FP$. Fig.~\Ref{FixedPoints}a illustrates the relevant torque directions and symmetry-determined direction $\FP$ in heavy metal/ferromagnet bilayers. The first two terms are called the \emph{dampinglike} and \emph{fieldlike} torques respectively, because the former damps the magnetization towards the fixed point while the latter incites precession about the axis parallel to the fixed point. In the language of vector spherical harmonics, the dampinglike and fieldlike terms are real vector spherical harmonics $\mathbf{Z}^{(\nu)}_{1,-1}(\mhat)$ with $l=1$ of type $\nu=1$ and 2, respectively \cite{PhysRevB.101.020407}. The third term $\tilde\torque(\mhat)$ includes higher-order harmonics with $l \geq 2$.

In ferromagnetic trilayers, the presence of one ferromagnetic layer lowers the symmetry of the spin-orbit torque on the other ferromagnetic layer (and vice-versa). In general, the spin-orbit torque on one ferromagnetic layer depends on the magnetization of \emph{both} layers ($\mhat_\A$ and $\mhat_\B$). In the following, our convention will be to use the index $\A$ for the top layer and $\B$ for the bottom layer (see Fig.~\Ref{FixedPoints}b). For example, given a generic orientation of $\mhat_\A$ and $\mhat_\B$, symmetry allows the torque on layer $\A$ to be written as
\begin{align}
\label{TorqueTrilayer}
    \torque_\A &= \mhat_\A \times (\FPdla(\mhat_\B)\times\mhat_\A) \nonumber \\ 
    &+ \mhat_\A \times \FPfla(\mhat_\B) + \tilde\torque_\A(\mhat_\A,\mhat_\B),
\end{align}
A convenient feature of this description is that the spin-orbit torque $\torque_a$ can still be described by dampinglike and fieldlike torques, as seen by the first two terms of \Eq{TorqueTrilayer}, which span the same space as a general linear combination of vector spherical harmonics $\mathbf{Z}^{(\nu)}_{1,m}(\mhat)$. In ferromagnetic trilayers, the dampinglike and fieldlike torques on layer $\A$ are defined relative to the fields $\FPdla$ and $\FPfla$, which have the following properties:
\begin{enumerate}
\item They depend on material properties of the entire trilayer
\item They only depend on the \emph{other} layer's magnetization direction (i.e. $\FPdla$ and $\FPfla$ are functions of $\mhat_\B$ and not $\mhat_\A$)
\item They point in different directions (i.e. $\FPdla \nparallel \FPfla$) in general
\end{enumerate}
The last property is in contrast to nonmagnet/ferromagnet bilayers, where, in this language, $\FPdla = \FPfla = \FP$. Thus, in trilayers, the dampinglike torque damps $\mhat_\A$ towards $\FPdla$ while the fieldlike torque incites precession about a generally different direction $\FPfla$.

The third term $\tilde\torque_{a}(\mhat_\A,\mhat_\B)$ includes all terms described by real vector spherical harmonics with $l \geq 2$. Our ab-initio calculations suggest that $l \geq 2$ terms, which have a qualitatively important effect on current-induced damping but are small compared to the dampinglike torque in bilayers \cite{PhysRevB.101.020407}, remain small in trilayers.

\begin{figure*}
    \centering
    \includegraphics[width=\textwidth]{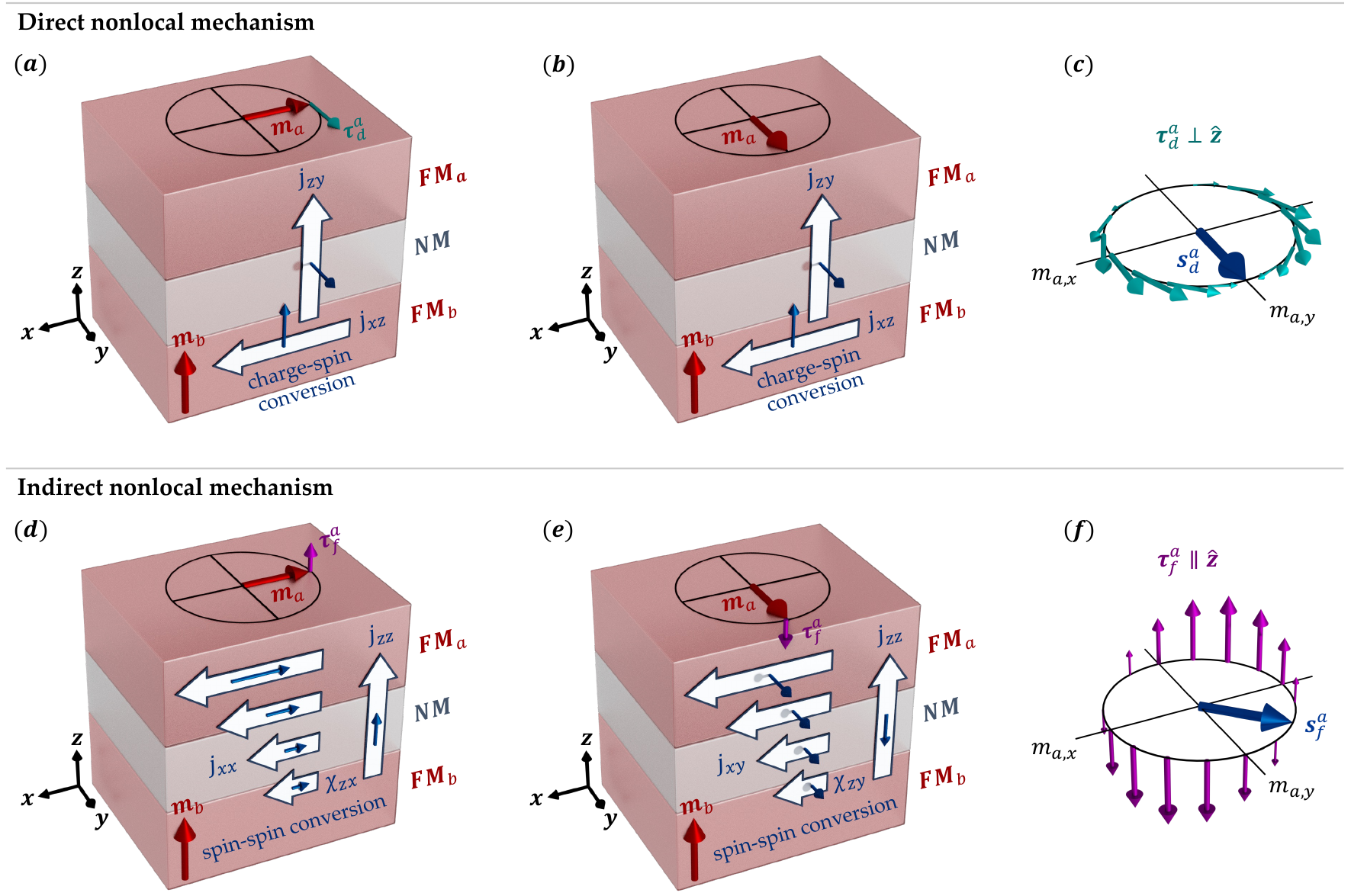}
    \caption{Depiction of nonlocal torques on the top ferromagnetic layer ($\text{FM}_\two$) when $\ImGm = 0$ for both interfaces. Panels (a)-(c) describe the direct nonlocal mechanism. The in-plane charge current in $\text{FM}_\one$ or near its interfaces results in an out-of-plane spin current. This spin current flows into $\text{FM}_\two$ and exerts a spin transfer torque. Panels (a) and (b) depict the torque (green arrow) for two different directions of $\mhat_\two$. The torque vanishes when $\mhat_\two$ points along the emitted spin current's polarization. (c) Spin torques from the direct nonlocal mechanism for the in-plane $\mhat_\two$ directions. The special field, here given by $\FPdltwo = \yhat$, points along the magnetization direction for which the dampinglike torque vanishes. Panels (d)-(e) describe the indirect nonlocal mechanism. The in-plane spin-polarized current in $\text{FM}_\two$ is also present in $\text{FM}_\one$ with reduced magnitude if the spacer layer thickness is comparable to its mean free path. The spin-spin conversion tensor $\chi_{zi}$ for $i \in {x,y,z}$, which captures processes such as the spin-orbit precession effect and/or spin swapping, relates the in-plane spin current $j_{xi}$ to an out-of-plane spin current $j_{zz}$. The out-of-plane spin current flows into $\text{FM}_\one$ and exerts a spin transfer torque. Unlike the direct nonlocal mechanism, the indirectly-generated spin current depends on both layer's magnetizations and, as shown in section \ref{mechanisms}, results in a primarily fieldlike torque.  (f) Spin torques from the indirect nonlocal mechanism for the in-plane $\mhat_\two$ directions. The direction that the fieldlike torque vanishes, i.e. $\FPfltwo$, can point in a different direction to $\FPdltwo$, owing to the fact that the spin-spin conversion does not have the same angular dependence as the direct charge-spin conversion.
    }
    \label{SayNoToImG}
\end{figure*}

The directions of $\FPdla$ and $\FPfla$ can be experimentally determined by measuring all spin-orbit torque components while rotating one layer's magnetization and leaving the other layer's magnetization fixed. The magnetization direction in which the dampinglike torque vanishes and the fieldlike torque vanishes determine $\FPdla$ and $\FPfla$ respectively. Note that experimentally determining the underlying physical mechanisms likely requires varying other material parameters in addition.

Symmetry-constrained definitions, based, for example, on the vector spherical harmonics (Appendix \ref{app:angular}), provide a representation for the special fields, which can also be used to infer the underlying physical mechanisms. Using hindsight from the results of \emph{ab initio} calculations, we expand them up to second order in the components of $\mhat_\B$, keeping only the terms that survive symmetrization with respect to the $C_{\infty v}$ group:
\begin{widetext}
\begin{align}
    \FPdla(\mhat_\B) &= \dsa\FP + \dxa(\mhat_\B\times\FP) + (\dma - \dsa)(\mhat_\B\cdot\FP)\mhat_\B + \dDa \hat \beta^D \mhat_\B + 4\dtwoa (m_{\B z})^2\FP + 4\dza m_{\B z}(\mhat_\B\cdot\FP)\hat z \label{FPdl}\\
    \FPfla(\mhat_\B) &= \fsa\FP + \fxa(\mhat_\B\times\FP) + (\fma - \fsa)(\mhat_\B\cdot\FP)\mhat_\B + \fDa \hat \beta^D \mhat_\B + 4\ftwoa (m_{\B z})^2\FP + 4\fza m_{\B z}(\mhat_\B\cdot\FP)\hat z.    \label{FPfl}
\end{align}
\end{widetext}
The $\hat\beta^D$ matrix is defined by its nonzero elements $\beta^D_{xz}=\beta^D_{zx}=1$. Additional terms are possible in a single-crystal sample, but they are not needed to fit the results of our first-principles calculations, and they are absent in typical polycrystalline samples studied in experiments.

The parameters appearing in \Eqsd{FPdl}{FPfl} are determined by the material system. As explained in Section \ref{connection-mechanisms}, each term in \Eqsd{FPdl}{FPfl} can be associated with one or more physical mechanisms. The terms in \Eq{FPdl}, which contribute to the dampinglike torque on $\mhat_\A$, are most simply associated with spin currents directly generated in layer $\B$, where the terms themselves are the various spin polarizations of those spin currents. Such spin currents include bulk spin currents from the spin Hall, spin anomalous Hall, and magnetic spin Hall effects and interface-generated spin currents. The terms in \Eq{FPfl} are associated with the indirect mechanism of spin current generation, which is described in the following section.

\section{Physical mechanisms of nonlocal spin-orbit torques}
\label{mechanisms}

In this section, we discuss two \emph{nonlocal} mechanisms by which the ferromagnetic layers influence each other (depicted in \Fig{SayNoToImG}). Then, we explain how these mechanisms contribute to the fields $\FPdla$ and $\FPfla$. Finally, we connect all the symmetry-allowed terms in $\FPdla$ and $\FPfla$ to possible microscopic mechanisms.

\subsection{Definitions}

Any physical mechanism responsible for differentiating $\FPdla$ and $\FPfla$ from $\FP$ and/or from each other has no analogue in nonmagnet/ferromagnet bilayers. We identify two broadly applicable nonlocal mechanisms):

\begin{enumerate}
\item \emph{Direct spin current generation}---An out-of-plane spin current generated in $\text{FM}_\B$ or its interfaces flows into $\text{FM}_\A$ and exerts a spin transfer torque (illustrated in \Fig{SayNoToImG}a-b). This well-studied mechanism leads to qualitatively different spin-orbit torques than in nonmagnet/ferromagnet bilayers because the spin current emitted by layer $\B$ can have spin components forbidden by symmetry in bilayers. This mechanism primarily results in dampinglike torques, where the field $\FPdla(\mhat_\B)$ is parallel to the spin polarization of the emitted spin current from layer $\B$ (shown in \Fig{SayNoToImG}c).

\item \emph{Indirect spin current generation}---Since spin-orbit torques are driven by an in-plane electric field, an in-plane spin polarized current forms in each ferromagnetic layer with a spin polarization antiparallel to its magnetization. If the nonmagnetic spacer layer thickness is similar to its mean free path, then the in-plane spin polarized current from each ferromagnetic layer is also present (with reduced magnitude) in the other layer. This "leakage" process is illustrated in \Fig{SayNoToImG}d-e, where the horizontal block arrows depict the in-plane spin-polarized currents leaked from $\text{FM}_\A$. What results is a two-step mechanism not present in bilayers: 1) the in-plane spin-polarized current from layer $\A$ leaks into layer $\B$, where it is converted (via the spin-orbit precession effect \cite{PhysRevLett.121.136805} or spin swapping \cite{PhysRevLett.103.186601,PhysRevLett.120.176802}) into an out-of-plane spin current; 2) the out-of-plane spin current flows back to layer $\A$ and exerts a spin transfer torque. Unlike the direct mechanism, the spin current emitted from layer $\B$ has a spin polarization that depends strongly on $\mhat_\A$.
As described in detail below, this results in fieldlike torques on layer $\A$ (shown in \Fig{SayNoToImG}f), which may be strong.
\end{enumerate}

In experimental systems, we expect that the special fields can be misaligned. If we ignore for a moment the traditional ``local'' spin-orbit torques (for example, disregarding the spin-orbit coupling on the given layer), then we expect, to a good approximation, that the special fields are only misaligned when both the direct and indirect nonlocal mechanisms are simultaneously present. Including the local spin-orbit torque contributions, the special fields may be misaligned in general. However, as we show below, the indirect nonlocal torque can cause strong fieldlike torques where $\FPfla \nparallel \FP$, even if the imaginary part of the spin mixing conductance is small. Thus, one plausible experimental signature of the indirect nonlocal mechanism is a fieldlike torque with the so-called \emph{spin-rotation symmetry}, which corresponds to $\FPfla(\mhat_\B) = \mhat_\B \times \FP$.

Before proceeding, we define the notation and terms used throughout this section. Spin currents are generally specified by the tensor elements $Q_{i\sigma}$, where $i$ denotes the flow direction and $\sigma$ denotes the spin direction ($i,\sigma \in [x,y,z]$). In what follows, \emph{out-of-plane spin current} and \emph{in-plane spin current} refer to flow direction and not spin direction. To denote spin currents, we use the notation $[\threevec{j}_i]_\sigma = (2/\hbar) Q_{i\sigma}$. For example, $\threevec{j}_z$ is a vector with components $(2/\hbar) Q_{z\sigma}$ whose magnitude equals the flux of spin angular momentum flowing out-of-plane and whose direction points along the spin polarization of the spin current. Finally, we use \emph{transverse} to denote spin orientations transverse to the \emph{magnetization} of the FM layer that receives the torque (always layer $\A$ unless otherwise noted).

\subsection{Circuit theory with injected spin currents}
\label{sec-circuit}

Interfaces can generate an out-of-plane spin current under an applied in-plane electric field through various mechanisms including spin filtering and spin precession \cite{PhysRevB.94.104420,PhysRevB.94.104419}. In this section we set up the equations for finding the resulting spin currents and torques within the magnetoelectronic circuit theory \cite{Brataas2000,Brataas2006,Flores2020}. In subsequent Sections \ref{sec-direct} and \ref{sec-indirect} we solve these equations for different kinds of mechanisms generating the spin current. Our main interest below will be in spin torques appearing when the thickness of the nonmagnetic layer $t_\mathrm{NM}$ is smaller than or comparable to the mean-free path $\lambda_\mathrm{NM}$. Although the circuit theory is formally inapplicable in this regime, we will see that it provides useful insight; we also expect its predictions should be approximately realized at $t_\mathrm{NM}\sim\lambda_\mathrm{NM}$.

\begin{figure}
    \centering
    \includegraphics[width=0.7\columnwidth]{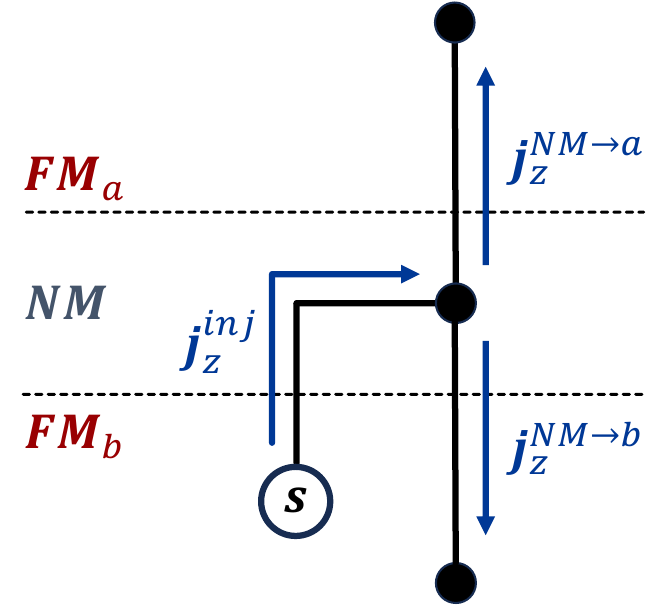}
    \caption{
    Magnetoelectronic circuit corresponding to the trilayer. Black circles represent nodes in the circuit. Dashed lines represent contacts, which include interfaces along with the adjacent portions of bulk regions (see text). The white circle represents the source that creates the injected spin current.  This spin current source is an additional feature beyond the conventional magnetoelectronic circuit theory which describes out-of-plane spin currents created by in-plane electric fields, as described in \cite{PhysRevB.94.104419,PhysRevB.94.104420}.
    }
    \label{MCT-graph}
\end{figure}

To avoid having to solve for the spatially resolved spin accumulation using spin-diffusion equations, we exploit the exact correspondence between the scattering theory and the spin-diffusion model established in Ref. \onlinecite{Flores2020}. Assuming the ferromagnetic layer thicknesses are much larger than the spin-diffusion length, we place the circuit theory nodes a few spin-diffusion lengths away from the interfaces inside these layers; the charge and spin accumulation in those nodes, therefore, vanishes. For the nonmagnetic layer, we introduce only one node somewhere in its middle. The circuit theory thus includes three nodes and two contacts, as shown in Fig. \ref{MCT-graph}. Each contact represents an interface along with the adjacent layers extending up to the nodes. With this formulation of the problem, there is no spin relaxation in the nodes. The unknown quantity is the spin accumulation $\boldsymbol{\mu}_s$ in the nonmagnetic node.

The circuit theory operates with spin currents flowing from nodes into contacts due to the deviations of the spin potentials from equilibrium. We will refer to such spin currents as \emph{backflow} spin currents and denote them as $\threevec{j}^{\mathrm{NM}\to\A}_z$ and $\threevec{j}^{\mathrm{NM}\to\B}_z$ (see Fig. \ref{MCT-graph}). Spin currents generated by interfaces act as sources produced directly by the applied electric field and are finite even when all charge and spin accumulations vanish. We will call such spin currents \emph{injected} spin currents. From the point of view of the circuit theory, an injection current flows into a given node, and it does not matter where it originates. We denote the spin current injected into the nonmagnetic node as $\threevec{j}^{\mathrm{inj}}_z$, although the subscript $z$ is superfluous within the circuit theory. Generally, the effects of spin currents injected into different nodes are additive, because the circuit theory equations are linear. Note that spin currents are not conserved in general when passing through layers and interfaces; the circuit shown in \Fig{MCT-graph} is meant to depict the magnetoelectronic circuit theory itself and should not be interpreted as spin current conservation through the layers and interfaces.

The boundary condition for the $\text{FM}_\B/\text{NM}$ contact gives the backflow spin current flowing from the nonmagnetic node into that contact:
\begin{align}
    \label{CTb}
    \threevec{j}^{\mathrm{NM}\to\B}_z
    &=\ReGmb \mhat_\B \times (\boldsymbol{\mu}_s
    \times \mhat_\B) \nonumber \\
     &+\ImGmb\mhat_\B \times \boldsymbol{\mu}_s
    \nonumber \\
     &+(\Gcb/2) (\boldsymbol{\mu}_s
    \cdot \mhat_\B) \mhat_\B
\end{align}
where $\Gcb$ and $G^{\B}_{\uparrow\downarrow}$ are the effective interfacial charge conductance and spin-mixing conductance, respectively. Equation (\ref{CTb}) gives the most general form of a boundary condition that does not depend on the orientation of the magnetizations and the spin accumulation relative to the plane of the interface. In particular, the effective conductance $\Gcb$ includes a contribution from an isotropic spin-memory loss conductance \cite{Flores2020}. The condition of isotropy with respect to the surface orientation can be relaxed for the FM$_\B$/NM contact without materially changing the conclusions.
An expression identical to Eq. (\ref{CTb}) with the index $\B$ replaced by $\A$ applies to the backflow spin current flowing from the nonmagnetic layer into the
FM$_\A$/NM contact.

The spin currents flowing into and out of the nodes are related by the spin Kirchhoff rule \cite{Brataas2000,Brataas2006}. Because there is no spin relaxation in the nonmagnetic node, we have simply
\begin{equation}
    \label{Kirchhoff}
    \threevec{j}^{\mathrm{inj}}_z=\threevec{j}^{\mathrm{NM}\to a}_z+\threevec{j}^{\mathrm{NM}\to b}_z
\end{equation}

We will further assume that spin-orbit coupling has no effect on the transport of spins across the \emph{physical} FM$_\A$/NM interface, and that the spin dephasing length for transverse spins in the FM$_\A$ layer is much shorter than the spin-diffusion length. Under these conditions, the total spin torque $\torque_a$ is equal to the part of $\threevec{j}^{\mathrm{NM}\to a}_s$ that is orthogonal to $\mhat_\A$, which is
\begin{align}
    \label{TorqueSC}
    \torque_a = &~\ReGma \mhat_\A \times (\boldsymbol{\mu}_s
    \times \mhat_\A) \nonumber \\
    + &~\ImGma \mhat_\A \times \boldsymbol{\mu}_s.
\end{align}
In addition, the form of Eq. (\ref{CTb}) for the FM$_\A$/NM contact is the most general in this case.

\subsection{Direct spin current generation mechanism}
\label{sec-direct}

\begin{figure}
    \centering
    \includegraphics[width=0.95\columnwidth]{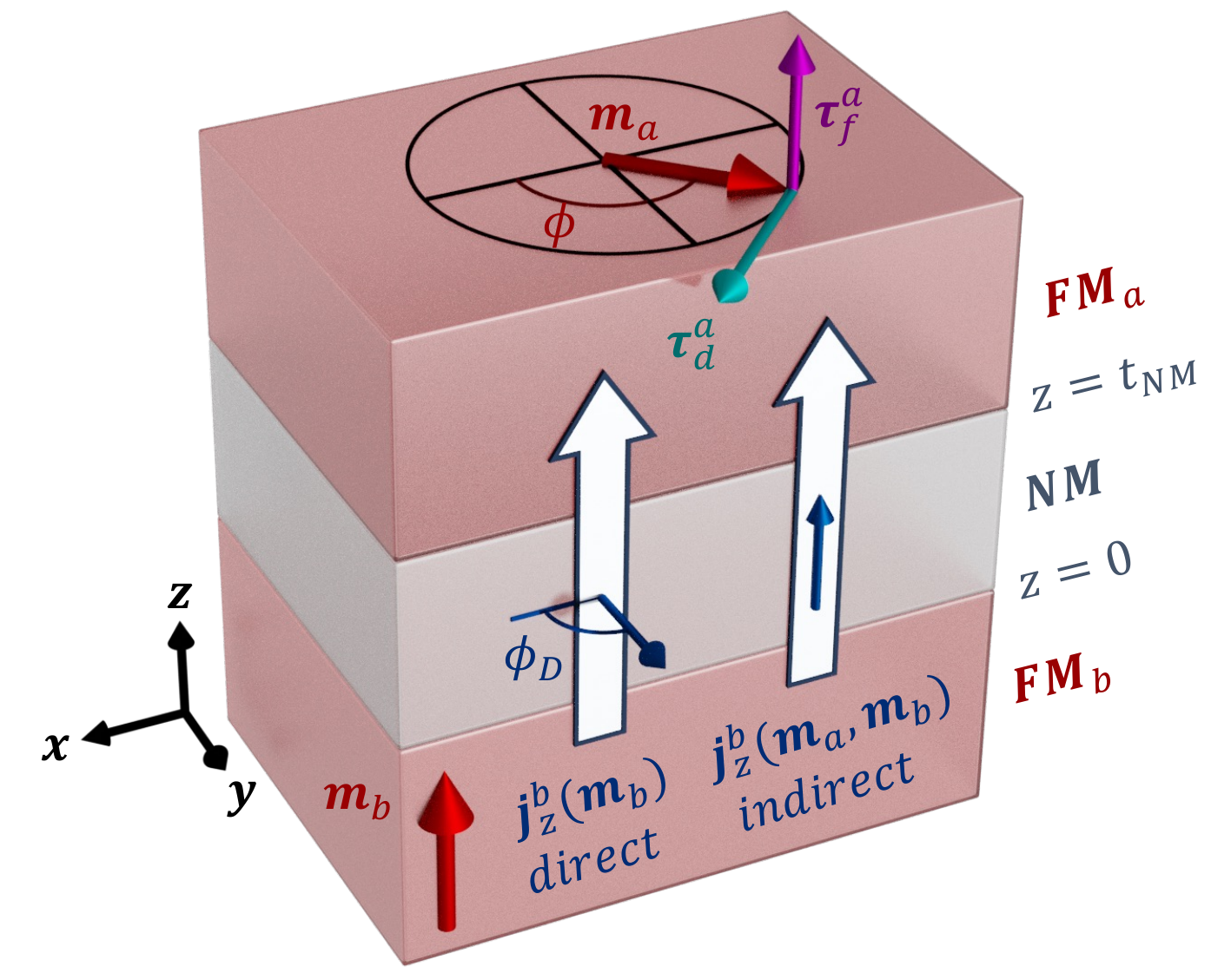}
    \caption{
    Schematic depicting the circuit theory solution for $\tautwodf$, which assumes that $\mhat_\one$ points either along the electric field or out-of-plane and $\mhat_\two$ is swept in the plane perpendicular to $\mhat_\one$ (the figure shows the case when $\mhat_\one \parallel \zhat$, i.e. out-of-plane). The spin current emitted by layer $\one$ or its interfaces is denoted by $\threevec{j}^\one_z$ and arises from both direct and indirect spin current generation. Symmetry arguments mandate that the spin polarization of the direct spin current points along the $\mhat_\two$-plane (where $\phi_D$ gives the polarization angle in that plane) while the spin polarization of the indirect spin current points along $\mhat_\one$. As $\mhat_\two$ is rotated through an angle $\phi$ in the $\mhat_\two$-plane, $\tautwod$ and $\tautwof$ will vanish at different $\phi$ values if both direct and indirect spin current generation are active.
    }
    \label{MCT}
\end{figure}

We now consider the case of direct spin current generation from layer $\B$, i.e., when the spin current $\threevec{j}^\mathrm{inj}_z$ injected into the nonmagnetic layer due to the in-plane electric field is generated exclusively in layer $\B$ or at its interfaces and, therefore, depends only on the orientation of $\mhat_\B$. We will denote $\threevec{j}^\mathrm{inj}_z=\threevec{j}^\B_z(\mhat_\B)$ in this case.
The dependence on $\mhat_\B$ is determined by the underlying mechanism, such as the spin Hall effect, spin anomalous Hall effect, magnetic spin Hall effect, or interface-generated spin current.

To derive an analytical solution from this circuit theory model, we focus on a magnetization configuration with experimental relevance: $\mhat_\B$ points either along the electric field or out-of-plane while $\mhat_\A$ is rotated in the plane perpendicular to $\mhat_\B$ (see Fig. \ref{MCT}). In this magnetization configuration, symmetry dictates that $\FPdla$, $\FPfla$, and $\threevec{j}^\B_z$ lie in the $\mhat_\A$-plane. The dampinglike and fieldlike torques on layer $\A$ are the first and second terms respectively of the following equation:
\begin{align}
\label{TorqueVSHl1}
    \torque_a &= \mhat_\A \times (\FPdla\times\mhat_\A) + \mhat_\A \times \FPfla.
\end{align}
It turns out that the circuit theory solution for these constrained magnetization directions is given exactly by \Eq{TorqueVSHl1}, enabling us to extract $\FPdla$ and $\FPfla$. However, if $\mhat_\B$ points along an arbitrary axis, finding $\FPdla$ and $\FPfla$ requires obtaining $\torque_\A$ for all magnetization directions and fitting the results to \Eq{TorqueTrilayer}. 

Let us further simplify \Eq{TorqueVSHl1}. Since by symmetry both $\FPdla$ and $\FPfla$ lie in the $\mhat_\A$-plane, the dampinglike torque also lies in the $\mhat_\A$-plane while the fieldlike torque is perpendicular to it, according to \Eq{TorqueVSHl1}. Combined with the fact that spin torques are perpendicular to the magnetization when saturated (i.e. $\torque_a \perp \mhat_\A$), the dampinglike torque is constrained along the vector $\mhat_\times \equiv \mhat_\B \times \mhat_\A$ and the fieldlike torque is constrained along $\mhat_\B$. Thus, in describing the torques, it is convenient to use basis vectors given by $\mhat_\A$, $\mhat_\B$, and $\mhat_\times$. In this magnetization-dependent basis, \Eq{TorqueVSHl1} becomes  
\begin{align}
\label{TorqueNewBasis}
    \torque_a &= \tauad(\phi) \mhat_\times + \tauaf(\phi) \mhat_\B
\end{align}
where $\tauadf(\phi)$ is the strength of the dampinglike/fieldlike torque and $\phi$ indicates the direction of $\mhat_\A$ in the $\mhat_\A$-plane, as illustrated in Fig. \ref{MCT}. Solving \Eqsd{CTb}{Kirchhoff} yields dampinglike and fieldlike torques which have the same angular dependence, given by
\begin{align}
    \label{Torquedl}
    \tauadf \propto~& (\ReGmb - \Gca/2) \sin(\phi - \phi_D)\nonumber \\
                            &+ \ImGmb \cos(\phi - \phi_D).
\end{align}
where $\phi_D$ is the angle of the injected spin current polarization $\threevec{j}^\B_z$ in the $\mhat_\A$-plane. It is clear from \Eq{Torquedl} that the torque does not vanish when $\mhat_\A$ points along the injected spin current polarization (i.e. $\phi = \phi_D$) unless $\ImGmb = 0$. Using the definition
\begin{align}
    \label{alphaab}
    \tan(\alpha_{\B\A}) \equiv \frac{\ImGmb}{\Gca/2 - \ReGmb},
\end{align}
\Eq{Torquedl} can be rewritten more compactly as
\begin{align}
    \label{TorquedlD}
    \tauadf \propto \sin(\phi - \phi_D - \alpha_{\B\A}).
\end{align}
\Eq{TorquedlD} shows that both the dampinglike and fieldlike torques vanish at the same $\mhat_\A$ direction given by $\phi'_D \equiv \phi_D + \alpha_{\B\A}$. Since $\FPdla$ and $\FPfla$ point along the magnetization direction of vanishing torque, the angle $\phi'_D$ gives the direction of $\FPdla$ and $\FPfla$ for the direct mechanism.

To first order in the spin mixing conductances of both interfaces, the ratio between the fieldlike and damplinglike torques is given by
\begin{align}
    \label{ratioDL}
    \frac{\tauaf}{\tauad} = -\frac{\Gcb\tan(\alpha_{\A\B})}{2\ReGma}
\end{align}
We can interpret \Eqsd{alphaab}{ratioDL} as follows. The quantity $\alpha_{\B\A}$ equals the separation angle between the injected spin current polarization $\threevec{j}^\B_z$ and the special fields $\FPdla$ and $\FPfla$ (where $\FPdla \parallel \FPfla$). When $\ImGmb = 0$, the separation angle $\alpha_{\B\A}$ vanishes and the injected spin current polarization aligns with the special fields, i.e. $\threevec{j}^\B_z \parallel \FPdla \parallel \FPfla$. On the other hand, the quantity $\alpha_{\A\B}$ determines the ratio of the fieldlike to dampinglike torque on $\mhat_\A$. The fieldlike torque vanishes when $\ImGma = 0$ (since $\tan(\alpha_{\A\B}) \propto \ImGma$). Thus, in the direct nonlocal mechanism, the imaginary part of the spin mixing conductance of each interface plays a unique role in determining the torque.

\begin{figure*}
    \centering
    \includegraphics[width=0.875\textwidth]{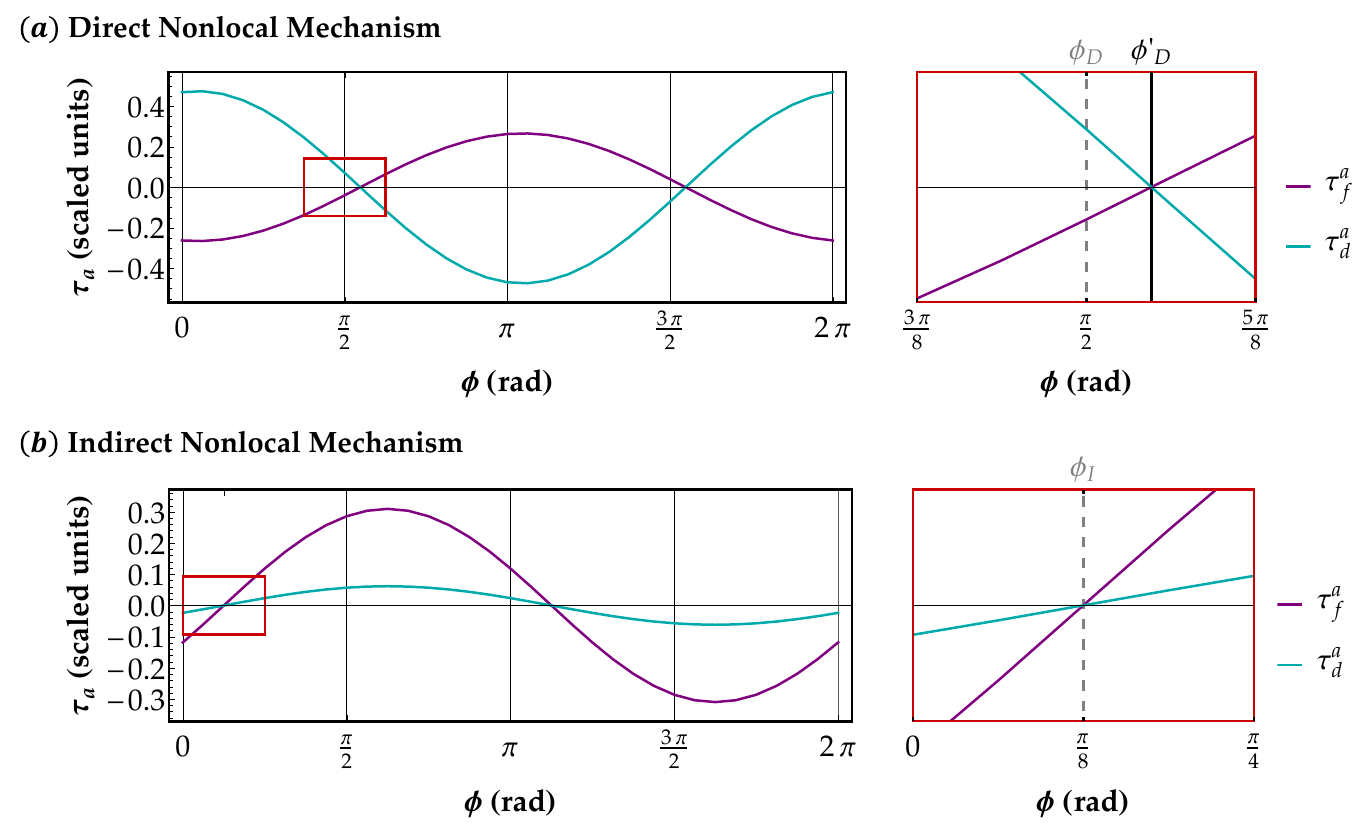}
    \caption{
    (a) Direct nonlocal mechanism. Plot of the dampinglike ($\tauad$) and fieldlike ($\tauaf$) torques on layer $\A$ as a function of magnetization angle $\phi$, obtained from magnetoelectronic circuit theory alone (solid lines) and from the spin diffusion model (dashed lines). As seen in the zoom panel (red box), both models show the dampinglike and fieldlike torques vanishing at the same magnetization direction. Thus, for direct nonlocal torques, the special fields $\FPdla$ and $\FPfla$ are parallel, since they point along the magnetization direction with vanishing torque. The circuit theory model predicts a shift between the spin direction of the injected spin current ($\phi_D = \pi/2$) and the direction of the special fields ($\phi'_D = \phi_D + \alpha_{\B\A}$). The spin diffusion model can yield a different shift in direction, but converges to $\phi'_D$ in the limit of infinite diffusion length. (b) Indirect nonlocal mechanism. Plot of the same torque components as in (a). The dampinglike and fieldlike torques vanish for the same magnetization direction ($\phi_I = \pi/8$), but this direction can be different than for the direct mechanism (i.e. $\phi'_D \neq \phi_I$). For vanishing $\ImGm$ at the $\text{NM}/\text{FM}_\A$ interface, $\tauaf = 0$ for the direct mechanism and $\tauad = 0$ for the indirect mechanism. Thus, if $\ImGm$ can be neglected, the direct mechanism determines the direction of $\FPdla$ and the indirect mechanism determines the direction of $\FPfla$.
    }
    \label{MCTDirectIndirect}
\end{figure*}

\subsection{Indirect spin current generation mechanism}
\label{sec-indirect}

A less straightforward mechanism which appears prominently in our ab-initio and semiclassical results occurs due to indirect spin current generation. This mechanism occurs because the in-plane, spin-polarized current from layer $\A$ is present with reduced magnitude in layer $\B$. Via spin-orbit coupling in layer $\B$ or its interfaces, this in-plane spin-polarized current is converted to an out-of-plane spin current. We refer to this out-of-plane spin current as the \emph{indirect spin current} because it is not directly generated from the electric field in layer $\B$ but rather from the electric field in layer $\A$. The remaining process is identical to direct spin current generation: the indirect, out-of-plane spin current flows into layer $\A$ and exerts a spin transfer torque.

We begin by assuming the spin current $\threevec{j}^\mathrm{inj}_z$ is injected into the nonmagnetic layer from layer $\B$
but depends on \emph{both magnetization directions}, i.e. $\threevec{j}^\mathrm{inj}_z=\threevec{j}^\B_z(\mhat_\A,\mhat_\B)$. This is necessary because the in-plane spin-polarized current originally created in layer $\A$ indirectly generates the out-of-plane spin current in layer $\B$, so the indirect spin current must depend on both magnetizations. The indirect spin current source can be written as
\begin{align}
    \threevec{j}^\B_z(\mhat_\A,\mhat_\B) &= \chi(\mhat_\B) \mhat_\A
\end{align}
where $\chi(\mhat_\B)$ is a $3 \times 3$ response tensor that can be separated into its symmetric ($\chi_\text{S}$) and antisymmetric ($\chi_\text{A}$) parts, i.e.
\begin{align}
    \chi(\mhat_\B) &= \chi_\text{S}(\mhat_\B) + \chi_\text{A}(\mhat_\B),
\end{align}
defined by $\chi_\text{S} = \chi_\text{S}^T$ and $\chi_\text{A} = -\chi_\text{A}^T$. We find that the symmetric spin-spin response tensor $\chi_\text{S}$ contributes to torques described by $l = 2$ vector spherical harmonics. Since, according to our ab-initio calculations, the $l = 2$ torques are small, we will focus on contributions from the antisymmetric spin-spin response tensor $\chi_\text{A}$.

Any antisymmetric matrix can be used to describe a cross product operation relative to some vector $\threevec{f} = (f_x, f_y, f_z)$, such that
\begin{align}
\label{chiA1}
    \chi_\text{A} \mhat &= \threevec{f} \times \mhat
\end{align}
for
\begin{align}
\label{chiA2}
    \chi_\text{A} &=
    \begin{pmatrix}
    0           &-f_{z}     &f_{y}    \\
    f_{z}      &0          &-f_{x}     \\
    -f_{y}       &f_{x}      &0          \\
    \end{pmatrix}
\end{align}
Then according to \Eq{chiA1}, the indirectly-generated spin current ejected from layer $\B$ vanishes when $\mhat_\A \parallel \threevec{f}$. In our magnetization configuration ($\mhat_\A \perp \mhat_\B$), symmetry arguments dictate that $\threevec{j}^\B_z(\mhat_\A,\mhat_\B)$ points along $\mhat_\B$, so $\threevec{f}$ must be parallel to the $\mhat_\A$-plane. Using $\phi_I$ to represent the angle in the $\mhat_\A$-plane pointing along $\threevec{f}$, the circuit theory solution for the indirect nonlocal mechanism can be written as:
\begin{align}
    \label{TorquedlI}
    \tauadf \propto \sin(\phi - \phi_I).
\end{align}
Thus, for the indirect mechanism, both the dampinglike and fieldlike torques vanish along the same magnetization direction $\phi_I$. Another way of stating this result is that the contributions to $\FPdla$ and $\FPfla$ from the indirect mechanism point along the vector $\threevec{f}$, which characterizes the spin-spin conversion at the $\text{FM}_\B/\text{NM}$ interface. 

The ratio of the dampinglike to fieldlike torque to first order in $\ImGma$ and $\ImGmb$ is
\begin{align}
    \label{ratioIL}
    \frac{\tauad}{\tauaf} = -\frac{(\Gcb - 2\ReGma)\Gcb\tan(\alpha_{\A\B})}{2\ReGma(\ReGma + \ReGmb)},
\end{align}
where, as a reminder, $\tan(\alpha_{\A\B}) \propto \ImGma$. Unlike the direct mechanism, the dampinglike torque is zero when $\ImGma$ vanishes. Given that $\ImGm/\ReGm$ is considered to be small for most material interfaces of interest, the indirect mechanism produces larger \emph{fieldlike} torques than dampinglike torques.

Thus, we find that when both the direct and indirect mechanisms are active, the special fields point in different directions. If $\ImGma$ can be neglected, then the direct mechanism yields zero fieldlike torque and the indirect mechanism yields zero dampinglike torque. Since direct and indirect spin current generation occur via distinct physical mechanisms, determining the special fields in experimental systems may provide clues about the underlying physical mechanisms of spin-orbit torques in ferromagnetic trilayers.

Since the conditions of circuit theory are not formally met if the spacer thickness is comparable with the mean free path, the complete description of spin transport in this case requires the use of the Boltzmann theory accounting for the details of the nonequilibrium distribution function. This complete treatment will be given below in Section \ref{semiclassical}, which also shows that $\FPdla \nparallel \FPfla$ when both the direct and indirect nonlocal mechanisms are active.

\subsection{Connection to known spin-orbit torque mechanisms}
\label{connection-mechanisms}

Let us discuss the physical meaning of different terms in \Eqs{FPdl}{FPfl}. For the sake of the argument, assume that all torques are linear in spin-orbit coupling, and first consider the torque generated in layer $\A$ by spin-orbit coupling in layer $\B\neq \A$. Also, let us assume that the absorption by layer $\A$ of spin current flowing across the spacer results in dampinglike torque with $\FPdla$ being equal to the polarization of that spin current. This amounts to assuming that the boundary condition is dominated by the real part of the spin-mixing conductance. Then, different terms in \Eq{FPdl} can be attributed to different mechanisms by which spin-orbit coupling in layer $\B$ generates spin currents emitted into the spacer layer. The mechanisms contributing to the first three terms are listed in Table \ref{SCGenTable}.

\begin{table}
\centering
\begin{tabular}{|l|l|}
    \hline
    \bf{Bulk mechanism}             &   \bf{angular dependence}  \\
    \hline
    spin Hall effect                &   $\FP = \zhat \times \Ehat$   \\
    \hline
    spin anomalous Hall effect      &   $(\mhat \cdot \FP)\mhat$    \\
    \hline
    spin swapping                   &   $\mhat \times \FP$   \\
    \hline
    magnetic spin Hall effect       &   $\mhat \times \FP$   \\
    \hline    \hline
    \bf{Interfacial mechanism}      &   \bf{angular dependence} \\
    \hline
    spin-orbit filtering            &   $\FP$   \\
    \hline
    spin-orbit precession           &   $\mhat \times \FP$   \\
    \hline
\end{tabular}
\caption{Angular dependence associated with various spin current generation mechanisms. Spin currents can be defined in general by a conductivity tensor $\sigma_{ijk}$ with spin flow index $i$, spin direction index $j$, and electric field index $k$, where $i,j,k \in [x, y, z]$. Assuming the electric field direction $\Ehat$ is in-plane and the spin flow direction is out-of-plane ($\zhat$), the remaining conductivity tensor elements form a vector $\threevec{p}$ where, for example, if $\Ehat = \xhat$ then $[\threevec{p}]_j = \sigma_{zjx}$. The vector $\threevec{p}$ is sometimes called the spin polarization of the spin current; here we refer to it as the spin direction. The table shows the angular dependence of $\threevec{p}$ for various spin current generation mechanisms as a function of the directions of the electric field $\Ehat$ and magnetization $\mhat$ of the generating layer or interface (if it is ferromagnetic).}
\label{SCGenTable}
\end{table}

The $\dsa$ term represents the spin-Hall effect, which can be generated in any part of the system, including layer $\B$, its interface with the spacer, and the spacer itself. The term with $\dma-\dsa$ has the same structure as the spin anomalous Hall (SAHE) effect \cite{SAHE}, but it can also be generated at the interface between layer $\B$ and the spacer \cite{PhysRevLett.121.136805}. We can interpret $\dsa$ as the effective spin-Hall conductivity when $\mhat_\B$ lies in the $xz$ plane and $\dma$ when it lies along the $\yhat$ axis; in both cases the spin current emitted by layer $\B$ is polarized along the $\yhat$ axis.

The $\fsb$ term is the conventional effective field due to the Rashba-Edelstein effect at the surface of layer $\B$. In contrast, the $\fsa$ term represents the \emph{nonlocal} $\mhat_\B$-independent effective field acting on layer $\A$ thanks to spin-orbit coupling on layer $\B$. Such field can be generated even by a nonmagnetic interface \cite{Geranton2017}. The additional $\mhat_\B$-dependent contribution is captured by the $\fma-\fsa$ term. To our knowledge, such nonlocal $\mhat_\B$-dependent fieldlike torque has not been discussed in the literature.

The $\dxa$ term can arise from the magnetic spin Hall effect (MSHE) \cite{Kimata2019} in layer $\B$ and from the spin-orbit filtering effect \cite{Humphries2017,Baek2018}. The $\fxa$ term describes an effective field with the rotated spin polarization, in the language of Ref.\ \onlinecite{Humphries2017} where it was denoted $h^\mathrm{R}_\mathrm{FL}$. This term was found \cite{Humphries2017} to be larger than $\dxa$ for a PML/Cu/Py trilayer, where PML is a perpendicularly magnetized CoFe/Ni multilayer. The large $\fxa/\dxa$ ratio was surprising because, within the spin-diffusion model, it requires an unrealistically large imaginary part of the spin-mixing conductance at the interface between layer $\A$ and the spacer layer. As we show in the following sections, the large $\fxa$ term can appear due to an indirect spin current generating mechanism facilitated by the crosstalk between the two ferromagnetic layers, which is reminiscent of the CIP-GMR effect and is absent in the conventional spin-diffusion theory.

An unconventional fieldlike torque with an out-of-plane spin orientation was observed in a Mn$_3$Sn/Ni-Fe bilayer and interpreted as a manifestation of MSHE in noncollinear antiferromagnetic Mn$_3$Sn \cite{Kondou2021}. Although our treatment is not directly applicable to this system, the fieldlike character of the torque with an unconventional spin polarization may indicate an indirect spin current generation mechanism.

The last three terms in \Eqs{FPdl}{FPfl} explicitly distinguish in-plane and out-of-plane components of the magnetization, reflecting interfacial anisotropies induced by spin-orbit coupling. As we find below in Section \ref{SectFP}, these terms are negligible in trilayers with a Cu spacer but appreciable with Pt.

\section{Semiclassical Calculations}
\label{semiclassical}

First we describe semiclassical calculations that qualitatively capture both direct and indirect nonlocal spin-orbit torques in ferromagnetic trilayers. Specifically, we solve the spin-dependent Boltzmann equation for a layered system with quantum coherent boundary conditions, which enables us to determine the spin accumulations, spin currents, and spin torques within each layer and at the interfaces. By studying this solution, we develop a model that can partially explain our first principles results.

\subsection{Computational Details}

For simplicity, all layers are described by free electron gases with the same spin-independent spherical Fermi surfaces, though other parameters like momentum relaxation times and spin diffusion lengths vary between layers. In the ferromagnetic layers, the momentum relaxation times are spin-dependent, which leads to spin-polarized currents. 

To proceed, we must first determine the Boltzmann distribution function throughout the layered system. We write the distribution function at position $\threevec{r}$ and crystal momentum $\threevec{k}$ as $f_\alpha(\bvec{r},\bvec{k})$, where $\alpha\in[x,y,z,c]$ denotes a \emph{spin/charge} index. For $\alpha = c$, the Boltzmann distribution function equals the number of carriers within the phase space volume $d^3r d^3k$. For $\alpha = x,y,z$, the Boltzmann distribution equals the difference in spin "up" and "down" carriers defined along the $\alpha$-axis per $d^3r d^3k$. To obtain $f_\alpha(\bvec{r},\bvec{k})$, we solve the spin-dependent Boltzmann equation given by
\begin{align}
\label{BE}
\frac{\partial f_\alpha}{\partial t} 
+ \threevec{v}(\threevec{k}) \cdot \pdf{f_\alpha}{\threevec{r}}
- e\threevec{E} \cdot \pdf{f_\alpha}{\threevec{k}}
&+ \gamma \epsilon_{\alpha \beta \gamma} H^{\text{ex}}_\beta f_\gamma \nonumber \\
&=
\pdf{f_\alpha}{t}_{\text{coll}}
\end{align}
where $\threevec{v} = \nabla_{\threevec{k}}\epsilon_{\bvec{k}}/\hbar$ denotes the electron velocity, $\epsilon_{\bvec{k}}$ is the $\bvec{k}$-dependent energy of carriers, $\threevec{E}$ equals the electric field, and the spin/charge indices ($\alpha, \beta \in [x,y,z,c])$ are implicitly summed over unless otherwise stated.  The fourth term on the L.H.S., however, describes spin precession in a ferromagnet and excludes the charge distribution from the implicit sums.  The collision term on the R.H.S. describes scattering with all $\threevec{k}$-states.

In what follows, we assume translational invariance along the interface plane ($x/y$ plane), so all position dependence is restricted to the out-of-plane direction ($z$ axis). In the linear response regime, the distribution function equals the spin-independent equilibrium distribution function $f_\text{eq}$ plus a spin-dependent perturbation $g_\alpha(z,\bvec{k})$, such that
\begin{align}
    f_\alpha(z,\bvec{k}) &= f_\text{eq}(\epsilon_{\bvec{k}}) \delta_{\alpha c} + \frac{\partial f_\text{eq}}{\partial \epsilon_{\bvec{k}}} g_\alpha(z,\bvec{k}).
\end{align}
Plugging $f_\alpha(z,\bvec{k})$ into \Eq{BE}, we obtain the linearized Boltzmann equation in steady-state,
\begin{align}
\label{LBEap}
&v_z(\threevec{k}) \pdf{g_\alpha}{z}(z,\threevec{k})
- e \threevec{E} \cdot v(\threevec{k}) \delta_{\alpha c}
+ \gamma \epsilon_{\alpha \beta \gamma} H^{\text{ex}}_\beta g_\gamma(z,\threevec{k}) 	\nonumber	\\
&= 
-R_{\alpha\alpha'}(\threevec{k}) g_{\alpha'}(z,\threevec{k})
+ \int_{\text{FS}} d\threevec{k}' P_{\alpha\alpha'}(\threevec{k}, \threevec{k}') g_{\alpha'}(z,\threevec{k}'),
\end{align}
where the first term on the R.H.S. describes scattering out to all other $\threevec{k}$-states, and the second scattering in to the given $\threevec{k}$ state. All $\threevec{k}$-vectors are constrained to the Fermi surface. 

Once the general solution to the Boltzmann equation is obtained in each layer, boundary conditions are required to obtain the full solution for the multilayer system. Here, boundary conditions are given by reflection ($R_I$) and transmission ($T_I$) matrices that relate the Boltzmann distributions on each side of the interface:
\begin{align}
\label{NEscat}
g_\alpha(z_I^-,\threevec{k}_{||}, -k_z) &= R^I_{\alpha\beta}(\threevec{k}_{||}) g_\beta(z_I^-,\threevec{k}_{||},k_z)    \nonumber \\
&+~T^I_{\alpha\beta}(\threevec{k}_{||}) g_\beta(z_I^+,\threevec{k}_{||},-k_z)
\\
g_\alpha(z_I^+,\threevec{k}_{||}, k_z) &= T^I_{\alpha\beta}(\threevec{k}_{||}) g_\beta(z_I^-,\threevec{k}_{||},k_z)    \nonumber \\
&+~R^I_{\alpha\beta}(\threevec{k}_{||}) g_\beta(z_I^+,\threevec{k}_{||},-k_z)
\end{align}
For the first and last layers of the heterostructure, we only assume the Boltzmann distribution does not diverge at $\pm\infty$. For the interfaces, the reflection and transmission matrices are adapted from the scattering matrix of plane waves scattering off a spin-dependent scattering potential given by:
\begin{align}
\label{IntPot}
V(\threevec{r})=\frac{\hbar^2 k_\text{F}}{2m} \delta(z) (\mu_0 + \mu_\text{ex} \boldsymbol{\sigma} \cdot \mhat + \mu_\text{soc} \boldsymbol{\sigma} \cdot (\threevec{k} \times \zhat))
\end{align}
where $k_\text{F}$ is the Fermi momentum and $\mu_0$, $\mu_\text{ex}$, and $\mu_\text{soc}$ are dimensionless parameters specifying the strength of the spin-independent potential barrier, exchange interaction, and Rashba spin-orbit coupling respectively. 

Even with the simple electronic structure assumed for each layer, obtaining a general analytical solution is intractable. Therefore, we solve for the nonequilibrium, spin-dependent Boltzmann distribution $g_\alpha(z,\bvec{k})$ within each layer using numerical methods detailed in Appendix \ref{AppendixBoltzmann}. Once the distribution function in each layer is obtained, the spin/charge accumulations $\mu_\alpha$ and nonequilibrium spin/charge currents $j_{i\alpha}$ are given by
\begin{align}
\label{accu}
\mu_\alpha(z)  &=  \frac{1}{(2\pi)^3 v_\text{F}} \int_\text{FS} g_\alpha(z, \threevec{k})   \\
\label{current}
j_{i\alpha}(z)  &=  \frac{1}{(2\pi)^3 v_\text{F}} \int_\text{FS} v_i g_\alpha(z, \threevec{k})
\end{align}
where $i \in [x,y,z]$ gives the flow direction of the nonequilibrium current and $v_\text{F}$ is the Fermi velocity. In the absence of spin-orbit coupling, the spin torque $\tau_\text{FM}$ is only determined by the spin current incident to the ferromagnet. For example, if the nonmagnet/ferromagnet interface is located at $z = 0$ with the nonmagnetic layer at $z \leq 0$, then
\begin{align}
\label{torque}
\torque_\text{FM} = \mhat \times (\threevec{j}_z(z = 0^-) \times \mhat)
\end{align}
where $[\threevec{j}_z]_\sigma = j_{z\sigma}$ is a vector giving the spin polarization of the z-flowing spin current and $\sigma \in [x,y,z]$ denotes the spin indices of $\alpha$. This expression is nothing more than the z-flowing spin current incident to the ferromagnet with spin polarization transverse to the magnetization, which represents the total angular momentum that can be transferred to the ferromagnetic layer. Due to our selective treatment of spin-orbit coupling, where spin-orbit coupling is absent in the layer which receives the torque, \Eq{torque} is exact. Note that if spin-orbit coupling is included in this layer, \Eq{torque} must be modified to include separate bulk and interfacial contributions \cite{PhysRevB.94.104420,PhysRevB.94.104419}.

\subsection{Semiclassical description of nonlocal spin-orbit torques}
\label{LeakageDetails}

\begin{figure*}
    \centering
    \includegraphics[width=0.95\textwidth]{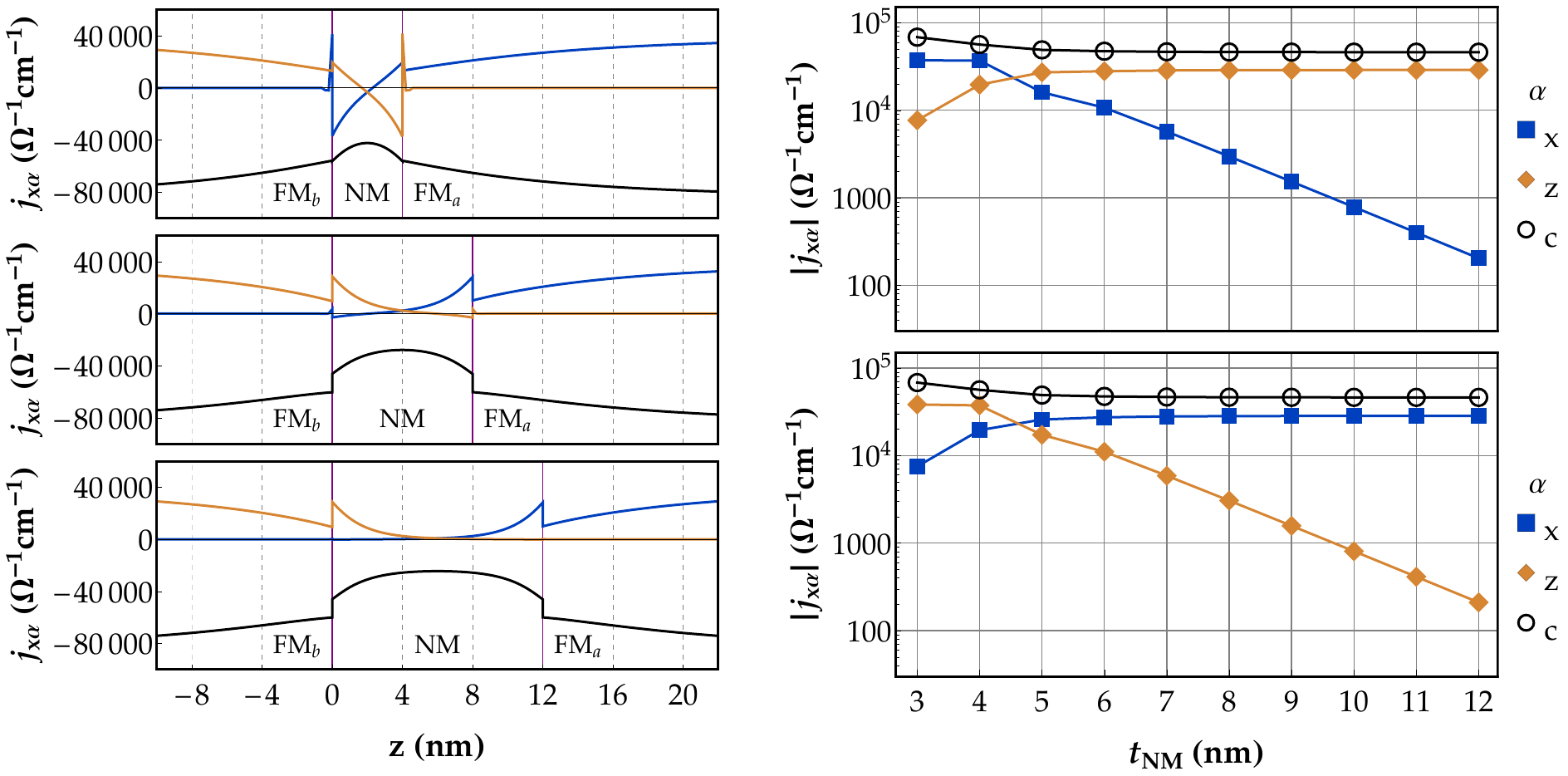}
    \caption{(a)-(c) In-plane charge and spin currents as a function of out-of-plane position ($z$-axis) under an applied electric field $\threevec{E} \parallel \xhat$. For both the charge and spin currents, the flow direction points along $\xhat$. For the spin currents, the spin polarization direction is shown along $x$ (blue curves) and $z$ (orange curves). Spin-orbit coupling is absent everywhere except the $\text{FM}_\B|\text{NM}$ interface. Both ferromagnetic regions are identical except $\mhat_\A = \xhat$ and $\mhat_\B = \zhat$. In-plane spin polarized currents form in each ferromagnetic layer with spin polarization antiparallel to the magnetization direction, as seen in the ferromagnetic regions. These in-plane spin-polarized current decay over out-of-plane distances on the order of the mean free path $\lambda_\text{NM} = 3$ nm, as expected within the relaxation time approximation. For $t_\text{NM} = 4$ nm, which is similar to $\lambda_\text{NM}$, the spin-polarized currents from each ferromagnetic layer are present in the other ferromagnetic layer and can even change sign. (d)-(e) In-plane charge currents at $z = 0^+$ (d) and $z = t_\text{NM}^-$ as a function of $t_\text{NM}$. For $t_\text{NM} > \lambda_\text{NM}$, the in-plane charge current and spin current polarized along the magnetization approach constant values while the spin currents polarized transversely to the magnetization (generated in the other layer) decay exponentially on the order of $\lambda_\text{NM}$.}
    \label{LeakageCurrents}
\end{figure*}

\begin{figure*}
    \centering
    \includegraphics[width=0.95\textwidth]{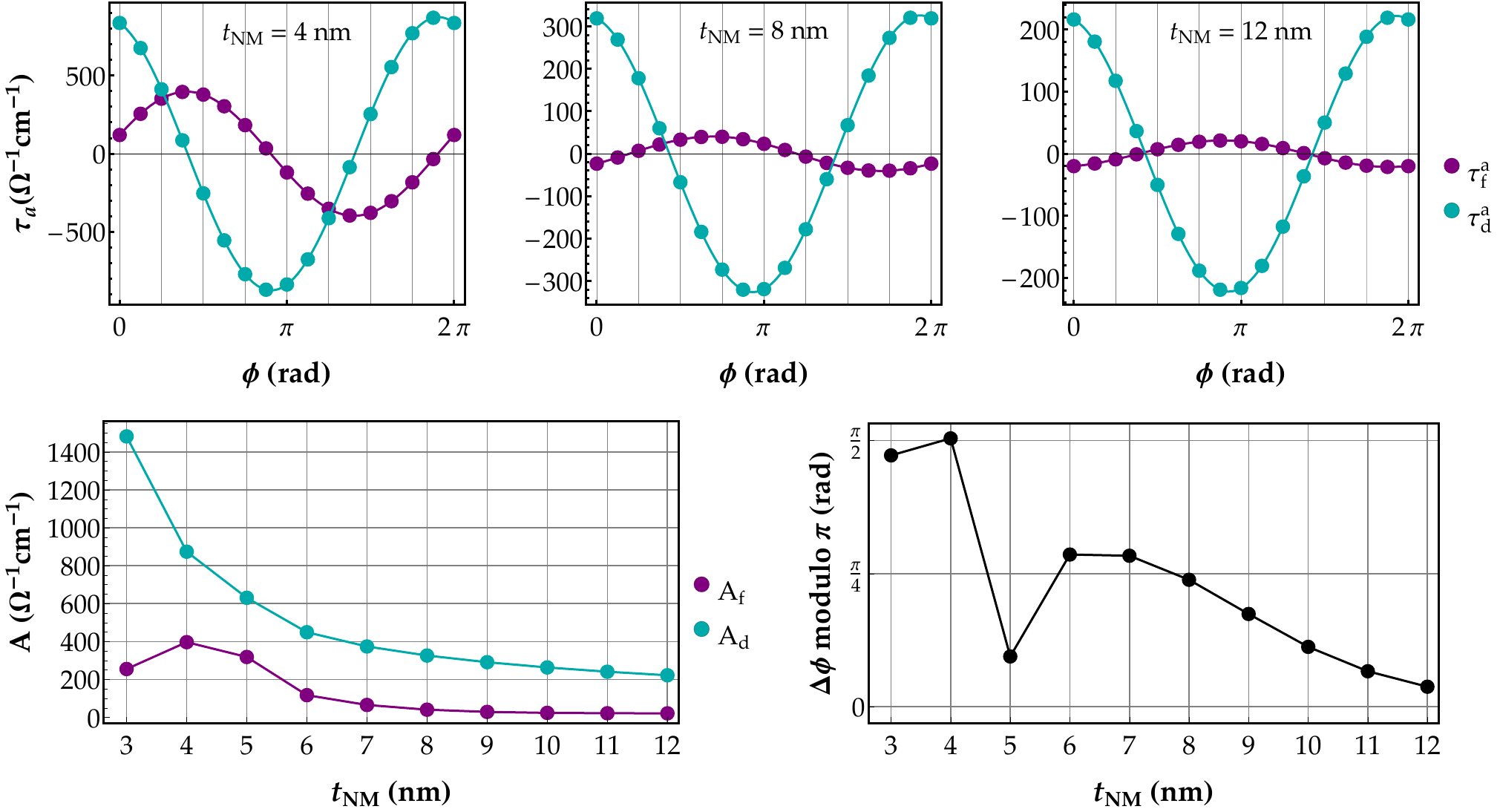}
    \caption{(a)-(c) Dampinglike $\tauad$ and fieldlike $\tauaf$ torques on layer $\A$ versus in-plane magnetization direction $\phi_a$ ($\mhat_\A = \xhat$ corresponds to $\phi_a = 0$). Here $\mhat_\B = \zhat$, so the two magnetization directions are always transverse to each other. In this configuration, both special fields $\FPdla$ and $\FPfla$ lie in-plane, so $\tauad$ and $\tauaf$ must vanish in-plane when pointing along $\FPdla$ and $\FPfla$ respectively. Thus, any phase difference in the angular dependence of $\tauad$ and $\tauaf$ indicates that $\FPdla$ and $\FPfla$ are misaligned, which can only occur due to indirect spin current generation in layer $\B$ (see section \ref{LeakageDetails}). For $t_\text{NM} = 4$ nm, the dampinglike and fieldlike curves have a phase difference $\Delta \phi = \phi_\dl - \phi_\fl \sim \pi/2$, indicating strong indirect spin current generation in layer $\B$. For $t_\text{NM} = 12$ nm, the dampinglike and fieldlike curves are in-phase, which can result from both direct and indirect spin current generation in layer $\B$. (d) Torque amplitudes and (e) torque phase difference $\Delta\phi$ for layer $\A$ as a function of $t_\text{NM}$. As $t_\text{NM}$ becomes greater than $\lambda_\text{NM}$ but remains less than the nonmagnetic layer's spin diffusion length, indirect spin current generation vanishes while direct spin current generation does not. In this limit, the torque phase difference approaches zero and the perpendicular torque arises solely due to spin rotation at the interface (parameterized by $\ImGma$).}
    \label{AmpPhase}
\end{figure*}

Using the multilayer Boltzmann formalism just described, we calculate the spin-orbit torque $\torque_a$ on ferromagnetic layer $\A$ assuming that spin-orbit coupling not present anywhere \emph{except} the $\text{FM}_\B/\text{NM}$ interface. This eliminates all spin-orbit torque contributions in ferromagnetic trilayers that are also present in nonmagnet/ferromagnet bilayers, isolating only the direct and indirect nonlocal contributions introduced in section \ref{SectOverview}. The semiclassical results give qualitative agreement with the ab-initio results presented in section \ref{SectFP} and further elucidate the mechanism underlying indirect spin-orbit torques. The parameters used in these calculations are listed in Appendix \ref{AppendixBoltzmann}.

We first discuss direct spin-orbit torques. Since spin-orbit coupling is only present at the $\text{FM}_\B/\text{NM}$ interface, any out-of-plane spin current directly generated by an in-plane electric field must occur at that interface. The boundary conditions used at this interface, described in detail in Appendix \ref{AppendixBoltzmann}, are sufficient to capture the spin-orbit filtering and spin-orbit precession effects \cite{PhysRevB.94.104420,PhysRevB.94.104419,PhysRevLett.121.136805} which cause spin current generation at interfaces.

Note that direct spin current generation could happen in the bulk ferromagnetic layer rather than its interfaces, via several mechanisms such as the spin Hall, spin anomalous Hall, spin planar Hall, and magnetic spin Hall effects. For simplicity, we do not include any of these effects in our semiclassical calculations because the interface is sufficient both to generate spin currents in the direct mechanism and to convert in-plane to out-of-plane spin currents in the indirect mechanism. Further, spin currents generated in the ferromagnetic layer can be modified when transmitting through the $\text{FM}_\B/\text{NM}$ interface, introducing additional complications in analyzing the charge-to-spin and spin-to-spin conversion processes. We emphasize that bulk contributions from the source ferromagnetic layer are likely quite important in experimental systems; however they are not necessary to qualitatively explain nonlocal torques within our semiclassical formalism.

Next, we discuss indirect spin-orbit torques. Fig.~\ref{LeakageCurrents} plots the in-plane charge and spin currents as a function of out-of-plane position ($z$-axis) in a ferromagnetic trilayer, as determined by our semiclassical calculations. The parameters used in these simulations are given in Appendix \ref{AppendixBoltzmann}. Both ferromagnetic layers are identical, except $\mhat_\A = \xhat$ and $\mhat_\B = \zhat$. We expect the in-plane spin currents are polarized antiparallel to the magnetization direction deep within each ferromagnetic layer, as clearly seen in Fig. \ref{LeakageCurrents}(a)-(c). However, near interfaces, where material properties change, the in-plane charge and spin currents must change as well; the decay length of the in-plane currents is roughly given by the mean free path, here $\lambda_\text{NM} = 3$nm. Thus, when $t_\text{NM} \sim \lambda_\text{NM}$, as is the case in Fig.~\ref{LeakageCurrents}(a), the in-plane spin-polarized current generated in each ferromagnetic layer is present in the other layer as well. As $t_\text{NM}$ increases, the overlap of in-plane spin-polarized currents vanishes. Fig.~\ref{LeakageCurrents}(d)-(e) show the in-plane charge and spin currents at $z = 0^+$ (i.e. at $\text{FM}_\B/\text{NM}$ on the NM side) and $z = t_\text{NM}^-$ (i.e. at $\text{NM}/\text{FM}_\A$ on the NM side). In both cases, the in-plane charge current and the in-plane spin current polarized antiparallel to the magnetization are independent of $t_\text{NM}$, while the in-plane spin current polarized along the other layer's magnetization decreases exponentially. Thus, we expect the nonlocal indirect mechanism, which requires the in-plane spin current from one ferromagnetic layer to be present in the other ferromagnetic layer, to only occur when $t_\text{NM} \sim \lambda_\text{NM}$.

We now discuss how to distinguish the direct and indirect nonlocal mechanisms within our simulation results, which in principle can be done for experimental measurements as well. We choose an orthogonal configuration for the magnetizations, where $\mhat_\B = \zhat$ and $\mhat_\A$ is rotated in the $xy$ plane. The angle $\phi$ specifies the direction of $\mhat_\A$, where $\phi = 0$ corresponds to $\mhat_\A = \xhat$. Since $\mhat_\A$ always points in plane and the spin-orbit torque $\torque_a$ is transverse to $\mhat_\A$, $\torque_a$ can be resolved into components along $\zhat$ and $\mhat_\times$, where $\mhat_\times$ is a unit vector pointing in-plane and transverse to $\mhat_\A$. Here we choose the definition $\mhat_\times = \mhat_\B \times \mhat_\A$, which works because $\mhat_\B$ is fixed out-of-plane. For $\mhat_\B = \zhat$, the special fields $\FPfla(\mhat_\B=\zhat)$ and $\FPdla(\mhat_\B=\zhat)$ must point in-plane but need not be parallel, as is verified by \Eqs{FPdl}{FPfl}. Rewriting \Eq{TorqueTrilayer} under these constraints (and ignoring the $l \geq 2$ vector spherical harmonics contributions), we have
\begin{align}
\label{TorqueTrilayerMod}
    \torque_a &= \mhat_\A \times (\FPdla(\mhat_\B=\zhat)\times\mhat_\A) \nonumber \\ 
    &+ \mhat_\A \times \FPfla(\mhat_\B=\zhat) \nonumber \\
    &= \tauad(\phi) \mhat_\times + \tauaf(\phi) \zhat,
\end{align}
where
\begin{align}
    \tauad(\phi)   &=  A_\dl \sin(\phi - \phi_\dl)     \label{DLOrtho} \\
    \tauaf(\phi)   &=  A_\fl \sin(\phi - \phi_\fl).    \label{FLOrtho}
\end{align}
According to \Eq{TorqueTrilayerMod}, the torque component along $\zhat$ is fieldlike and the torque component along $\mhat_\times$ is dampinglike. In this formulation, $\FPfla$ points along the magnetization direction for which $\tauaf(\phi) = 0$ and $\FPdla$ points along the magnetization direction for which $\tauad(\phi) = 0$. In this way, we can determine the orientation of the special fields.

Under the direct nonlocal mechanism, $\FPdla \parallel \FPfla$, so $\tauad(\phi)$ and $\tauaf(\phi)$ must vanish at the same magnetization angle, i.e. $\phi_\dl = \phi_\fl$ (modulo $\pi$). Thus, $\phi_\dl \neq \phi_\fl$ indicates the presence of indirect nonlocal spin-orbit torques. Fig.~\ref{AmpPhase}(a)-(c) show the semiclassical calculations (indicated by the symbols) of $\tauaf(\phi)$ and $\tauad(\phi)$ for three different spacer thicknesses. These data are then fit to \Eqs{DLOrtho}{FLOrtho} (indicated by the curves). As seen in the plots, for small spacer thickness ($t_\text{NM} \sim \lambda_\text{NM}$), there is a phase difference $\Delta\phi = \phi_\dl - \phi_\fl$ of nearly $\pi/2$, indicating the indirect mechanism, while for larger spacer thicknesses, the phase difference vanishes, which does not rule out the indirect mechanism but can be solely caused by the direct mechanism.

Panels (d)-(e) show the amplitudes ($A_\dl$ and $A_\fl$) and phase differences $\Delta\phi$ (modulo $\pi/2$) of the spin-orbit torques as a function of $t_\text{NM}$. Note that as spacer thickness increases, we expect a $1/t_\text{NM}$ thickness dependence due to the increased spin resistance of the spacer layer. At $t_\text{NM} = 4$ nm, the fieldlike torque amplitude is nearly half the dampinglike, which would require $\ImGma/\ReGma \sim 0.5$, already improbable given $\ImGma/\ReGma \sim 0.1$ or less for most $\text{NM}/\text{FM}$ systems of interest. However, more decisive than the torque amplitudes, the phase difference $\Delta\phi \approx \pi/2$ at $t_\text{NM} = 4$ nm, as seen in Fig.~\ref{AmpPhase}(e), proves only the indirect mechanism could cause this large fieldlike torque. As the thickness increases, the fieldlike torque decreases relative to the dampinglike torque and the phase difference approaches zero, consistent with direct spin current generation.

To summarize this section, we performed semiclassical Boltzmann calculations for a ferromagnetic trilayer and demonstrated that, for nonmagnetic spacer thicknesses comparable to the mean free path, the special fields $\FPfla$ and $\FPdla$ do indeed point in different directions as predicted by symmetry arguments. Only the indirect nonlocal mechanism explains this result, as supported by our semiclassical calculations of the in-plane charge and spin currents and the spin-orbit torques on layer $\A$. In the next section, we present ab-initio calculations that also demonstrate strong phase differences $\Delta \phi$ in orthogonal magnetization configurations, consistent with the semiclassical calculations. 

\section{First-principles calculations}
\label{SectFP}

\subsection{Computational details}

We consider several FM/NM/FM trilayer systems, including Co/Cu/Co, Py/Cu/Py (where Py stands for permalloy, the random face-centered cubic Ni$_{80}$Fe$_{20}$ alloy), and Co/Pt/Co. The electronic structure and transport properties are described within the tight-binding linear muffin-tin orbital (TB-LMTO) method in the atomic sphere approximation implemented in the Questaal package \cite{Questaal}. The atomic potentials for permalloy are obtained self-consistently for the minimal unit cell of the given trilayer using the coherent potential approximation (CPA), adjusting the atomic sphere radii in each Py monolayer to make the sphere charges of Fe and Ni equal to each other.

The spin torques are calculated using the nonequilibrium Green's function (NEGF) technique \cite{Faleev2005,Belashchenko2019,Belashchenko2020} with explicit averaging over disorder configurations.
The NEGF method properly treats charge and spin diffusion \cite{belashchenko2023breakdown} and includes all microscopic mechanisms of spin torque generation on the same footing, which in the Kubo linear response technique would require a consistent inclusion of vertex corrections.

In the systems with Py, the Fe and Ni atoms are randomly distributed over the lattice sites in the corresponding monolayers. The use of atomic spheres with equal charges makes the Madelung potentials in the substitutional alloy nonrandom and makes it possible to avoid computationally prohibitive self-consistent calculations for the entire disordered supercell.

Disorder is simulated by adding a uniformly distributed random Anderson potential $-V_m<V_i<V_m$ at each lattice site $i$. This is also done for the Py/Cu/Py system, because substitutional disorder alone does not generate enough scattering to make the mean-free path sufficiently short. To obtain bulklike torquances, we used a correction for the potential drop in the contacts \cite{belashchenko2023breakdown}.

Ideal face-centered cubic lattices were assumed with lattice constants of 3.56 \AA\ for Py/Cu/Py and 3.75 \AA\ for Co/Cu/Co and Co/Pt/Co trilayers.
Co/Cu/Co and Co/Pt/Co trilayers were set up with (001) interfaces and current flowing in the [$1\bar10$] direction. Py/Cu/Py trilayers we set up with (111) or (001) interfaces and current flowing in the [$1\bar10$] direction in both cases. All supercells had 400 monolayers (ML) in the direction of the current flow, which is much longer than the effective mean-free path. The periodic setup is completed by vacuum separating the two ferromagnetic layers, which is represented by 4 monolayers of empty spheres. Supercells are periodic in the third dimension with the thickness of 4 or 6 ML for supercells with (001) or (111) interfaces, respectively. Subscripts in the specification of the trilayer, such as Co$_4$Cu$_4$Co$_4$, indicate the thicknesses of the corresponding layers in ML.

Disorder strength $V_m$ was chosen as 40 mRy or 80 mRy. The resistivities of bulk materials are: 0.4 or \SI{1.5}{\micro\ohm\centi\meter} in Cu, 2.5 or \SI{9.4}{\micro\ohm\centi\meter} in Co, and 6.4 and \SI{40}{\micro\ohm\centi\meter} in Py at $V_m=40$ or 80 mRy, respectively. The experimental resistivities of bulk Cu and Co at room temperature are 1.7 and 5.6 $\mu\Omega\cdot$cm, respectively. Thus, $V_m=80$ mRy for these materials appears to be reasonable for room-temperature properties in high-quality samples, while 40 mRy may correspond to extremely clean samples at low temperatures.

Temperature-dependent resistivity of bulk Py was calculated by Starikov \emph{et al.} \cite{Starikov2018} who found that the residual resistivity of about \SI{2.3}{\micro\ohm\centi\meter} is dominated by spin-orbit coupling and is about twice smaller compared to experiment, which they attributed to grain boundary scattering in polycrystalline samples. Taking thermal lattice and spin disorder, they found room-temperature resistivity of about \SI{15}{\micro\ohm\centi\meter} in good agreement with experiment \cite{Py-expt}. However, thin Py films used in spin-orbit torque measurements have significantly larger resistivities, for example, about \SI{30}{\micro\ohm\centi\meter} in Ref. \cite{Wang2019}. Thus, Anderson disorder amplitudes of 40 and 80 mRy bracket the typical range of resistivities in thin Py films.

The Fermi sea term \cite{PhysRevMaterials.3.011401} can contribute to the torques that are even under time reveral, such as those described by coefficients $\ds$, $\dm$, and $\fx$. Calculation of this term for a disordered system is computationally very expensive. Our estimates for the disorder-free system show that it is rather small (of order \SI{0.1e5}{\per\ohm\per\meter}).

\subsection{Giant magnetotorquance}

\begin{figure}
    \includegraphics[width=0.9\columnwidth]{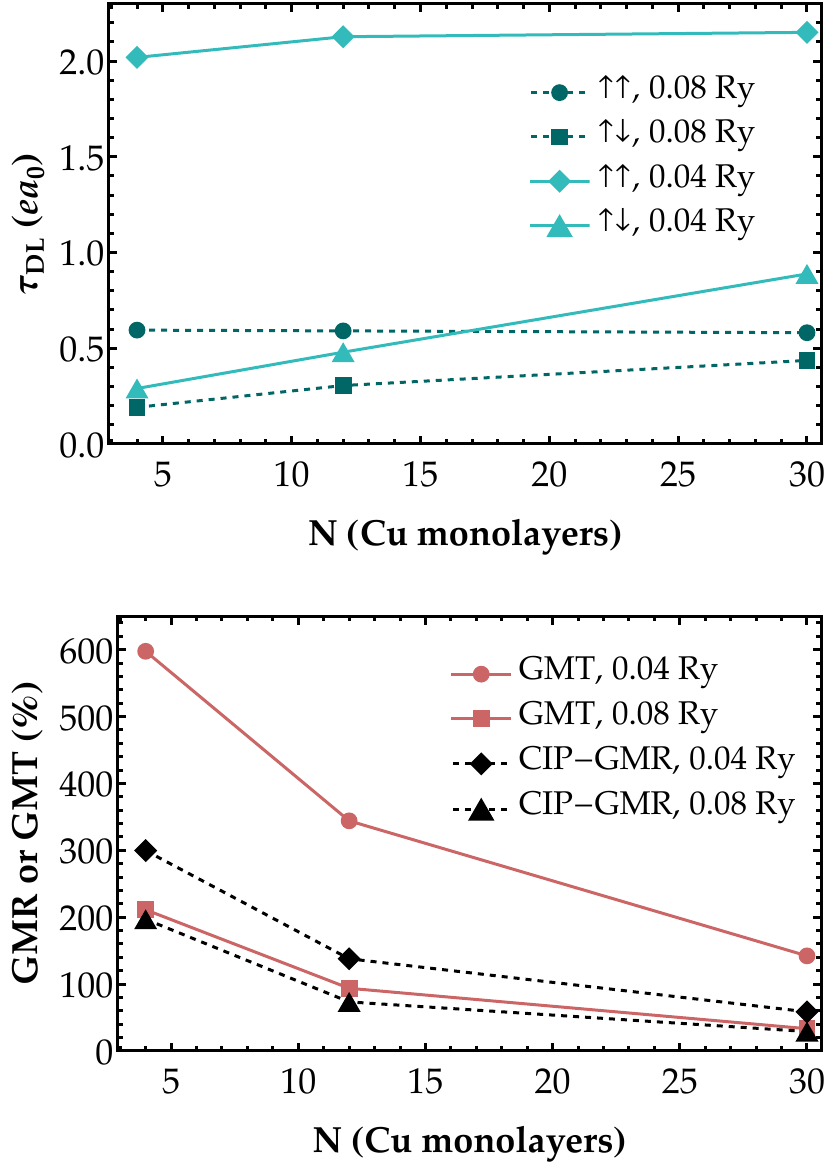}
    \caption{(a) Torquance with dampinglike angular dependence acting on the top Co layer in Co$_4$Cu$_N$Co$_4$ trilayers in the parallel or antiparallel configurations. (b) Giant magnetotorquance and current-in-plane giant magnetoresistance for the same systems.}
    \label{fig:GMT}
\end{figure}

First, we consider the configurations in which the magnetizations of the two ferromagnetic layers are either parallel (P) or antiparallel (AP). The in-plane conductivity of such a trilayer depends on the relative orientation of the magnetizations (P or AP), which is the essence of the CIP-GMR effect. Figure \ref{fig:GMT}(b) shows the magnetoresistance ratio $R=(G_{\text{P}}-G_{\text{AP}})/G_{\text{AP}}$ for the Co$_4$Cu$_N$Co$_4$ trilayer as a function of the Cu layer thickness (green symbols) for $V_m=40$ or 80 mRy. The CIP-GMR exceeds 300\% in the system with $N=4$ and $V_m=40$ mRy and declines with increasing $N$ and increasing disorder strength, as expected.

We then calculate the spin-orbit torque acting on the top layer in the P or AP configuration of this system as a function of the orientation of the magnetizations, which are always kept collinear such that $\mhat_\two=\pm\mhat_\one=\mhat$. 
Because none of the layers has a fixed orientation in this case, the torques are characterized by Eq. (\ref{TorqueBilayer}) with the conventional orientation of $\FP$; for example, the torque acting on layer $\two$ is $\torque_\two=\tau_\mathrm{DL}\mhat\times(\FP\times\mhat)+\tau_\mathrm{FL}\mhat\times\FP$, where $\tau_\mathrm{DL}$ and $\tau_\mathrm{FL}$ are different in the P and AP configurations. By analogy with GMR, we define the \emph{giant magnetotorquance} (GMT) as $R_T=(\tau^\text{P}_\mathrm{DL}-\tau^{\text{AP}}_\mathrm{DL})/\tau^{\text{AP}}_\mathrm{DL}$.
Experimentally, GMT could be measured by using two ferromagnetic layers of different thickness with sufficiently strong antiferromagnetic interlayer coupling and an external magnetic field to enforce both the orientation and the P or AP configuration. 

Keeping only the first three terms in Eq. (\ref{FPdl}) and (\ref{FPfl}) for simplicity, the observed coefficients are $\tau_\mathrm{DL}=\dsa\mp \fxa$ and $\tau_\mathrm{FL}=\fsa\pm\dxa$, where the upper and lower signs correspond to the P and AP confugurations, respectively. This means that $\tau_\mathrm{DL}$ and $\tau_\mathrm{FL}$ do not have pure dampinglike and fieldlike character and are defined solely based on the angular dependence of $\torque_\two(\mhat)$.

Large CIP-GMR in Co/Cu/Co trilayers indicates that the effective momentum relaxation time for electrons near the interfaces depends strongly on the angle between the two magnetic layers. Interfacial mechanisms of spin current emission and conversion depend on the scattering rates and may, therefore, be affected by the magnetic configuration. As a result, the coefficients $\dsa$, $\fxa$, etc., may themselves be different in the P and AP configurations. 

Figure \ref{fig:GMT}(a) shows $\tau_\mathrm{DL}$ as a function of Cu layer thickness and disorder strength in Co/Cu/Co trilayers.
In the parallel configuration $\tau_\mathrm{DL}$ for the thinnest Cu layer is about \SI{7.4e5}{\per\ohm\per\meter} at $V_m=40$ mRy, which is reduced to 2.2 at $V_m=80$ mRy. This reduction follows the reduction of the conductivity, suggesting that this torque comes from processes proportional to the relaxation time $\tau$, such as the spin currents generated at the interfaces. Figure \ref{fig:GMT}(b) shows the GMT, which behaves qualitatively similar to CIP-GMR. Clearly, the crosstalk between the scattering in the two ferromagnetic layers strongly affects the generation of the spin-orbit torque. This can happen in part due to the variation of the relevant relaxation times in the regions where spin currents are generated; these processes are well understood in the context of CIP-GMR \cite{Tsymbal2001}. However, a finite $\fxa$ could also contribute to the difference between $\tau^\text{P}_\mathrm{DL}$ and $\tau^\text{AP}_\mathrm{DL}$. To isolate this contribution, in the following we focus on the orthogonal configurations.

\subsection{Orthogonal magnetizations}
\label{Orthogonal}

To exclude the indirect influence of CIP-GMR on the torquances, we now focus on magnetic configurations with orthogonal magnetizations, $\mhat_\one\perp\mhat_\two$.
The sheet resistance of the trilayer in all such configurations is the same apart from small relativistic effects, ensuring that the angular dependence of the spin-orbit torque is not overwhelmed by strong CIP-GMR effects.

\subsubsection{Geometry}

For each trilayer system, we perform a series of calculations in which the magnetization of one layer is fixed while the other is rotated in the orthogonal plane. For example, we can set $\mhat_\one=-\xhat$ and rotate $\mhat_\two$ in the $yz$ plane, or we can set $\mhat_\two= -\zhat$ and rotate $\mhat_\one$ in the $xy$ plane. We use $-\xhat$, $-\yhat$, or $-\zhat$ for the fixed direction. The orientation of the rotating magnetization will be encoded by angle $\phi$ measured with respect to $-\zhat$ unless the fixed layer is aligned with $-\zhat$; in the latter case, the angle $\phi$ is measured with respect to $-\xhat$.

\begin{figure*}
    \centering
    \includegraphics[width=0.95\textwidth]{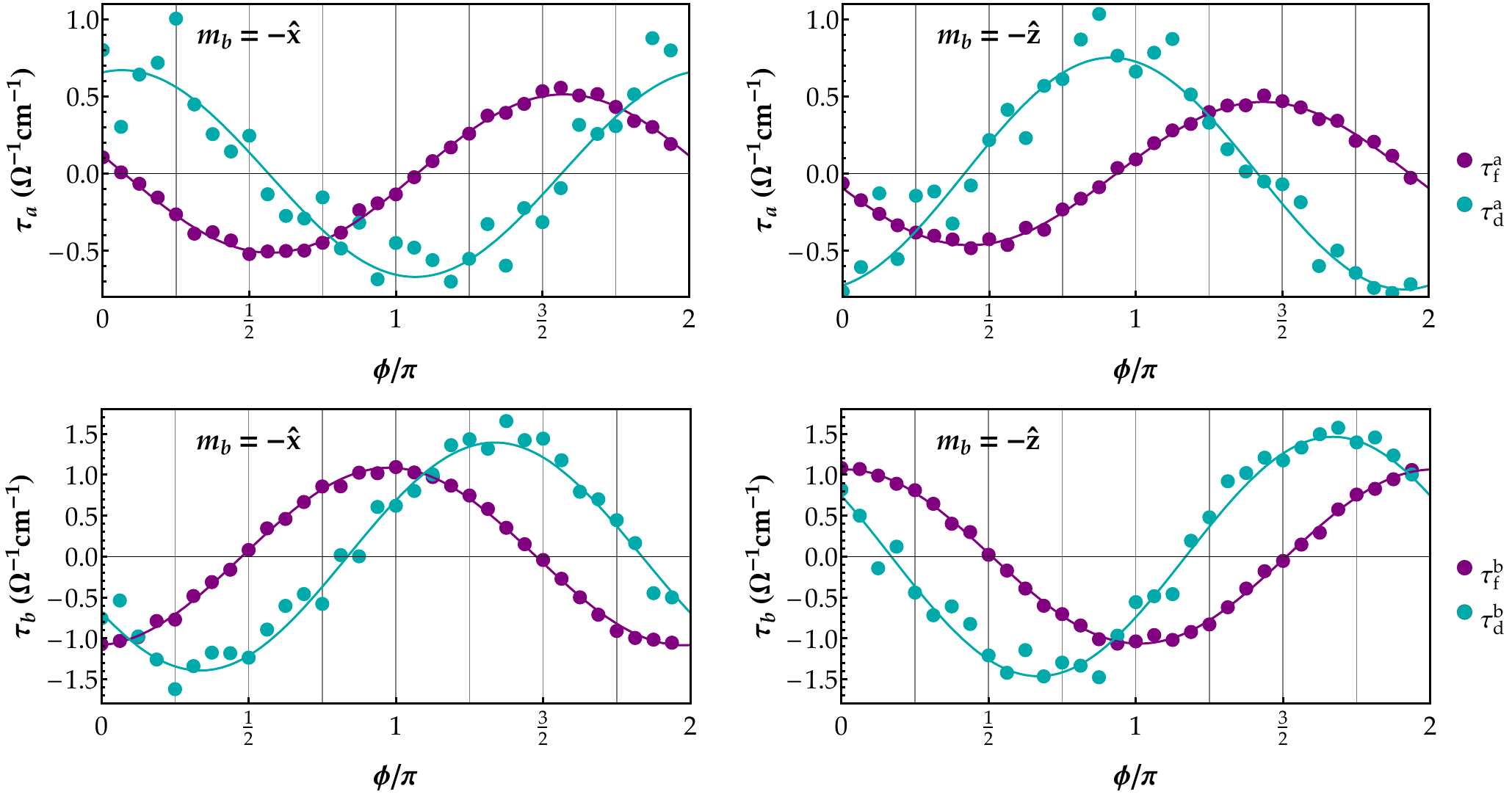}
    \caption{Spin-orbit torques $\tautwodf$ and $\tauonedf$ in Co$^\prime_4$Cu$_4$Co$_4$ with $V_m=40$ mRy. Spin-orbit coupling is included only in the bottom fixed layer 1 and two adjacent monolayers of Cu. The torques are plotted against the angle made by $\mhat_\two$ with the $-\zhat$ axis [for $\mhat_\one=-\xhat$, panels (a)-(b)] or with the $-\xhat$ axis [for $\mhat_\one=-\zhat$, panels (c)-(d)]. Solid lines are fits to a harmonic function $C+A_1\cos\phi+B_1\sin\phi$. Vectors $\mhat_\one$ and $\mhat_\two$ represent the directions of the spin moments rather than the magnetizations.}
    \label{fig:trq444_04}
\end{figure*}

The disorder-averaged torquance on each ferromagnetic layer $\mhat_\A$ is projected onto the ($\phi$-dependent) basis of $\mhat_{\B}$ ($\B\ \neq \A$) and $\mhat_\times=\mhat_\one\times\mhat_\two$ and represented in terms of a Fourier series. As an example, Fig.\ \ref{fig:trq444_04} shows the calculated angular dependence of $\boldsymbol{\tau}_\A\cdot\mhat_\B$ and $\boldsymbol{\tau}_\A\cdot\mhat_\times$ for the Co$^\prime_4$Cu$_4$Co$_4$ trilayer with spin-orbit coupling turned off in the upper half of the trilayer (as further discussed below).

\begin{figure*}
    \centering
    \includegraphics[width=0.95\textwidth]{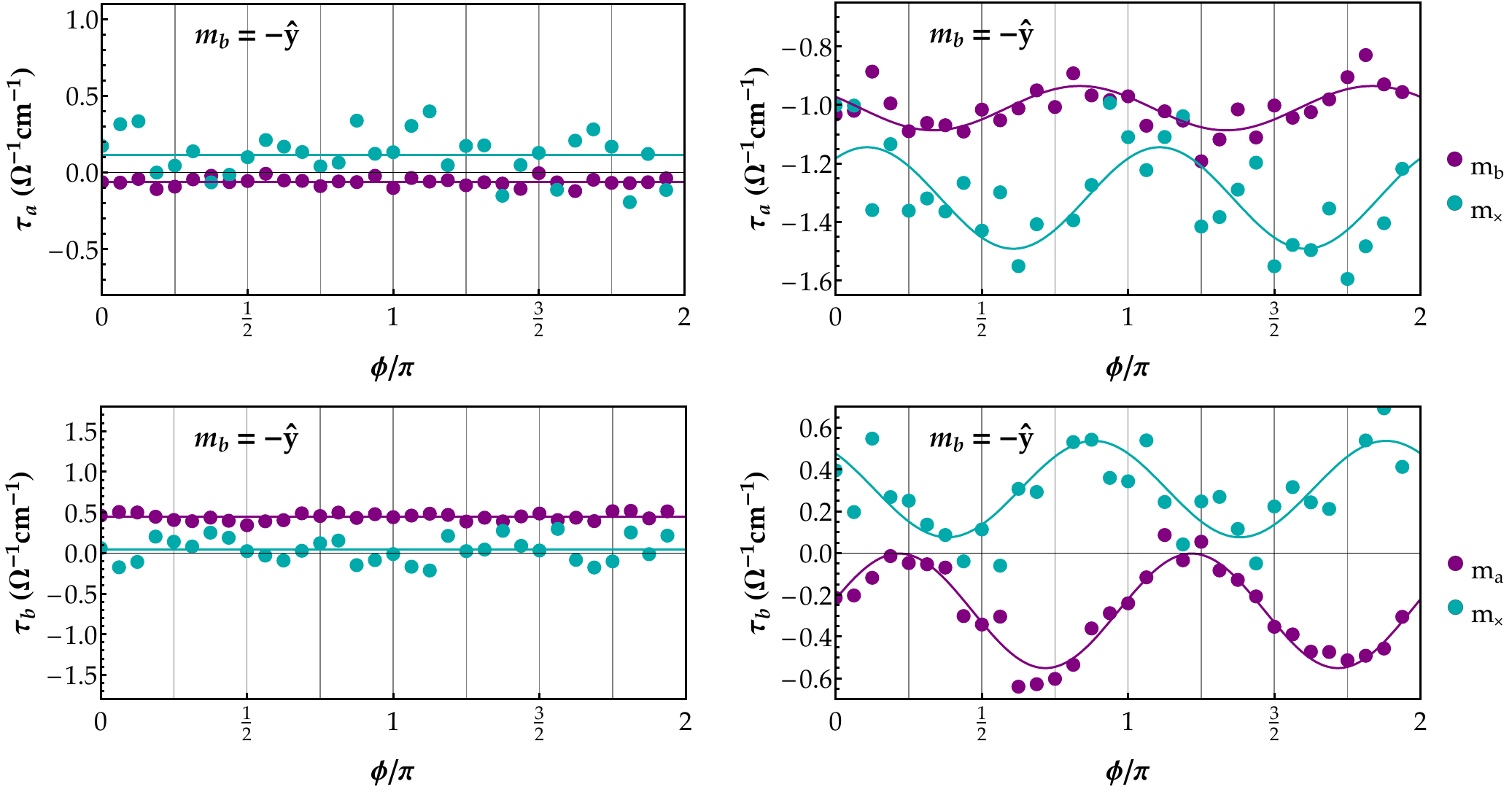}
    \caption{(a)-(b) Spin-orbit torques $\torque_\two$ and $\torque_\one$ in Co$^\prime_4$/Cu$_4$/Co$_4$ at $V_m=40$ mRy, projected onto $\mhat_\two$ and $\mhat_\times=\mhat_\one\times\mhat_\two$ (as indicated on the right). (c)-(d) Same for Co$_4$/Pt$_4$/Co$_4$.}
    \label{fig:2nd}
\end{figure*}

When the fixed magnetization is parallel to $\xhat$ or $\zhat$, as in Fig.\ \ref{fig:trq444_04} for a Co/Cu/Co trilayer, the $\cos(n\phi)$ and $\sin(n\phi)$ terms with even $n$ are not allowed by symmetry, and the angular dependence of all torque components $\boldsymbol{\tau}_\A\cdot\mhat_\B$ and $\boldsymbol{\tau}_\A\cdot\mhat_\times$ is captured by a linear combination of $\cos(\phi)$ and $\sin(\phi)$. Note that when the fixed magnetization is parallel to $\yhat$, the torque components are nearly independent of $\phi$ for Co/Cu/Co and Py/Cu/Py trilayers (not shown in Fig.\ \ref{fig:trq444_04}). However, for Co/Pt/Co trilayers, there are also sizeable second harmonics $\cos(2\phi)$ and $\sin(2\phi$) in this configuration, as shown in Fig.\ \ref{fig:2nd}.

The calculated Fourier coefficients for all systems are given in Appendix \ref{app:fp} (Tables \ref{tab:trq}, \ref{tab:2nd}).
They are related to the $d$- and $f$-coefficients in Eq. (\ref{FPdl})-(\ref{FPfl}), as shown in Table \ref{tab:corr}, which are subsequently obtained by a least-squares fit. The results are listed in Table \ref{tab:params} for trilayers with a Cu spacer and in Table \ref{tab:paramsPt} for Co/Pt/Co. We will discuss them in the following subsections.

\begin{table*}
\caption{Coefficients in Eqs.\ (\ref{FPdl})-(\ref{FPfl}) for systems with a Cu spacer obtained by fitting of the data from Table \ref{tab:trq}. The superscript denotes the layer on which the torque acts. A prime next to Co or Py indicates that spin-orbit coupling is turned off in the top ferromagnetic layer and in the adjacent half of the spacer. A prime on Pt indicates that spin-orbit coupling is also turned off in the Pt spacer. Coefficients are shown for torque efficiencies $\xi^E$ in units of $10^{5}\Omega^{-1}m^{-1}$. The last column lists the phase difference $\Delta \phi = \phi^\two_\dl - \phi^\two_\fl = \arctan\left[\dxtwo/\dstwo\right]-\arctan\left[\fxtwo/\fstwo\right]$.}
\centering
\setlength{\tabcolsep}{4.45pt}
{%
\begin{tabular}{|l|c|c|c|c|c|c|c|c|c|c|c|c|c|c|}
\hline
System &  $V_m$ &  $\dstwo$ &  $\dsone$ &  $\dmtwo$ &  $\dmone$ &  $\dxtwo$ &  $\dxone$ &  $\fstwo$ &  $\fsone$ &  $\fmtwo$ &  $\fmone$ &  $\fxtwo$ &  $\fxone$ & $\Delta\phi/\pi$ \\
\hline
 \text{Co$_4$Cu$_4$Co$_4$} & 40 & 1.45 & -1.45 & -0.16 & 0.14 & -0.18 & 0.23 & 0.92 & -0.93 & -1.04 & 1.13 & -0.94 & 0.86 & --- \\
\hline
\text{Co$^\prime_4$Cu$_4$Co$_4$} & 40 & 0.69 & -0.72 & -0.06 & 0.06 & -0.17 & 0.04 & -0.11 & -1.07 & 0.12 & 1.24 & -0.48 & 0.45 & 0.50 \\
\hline
\text{Co$^\prime_4$Cu$_4$Co$_4$} & 80 & 0.55 & -0.56 & 0.01 & 0.05 & 0.14 & 0.13 & -0.16 & -0.43 & 0.07 & 0.34 & -0.34 & 0.29 & 0.28\\
\hline
\text{Co$^\prime_4$Cu$_{12}$Co$_4$} & 40 & 1.05 & -1.24 & -0.04 & 0.15 & -0.22 & -0.05 & -0.11 & -2.46 & 0.05 & 2.67 & -0.86 & 0.71 & 0.52\\
\hline
\text{Co$^\prime_4$Cu$_{12}$Co$_4$} & 80 & 0.55 & -0.69 & 0.01 & 0.12 & 0.19 & 0.08 & -0.13 & -1.0 & 0.02 & 0.79 & -0.46 & 0.41 & 0.31\\
\hline
\text{Py$_4$Cu$_4$Py$_4$} & 40 & 2.42 & -2.40 & -0.44 & 0.44 & -0.02 & -0.15 & 0.55 & -0.6 & -0.51 & 0.65 & -1.53 & 1.48 & ---\\
\hline
\text{Py$^\prime_4$Cu$_4$Py$_4$} & 40 & 1.12 & -1.30 & -0.06 & 0.29 & 0.98 & 0.87 & -1.19 & -1.89 & 0.42 & 0.93 & -0.83 & 0.96 & -0.03 \\
\hline
\text{Py$^\prime_4$Cu$_4$Py$_4$} & 80 & 0.46 & -0.54 & -0.08 & 0.10 & 0.42 & 0.24 & -0.24 & -0.65 & 0.08 & 0.28 & -0.07 & 0.08 & -0.14\\
\hline
\text{Py$^\prime_4$Cu$_4$Py$_4$ (001)} & 40 & 0.92 & -1.04 & -0.06 & 0.09 & 0.97 & 0.98 & -0.92 & -0.89 & 0.02 & -0.03 & -0.71 & 0.72 & -0.05\\
\hline
\text{Py$^\prime_{12}$Cu$_4$Py$_{12}$} & 40 & 0.62 & -0.66 & -0.05 & 0.11 & 0.39 & 0.57 & -0.40 & -1.56 & 0.14 & 1.06 & -0.34 & 0.52 & 0.04\\
\hline
\end{tabular}}
\label{tab:params}
\end{table*}

\begin{table*}[htb]
 \caption{Same as in Table \ref{tab:params} but for Co$_4$Pt$_N$Co$_4$ systems with $V_m=40$ mRy. Spin-orbit coupling is included everywhere. The (signs of) coefficients for symmetric systems are for the top layer $\two$. The Co$^\prime_4$Pt$_{12}$Co$_4$ system is denoted $12^\prime$, and the torques acting on the two layers are listed in separate lines. For this system it is assumed that $\dz$ and $\fz$ coefficients are zero to resolve the indeterminacy of the fit.}
\setlength{\tabcolsep}{4.45pt}
    \centering
    \begin{tabular}{|c|c|c|c|c|c|c|c|c|c|c|c|c|c|c|}
    \hline
$N$ & $\ds$ & $\dm$  & $\dx$ & $\fs$ & $\fm$ & $\fx$ & $P_{1}$ & $P_{2}$ & $\dD$ & $\dtwo$ & $\dz$ & $\fD$ & $\ftwo$ & $\fz$ \\
 \hline
 \text{ 4 }  &  2.25 & -1.03 & -0.18 &  1.17 & -1.34 & 0.28 & -0.09 & -0.06 & 0.261 & -0.01 & -0.07 & -0.08 & -0.11  & 0.03 \\
 \hline
 \text{ 12 } & 2.80 & -1.68 & 0.21 & 1.70 & -1.37 & -0.19 & -0.11 & 0.08 & 0.20 & 0.00 & 0.00 & 0.02 & -0.09 & -0.01 \\
     \hline
\text{ 12$^\prime$} (layer $\two$) & 1.90 & -1.01 & -0.19 & 0.88 & -0.50 & -0.16 & -0.03 & 0.05 & 0.43 & -0.08 & --- & -0.07 & -0.10 & --- \\
\text{ 12$^\prime$} (layer $\one$) & -2.52 & 1.83 & -0.51 & -0.98 & 1.12 & 0.01 & -0.06 & 0.13 & -0.40 & 0.01 & --- & 0.05 & 0.20 & --- \\
\hline
\end{tabular}
\label{tab:paramsPt}
\end{table*}

\subsubsection{Selective treatment of spin-orbit coupling}
\label{sec:SOC}

As explained in Sections \ref{mechanisms} and \ref{semiclassical}, the phase shift between the dampinglike and fieldlike torque components in a ferromagnetic layer without spin-orbit coupling is a manifestation of an indirect spin current generation in the trilayer. This litmus test is not available in experimental measurements if both ferromagnetic layers have comparable spin-orbit coupling, which, in particular, is the case for symmetric trilayers. However, it is readily available in theoretical simulations where spin-orbit coupling can be turned on or off in different parts of the system. 
Therefore, we perform auxiliary calculations with SOC turned off in one of the ferromagnetic layers and the adjacent half of the spacer. Such an auxiliary trilayer with spin-orbit coupling turned off in the top half of the structure will be labeled with a prime on the top layer, e.g., Co$^\prime_4$Cu$_4$Co$_4$.

Direct comparison shows that the spin-orbit torque in trilayers with a Cu spacer are approximately additive: the spin-orbit torque in FM$_\two$/Cu/FM$_\one$ is equal to the sum of the spin-orbit torques in FM$^\prime_\two$/Cu/FM$_\one$ and FM$_\two$/Cu/FM$^\prime_\one$. This is reasonable because spin-orbit coupling in these systems is relatively weak, and the layer thicknesses are small compared to the spin-diffusion lengths; as a result, the torquances are dominated by the first-order term in spin-orbit coupling.
To see this additivity explicitly, first note that Table \ref{tab:params} lists the coefficients from \Eqsd{FPdl}{FPfl} individually for each layer, which is identified by the subscript. For our symmetric trilayers with spin-orbit coupling included everywhere (no primes in the system name) symmetry requires $\gtwo=-\gone$, where $g$ stands for any of the $d$- or $f$-coefficient. This relation is approximately satisfied by disorder-averaged quantities. If the spin-orbit torque is additive, we should have $\gtwo=\gtwo_{\one}-\gone_{\one}$, where the subscript ${\one}$ indicates that spin-orbit coupling is only turned on in half of the trilayer including layer $\one$. This relation holds well for all terms in Co$_4$Cu$_4$Co$_4$ and somewhat less accurately in Py$_4$Cu$_4$Py$_4$.

Thus, for systems with the Cu spacer, calculations of torques with partially turned-off spin-orbit coupling allow us to separate the torques generated by spin-orbit coupling in different layers, which is necessary for identifying the indirect generation mechanism.
On the other hand, there are large deviations from additivity in Co$_4$Pt$_{12}$Co$_4$, as evident from Table \ref{tab:paramsPt}. These deviations are expected because strong spin-orbit coupling in Pt leads to a nonlinear interplay between the generation and relaxation of spin currents. Even for a Co/Pt bilayer, non-additivity follows from the nonlinear dependence of the dampinglike torques on the thickness of the Pt layer. 

\subsubsection{Co/Cu/Co and Py/Cu/Py trilayers}

For Co/Cu/Co and Py/Cu/Py trilayers, the first-principles results listed in Table \ref{tab:trq} are well described by the first three terms in \Eqs{FPdl}{FPfl}. The values of these coefficients obtained by least-squares fits of the expressions in Table \ref{tab:corr} to the data in Table \ref{tab:trq} are listed in Table \ref{tab:params}.

First, we consider the phase shift between the fieldlike and dampinglike torque components, which, as explained in Section \ref{LeakageDetails}, signals the importance of indirect spin current generation mechanism. This phase shift $\Delta \phi = \phi^\two_\dl - \phi^\two_\fl$ is listed in the last column of Table \ref{tab:params} for systems with spin-orbit coupling turned off in layer $\two$. We see that $\Delta \phi$ is close to $\pi/2$ for Co$'$/Cu/Co systems with $V_m=40$ mRy and decreases to about $0.3\pi$ at $V_m=80$ mRy. We interpret this large phase shift as an indication of a strong indirect spin current generation. On the other hand, $\Delta\phi$ is rather small in the Py$'$/Cu/Py systems for all parameter values. 

The $\dstwo$ coefficient gives the dampinglike torque polarized along $\FP$ in layer $\two$ generated by spin-orbit coupling in layer $\one$ with $\mhat_\one$ lying in the $xz$ plane. This is the direct nonlocal spin current generation mechanism discussed in Sec. \ref{sec-direct}. The magnitude of this term is comparable to the dampinglike torque in Co/Pt bilayers \cite{PhysRevMaterials.3.011401,PhysRevB.101.020407,PhysRevB.105.054405}.
It decreases with increasing $V_m$ consistent with its origin in interfacial scattering. Interestingly, in asymmetric Co$'$/Cu/Co and Py$'$/Cu/Py systems listed in Table \ref{tab:params} the coefficient $\dstwo$ is close to the negative of $\dsone$, which describes the corresponding torque acting on the source layer 1. This correspondence suggests that this nonlocal mechanism leads to an exchange of angular momentum between the two layers with a negligible loss of angular momentum to the lattice.

We note that the $\dm$ terms are much smaller than $\ds$ in Table \ref{tab:params}, which means the spin current polarized along $\FP$ generated by the Co/Cu and Py/Cu interfaces is strongly suppressed at $\mhat\parallel\FP$ compared to $\mhat\perp\FP$.
A similar trend was found for the intrinsic spin-Hall effect in Co and Fe, and, to a lesser degree, Ni \cite{SAHE}. 

The $\dxtwo$ term describes dampinglike torque with spin polarization $\mhat_\one\times\FP$ (i.e., the spin-rotated polarization, in the language of Ref.~\cite{Humphries2017}). In the typical case $\ImGm\ll\ReGm$, it corresponds to the absorption of spin current with spin polarization $\mhat_\one\times\FP$ emitted from the interface of layer $\one$. In the Co$'$/Cu/Co system this term is a few times smaller than $\dstwo$, as it usually expected. However, in Py$'$/Cu/Py the $\dxtwo$ and $\dstwo$ terms are of similar magnitude. 

The $\fsone$ parameter describes the fieldlike torquance due to the Rashba-Edelstein effect at the interface with spin-orbit coupling. This torquance is arguably the clearest case of an extrinsic process proportional to the effective momentum relaxation time $\tau_\mathrm{int}$ for electrons moving through the interfacial region. The large CIP-GMR seen in Co/Cu/Co trilayers (Fig. \ref{fig:GMT}) can be interpreted as a strong dependence of $\tau_\mathrm{int}$ on the angle between $\mhat_\one$ and $\mhat_\two$. This dependence (and hence the CIP-GMR ratio) becomes weaker with increasing Cu spacer thickness $t_\text{Cu}$, on the scale of its mean-free path. In the present case of orthogonal $\mhat_\one$ and $\mhat_\two$, the presence of the second ferromagnetic layer leads to additional scattering, and hence $\tau_\mathrm{int}$ is expected to increase with increasing $t_\text{Cu}$. In agreement with this expectation, the parameter $\fsone$ increases, by more than a factor of 2, when $t_\text{Cu}$ in the Co/Cu/Co trilayer is increased from 4 to 12 monolayers, at both $V_m=40$ mRy and 80 mRy.

The coefficient $\fstwo$ is rather small in Co$'$/Cu/Co trilayers, 
but in Py$'$/Cu/Py trilayers $\fstwo$ is comparable with $\dstwo$. As noted above, the phase shift $\Delta\phi$ is small in this system, which means the $\fxtwo/\fstwo$ and $\dxtwo/\dstwo$ ratios are not very far from each other. These features could be explained within the circuit theory in the presence of a large imaginary part of the spin-mixing conductance $\ImGm$. 
Calculations of Ref. \cite{CarvaSMC} indeed found a rather large $\ImGm$ for thin layers of Ni and Py inserted between copper leads due to incomplete dephasing of the transverse component of the spin current, but increasing the thickness of Py to 8-12 ML of Py was enough to strongly suppress $\ImGm$. The fact that we see $\fstwo/\dstwo\approx0.65$ even with 12-ML Py layers in Py$'_{12}$/Cu$_4$/Py$_{12}$ suggests that the large fieldlike terms $\fstwo$ and $\fxtwo$ in Py$'$/Cu/Py trilayers are, in fact, both due to indirect spin current generation (i.e., to the $a_0$, and $a_x$/$a_z$ terms in Appendix \ref{AppendixLeakage}), while the small values of the phase shift $\Delta\phi$ are rather accidental. Note that nonlocal fieldlike torque corresponding to $\fstwo$ was discussed in Ref. \cite{Geranton2017} for a system where the generating interface is nonmagnetic.

\subsubsection{Co/Pt/Co trilayers}

Table \ref{tab:paramsPt} shows that the $\dx$ and $\fx$ terms in Co$_4$/Pt$_N$/Co$_4$ at both $N=4$ and $N=12$ are a few times smaller compared to $\ds$ and $\fs$, respectively. This means that the spin-rotated \cite{Humphries2017} dampinglike and fieldlike torques are relatively small in this system.

Because, as mentioned in Section \ref{sec:SOC}, the torques in Co/Pt/Co trilayers are not additive, the expansion coefficients and the phase shift $\Delta\phi$ for the Co$'$/Pt/Co system are not necessarily applicable to the physical Co/Pt/Co system. Nevertheless, we can consider Co$'_4$/Pt$_{12}$/Co$_4$ as a hypothetical  system to gain some insight about the indirect spin current generation mechanism. Using the data in Table \ref{tab:paramsPt}, we find the phase shift $\Delta\phi (\mathrm{mod}\; \pi) \approx 0.03\pi$. Thus, the dampinglike and fieldlike torques in Co$'_4$/Pt$_{12}$/Co$_4$ are almost exactly out-of-phase with each other. The spin-rotated components $\dx$ and $\fx$ on Layer $\two$ (with turned-off spin-orbit coupling) are small compared to their non-rotated counterparts $\ds$ and $\fs$, just as in the symmetric Co$_4$/Pt$_{12}$/Co$_4$ system.

We also note rather large values of some coefficients beyond the first three terms in \Eqs{FPdl}{FPfl}, particularly $\dD$. The $\dD$ term reflects spin-rotated current generation at the Pt/Co surface with Dresselhaus-like, rather than Rashba-like, angular dependence.

\section{Conclusions}
\label{conclusions}

The central result of this paper is the identification of the indirect mechanism of spin current generation in $\text{FM}_\two$/N/$\text{FM}_\one$ trilayers, in which the in-plane longitudinal spin current originating from $\text{FM}_\two$ is converted, at the N/$\text{FM}_\one$ interface, into an out-of-plane spin current with a different spin polarization. The absorption of this spin current by $\text{FM}_\two$ results in a predominantly fieldlike spin torque with a spin polarization that depends on the magnetization of $\text{FM}_\one$ and is generally unconventional, i.e., not parallel to the polarization of the spin-Hall current. Similar to the current-in-plane giant magnetoresistance,
the indirect mechanism is only possible if the thickness of the nonmagnetic spacer is smaller than or comparable to the mean-free path and can not be described within the conventional spin-diffusion model.

Our semiclassical simulations and first-principles calculations show that unconventional fieldlike torques can be similar in magnitude to unconventional dampinglike torques, providing a plausible interpretation of the large fieldlike torque with a ``spin-rotated'' symmetry observed experimentally in Ref. \onlinecite{Humphries2017}. Such unconventional fieldlike torques can play an important role in the field-free switching of perpendicularly magnetized ferromagnetic layers.

\acknowledgments

We thank Mark Stiles, Satoru Emori, Xin Fan, and Ilya Krivorotov for useful comments about the manuscript.
This work was supported by the National Science Foundation under Grants No. DMR-2105219 (VPA) and DMR-1916275 and DMR-2324203 (GGBF, KDB).
AAK was supported by the U.S. Department of Energy, Office of Science, Basic Energy Sciences, under Award No. DE-SC0021019. Calculations were performed utilizing the Holland Computing Center of the University of Nebraska, which receives support from the Nebraska Research Initiative.

\appendix

\section{Angular dependence}
\label{app:angular}

Equation (\ref{TorqueTrilayer}) defines the dampinglike and fieldlike torques acting on $\mhat_\A$ which are parametrized by two axial polarization vectors $\FPdla$ and $\FPfla$. Each of those vectors is a function of $\mhat_\B$ and linear in $\mathbf{E}$. Just as in the analysis of the angular dependence of the spin-orbit torque in a bilayer \cite{PhysRevB.101.020407}, we can write an axial vector $\mathbf{A}$ (which stands for $\FPdla$ or $\FPfla$) in terms of a tensor $\hat K$:
\begin{equation}
\mathbf{A}=\hat K(\mhat_\B)\cdot\mathbf{E}
\end{equation}
which can be expanded using vector spherical harmonics (VSH):
\begin{equation}
\hat K(\mhat)=\sum_{lm\nu}\mathbf{Y}_{lm}^{(\nu)}(\mhat)\otimes\mathbf{K}_{lm}^{(\nu)}.
\label{tensorK}
\end{equation}
Here $\mathbf{K}_{lm}^{(\nu)}$ are constant vectors, and symmetry considerations can be used to determine their nonzero components. In the expansion for the spin-orbit torque in bilayers, the torque given by $\mathbf{A}$ is always orthogonal to $\mhat$, and the index $\nu$ takes only two values corresponding to VSH $\mathbf{\Psi}_{lm}$ and $\mathbf{\Phi}_{lm}$. The corresponding real VSH were denoted as $\mathbf{Z}^{(1)}_{lm}$, $\mathbf{Z}^{(2)}_{lm}$ in Ref. \onlinecite{PhysRevB.101.020407}. Here the axial vector $\mathbf{A}$ is not restricted to be orthogonal to $\mhat_\B$, and we need to add the third VSH type, which we denote $\mathbf{Z}^{(3)}_{lm}$, which corresponds to $\mathbf{Y}_{lm}(\mhat)=Y_{lm}(\mhat)\mhat$. The set of $\mathbf{Z}^{(\nu)}_{lm}$ is then a complete orthonormal basis for the expansion of any vector field defined on a sphere. The sum of direct products (\ref{tensorK}) defines the expansion of $\mathbf{A}(\mhat_\B)$ for any value and orientation of $\mathbf{E}$.

We restrict our consideration to the angular dependence allowed by the $C_{\infty v}$ symmetry, even though trilayers under consideration have lower crystallographic symmetries which may allow additional terms. As shown in Ref. \cite{PhysRevB.101.020407}, 
$C_{\infty v}$ only allows VSH with $m=(-1)^l$ if the $x$ axis is aligned with the electric field.

At order $l=1$, we have
\begin{align}
\mathbf{Z}_{1,-1}^{(1)}(\mhat)&\sim\mhat\times(\mathbf{s}\times\mhat)\\
\mathbf{Z}_{1,-1}^{(2)}(\mhat)&\sim\mhat\times\mathbf{s}\\
\mathbf{Z}_{1,-1}^{(3)}(\mhat)&\sim(\mhat\cdot\mathbf{s})\mhat.
\end{align}
The first three terms in Eq. (\ref{FPdl}) and (\ref{FPfl}) represent a linear combination of these terms.

At order $l=2$ the VSH are
\begin{align}
\mathbf{Z}_{2,1}^{(1)}(\mhat)&\sim\mhat\times\mathbf{Z}_{2,1}^{(2)}(\mhat)\\
\mathbf{Z}_{2,1}^{(2)}(\mhat)&\sim(m_xm_y,m_z^2-m_x^2,-m_ym_z)\\
\mathbf{Z}_{2,1}^{(3)}(\mhat)&\sim m_xm_z\mhat.
\end{align}
We can form a linear combination of $\mathbf{Z}_{2,1}^{(1)}$ and $\mathbf{Z}_{2,1}^{(3)}$ which, on account of $\mhat^2=1$, is first order in the components of $\mhat$:
\begin{equation}
\mathbf{Z}_{2,1}^{(1)}(\mhat)+\sqrt{\frac23}\mathbf{Z}_{2,1}^{(3)}(\mhat)\sim(m_z,0,m_x).
\end{equation}
This linear combination is represented by the Dresselhaus terms $d_D$ and $f_D$ in Eq. (\ref{FPdl}) and (\ref{FPfl}).

\section{Numerical solution of the multilayer spin-dependent Boltzmann equation}
\label{AppendixBoltzmann}

To solve the multilayer Boltzmann equation with quantum coherent boundary conditions, we proceed numerically by discretizing the Fermi surface into a mesh of $N_k$ $\threevec{k}$-vectors labeled by index $i$ and $\int_{FS} d\threevec{k}' \longrightarrow \sum_{i} w_i$. Then \Eq{LBEap} can be rewritten as
\begin{align}
\label{LBEDap}
&v_{z,i} \pdf{g_{i\alpha}}{z}
- e E v_{x,i} \delta_{\alpha c}
+ \gamma \epsilon_{\alpha \beta \gamma} H^{\text{ex}}_\beta g_{i\gamma} 	\nonumber	\\
&= 
-R_{i,\alpha\alpha'} g_{i\alpha'}
+ \sum_{j} w_j P_{ij,\alpha\alpha'} g_{j\alpha'}
\end{align}
for an electric field along $x$. Defining
\begin{align}
\label{BMatrix}
B_{ij,\alpha\alpha'} &= \frac{1}{v_{z,i}}(
\gamma \epsilon_{\alpha \beta \gamma} H^{\text{ex}}_\beta \delta_{ij}
+ R_{i,\alpha\alpha'} \delta_{ij}
- w_i P_{ij,\alpha\alpha'}
) 
\end{align}
we may write
\begin{align}
\label{LBEindices}
\pdf{g_{i\alpha}}{z}
+ \sum_j B_{ij,\alpha\alpha'} g_{j\alpha'}
&= 
e E \frac{v_{x,i}}{v_{z,i}} \delta_{\alpha c}.
\end{align}
We define the scattering integrals for a ferromagnetic layer. A nonmagnetic layer is obtained by setting parameters describing majority and minority carriers equal to each other. To proceed, we use the following definitions
\begin{align}
\label{DefineGamma}
\Gamma^{0}_{ij} &= \frac{\delta_{ij}}{v_{z,i}}, \\
\Gamma^{\pm}_{ij} &= -\frac{1}{v_{z,i}} \Big{(} \frac{w_j}{A_\text{FS}} \pm \delta_{ij} \Big{)}, \\
R^\text{rot}_{\alpha\alpha'} &= \epsilon_{\alpha z \alpha'}   \\
\label{Rscat}
R^\text{scat} &=    
    \begin{pmatrix}
    R_\perp          &0     &0     &0    \\
    0      &R_\perp      &0       &0     \\
    0      &0      &R^+_\parallel       &R^-_\parallel     \\
    0      &0      &R^-_\parallel       &R^+_\parallel     \\
    \end{pmatrix},
\end{align}
where $A_\text{FS}$ is the area of the Fermi surface, $R^\text{rot}_{\alpha\alpha'}$ captures spin precession about the magnetization (which leads to dephasing), and $R^\text{scat}$ captures spin-dependent impurity scattering. The scattering rates in \Eq{Rscat} are given by
\begin{align}
\label{Rpar}
R^\pm_\parallel &= \frac{1}{2}\Big{(} \frac{1}{\tau_\uparrow} \pm \frac{1}{\tau_\downarrow} \Big{)}   \\
\label{Rperp}
R_\perp &= \frac{1}{\sqrt{\tau_\uparrow \tau_\downarrow}} + \frac{1}{\tau_\text{ex}}
\end{align}
where $\tau_{\uparrow/\downarrow}$ is the momentum relaxation time for majority/minority carriers, $\tau_\text{ex}$ is the time scale associated with spin precession, and $R_\perp$ is the scattering rate for spins transverse to the magnetization. Using these definitions, we have
\begin{align}
\label{BCode}
B_{ij,\alpha\alpha'} &= \Gamma^{-}_{ij} U^\text{T}_{\alpha\beta}(\mhat) R^\text{scat}_{\beta\beta'} U_{\beta'\alpha'}(\mhat) \nonumber \\
&+ \frac{1}{\tau_\text{sf}} (\Gamma^{-}_{ij} \delta_{\alpha c} + \Gamma^{+}_{ij} (\delta_{\alpha c} - 1)) \delta_{\alpha\alpha'}  \nonumber  \\
&+ \frac{1}{\tau_\text{ex}} \Gamma^{0}_{ij} U^\text{T}_{\alpha\beta}(\mhat) R^\text{rot}_{\beta\beta'} U_{\beta'\alpha'}(\mhat),
\end{align}
where $\tau_\text{sf}$ is the spin-flip lifetime and $U_{\alpha\beta}(\mhat)$ is a rotation matrix in spin space that rotates $\mhat$ to $\zhat$. Note that we have used the relaxation time approximation to simplify the scattering integrals \cite{BoltzmannStiles2}. Dephasing occurs due to spin precession, which is captured by the last term in \Eq{BCode}. However, dephasing is underestimated in systems with spherical Fermi surfaces as compared to systems with realistic electronic structures, so we have added an extra term $1/\tau_\text{ex}$ in \Eq{Rperp} to enhance dephasing.

If the $k$-index $i$ and the spin/charge index $\alpha$ are combined into a single index:
we may write
\begin{align}
\label{LBEmatrix}
\pdf{g}{z} + B g &= E v^*_x
\end{align}
for
\begin{align}
\label{vstardef}
[v^*_x]_{i\alpha} = e \frac{v_{x,i}}{v_{z,i}} \delta_{\alpha c},
\end{align}
where $g$ and $v^*_x$ are vectors of length $4 N_k$ and $B$ is a $4 N_k \times 4 N_k$ matrix. The solution of \Eq{LBEmatrix} can be written as
\begin{align}
\label{solansatz}
g(z) = \sum_n c_n e^{\lambda_n z} \Bar{g}_n + g_\text{P}
\end{align}
where the first term on the right hand side is the sum of homogeneous solutions and the second term is the particular solution. Both the set of vectors $\Bar{g}_n$ and $g_\text{P}$ are determined by solving
\begin{align}
\label{particular}
B \Bar{g}_n &= -\lambda_n \Bar{g}_n   \\
B g_\text{P} &= E v^*_x.     
\end{align}
Techniques to solve these equations can be found in \cite{BoltzmannStiles1,BoltzmannStiles2}. Note that while the Boltzmann distribution function is real-valued, the precession term introduces complex eigenvalues \cite{PhysRevB.87.174411,PhysRevB.94.104420, PhysRevB.94.104419}. However, the homogeneous solutions with complex eigenvalues come in pairs whose eigenvalues are complex conjugates of each other, such that the appropriate linear combinations may be real-valued. When solving the multilayer Boltzmann equation with the boundary conditions specified in this section, the solutions are always real-valued.

The solution in each layer is specified once the $4 N_k$ coefficients $c_n$ for each layer are determined by boundary conditions. For each interface located at $z = z_I$ (where $I$ labels the interface), we define reflection ($R_I$) and transmission ($T_I$) matrices that connect the Boltzmann distributions on each side of the interface in accordance with \Eq{NEscat}. Note that in principle, the reflection matrices corresponding to the left and right sides of the interface can be different, though this is not the case in \Eq{NEscat}. This is because we obtain the scattering matrices at each interface by calculating the reflection and transmission coefficients for plane waves in free space scattering off the delta function potential given by \Eq{IntPot}. Thus, the differences in material properties on each side of the interface are included in the bulk solutions and in the incoming distributions but in our simple model do not affect the reflection and transmission coefficients themselves. Switching to the vector and matrix notation
\begin{align}
g(z_I^\pm)_\text{in} &\longrightarrow g_\alpha(z_I^\pm,\threevec{k}_{||},\mp k_z)   \\
g(z_I^\pm)_\text{out} &\longrightarrow g_\alpha(z_I^\pm,\threevec{k}_{||},\pm k_z)   \\
\mathrm{R}^I &\longrightarrow R^I_{\alpha\beta}(\threevec{k}_{||})  \\
\mathrm{T}^I &\longrightarrow T^I_{\alpha\beta}(\threevec{k}_{||})
\end{align}
where $g(z_I^\pm)_\text{in}$ and $g(z_I^\pm)_\text{out}$ are $2N_k$-length vectors representing the incoming and outgoing portions of the nonequilibrium Boltzmann distribution on each side of interface $I$ and $\mathrm{R}^I$ and $\mathrm{T}^I$ are $2N_k \times 2N_k$ matrices, we may write the boundary conditions at interface $I$ as:
\begin{align}
\label{BC}
\begin{pmatrix}
g(z_I^-)_\text{out} \\
g(z_I^+)_\text{out}
\end{pmatrix}
=
\begin{pmatrix}
\mathrm{R}^I 		&		 \mathrm{T}^I		 \\
\mathrm{T}^I    	&		 \mathrm{R}^I
\end{pmatrix}
\begin{pmatrix}
g(z_I^-)_\text{in} \\
g(z_I^+)_\text{in}
\end{pmatrix}
\end{align}
Finally, we assume that the first and last layers of our heterostructure are capping layers, such that:
\begin{align}
\label{BC2App}
g(z \rightarrow \pm\infty) = 0
\end{align}
For a system with $L$ layers and $L - 1$ interfaces, \Eq{BC} for each interface plus \Eq{BC2App} gives $4 L N_k$ equations, which is sufficient to solve for the $4 L N_k$ unknown coefficients $c_n$, thus fully determining the nonequilibrium Boltzmann distribution in the system. The material parameters used in the Boltzmann calculations are found in Table \ref{BMP}.

\begin{table}
    \centering
    \begin{tabular}{|l|l|l|}
    \hline
    \bf{Bulk parameters} & \bf{Value} & \bf{Units}  \\
    \hline
    $k_{F}$ & 13.58 & nm$^{-1}$    \\
    \hline
    $l^\text{NM}$ & 3.00 & nm    \\
    \hline
    $l^\text{FM}_{\uparrow}$ & 16.25 & nm    \\
    \hline
    $l^\text{FM}_{\downarrow}$ & 6.01 & nm    \\
    \hline
    $l^\text{FM}_\text{sf}$ & 3280 & nm   \\
    \hline
    $l^\text{FM}_\text{ex}$ & 0.258 & nm   \\
    \hline \hline
    \bf{Interfacial parameters} & \bf{Value} & \bf{Units} \\
    \hline
    $\mu_0$ & 0.50 & dimensionless  \\
    \hline
    $\mu_\text{ex}$ & 0.10 & dimensionless  \\
    \hline
    $\mu_\text{soc}$ at FM$_\B$/NM & 0.01 & dimensionless  \\
    \hline
    $\mu_\text{soc}$ at NM/FM$_\A$ & 0.00 & dimensionless  \\
    \hline
\end{tabular}
\caption{Parameters used in the Boltzmann calculations shown in Section \ref{semiclassical}. Each length parameter is converted to the corresponding lifetime by dividing by the Fermi velocity $v_\text{F} = \hbar k_\text{F}/m$. For example, the spin-dependent scattering times in the ferromagnetic layers are given by $\tau_{\uparrow/\downarrow} = l^\text{FM}_{\uparrow/\downarrow}/v_\text{F}$. Parameters labeled FM are used for both ferromagnetic layers while those labeled NM are used for the nonmagnetic spacer layer. The interfacial parameters $\mu_0$ and $\mu_\text{ex}$ are used in the scattering potential given by \Eq{IntPot} for all interfaces.}
\label{BMP}
\end{table}

\section{Symmetry-allowed contributions to $\FPdla$ and $\FPfla$ from the indirect nonlocal mechanism}
\label{AppendixLeakage}

The terms in \Eqs{FPdl}{FPfl} can be related to the antisymmetric spin-spin response tensor $\chi_\text{A}$ by expanding the response tensor in terms of its symmetry-allowed contributions. Here we assume $\Ehat = \xhat$, which sets $\FP = \yhat$. To first order in magnetization $\mhat$, we have 
\begin{align}
\label{chi1storder}
    \chi_\text{A}(\mhat) &=
    \begin{pmatrix}
    0           &a_x m_x          &-a_0    \\
    -a_x m_x           &0          &-a_z m_z    \\
    a_0         &a_z m_z          &0    \\
    \end{pmatrix}
\end{align}
where $a_0$, $a_x$, and $a_z$ are free parameters. Using \Eq{chiA2}, the first-order (in $\mhat$) contribution to the vector $\threevec{f}$ is
\begin{align}
\label{chi2ntorder}
    \threevec{f}^1(\mhat) &=
    \begin{pmatrix}
    -a_z m_z           \\
    a_0            \\
    a_x m_x        \\
    \end{pmatrix}   \nonumber \\
    &= a_0 \yhat + \frac{1}{2}(a_x + a_z) \mhat \times \yhat \nonumber \\
    &+ \frac{1}{2}(a_x - a_z) \beta^D \mhat.
\end{align}
Noting again that $\FP = \yhat$ and making the following substitutions
\begin{align}
\label{sub1}
    \fs &= a_0 \\ 
    \fx &= \frac{1}{2}(a_x + a_z) \\
    \fD &= \frac{1}{2}(a_x - a_z)
\end{align}
we reproduce the first order in $\mhat$ terms in \Eq{FPfl}. 

The spin-spin response tensor to second order in $\mhat$ is
\begin{align}
\label{f1st}
    &\chi_\text{A}(\mhat) = \nonumber \\
    &\begin{pmatrix}
    0           &b_{z} m_z m_y          &-b_{x} m_y^2 - b_{zz} m_z^2    \\
    -b_{z} m_z m_y           &0          &b_{x} m_x m_y    \\
    b_{x} m_y^2 + b_{zz} m_z^2         &-b_{x} m_x m_y          &0    \\
    \end{pmatrix}
\end{align}
Using \Eq{chiA2}, the second order in $\mhat$ contribution to the vector $\threevec{f}$ is
\begin{align}
\label{f2nd}
    \threevec{f}^2(\mhat) &=
    \begin{pmatrix}
    b_{x} m_x m_y           \\
    b_{x} m_y^2 + a_{zz} m_z^2            \\
    b_{z} m_z m_y        \\
    \end{pmatrix}  
\end{align}
which can be rewritten as
\begin{align}
\label{f2ndRW}
    \threevec{f}^2(\mhat) &=
    b_{x} (\mhat \cdot \yhat) \mhat + b_{zz} m_z^2 \yhat \nonumber \\
    &+ (b_{z}-b_{x}) (\mhat \cdot \yhat) m_z \zhat
\end{align}
Making the following substitutions
\begin{align}
\label{sub2}
    \fm - f_s &= b_x \\ 
    4\ftwo &= b_{zz}   \\ 
    4\fz &= b_{z}-b_{x}
\end{align}
we reproduce the second order in $\mhat$ terms in \Eq{FPfl}. 

\section{Results of first-principles calculations}
\label{app:fp}

Table \ref{tab:corr} shows how various terms in \Eqs{TorqueBilayer}{FPdl} contribute to the expansion parameters for the configurations appearing in Tables \ref{tab:trq} and \ref{tab:2nd}. In addition to the terms appearing in \Eqs{FPdl}{FPfl}, we include the $\mathbf{Z}^{(1)}_{2,1}$ and $\mathbf{Z}^{(2)}_{2,1}$ harmonics in the $\tilde\torque_a(\mhat_\A)$ term in \Eq{TorqueTrilayer} with coefficients $P_1$ and $P_2$, respectively. Among the systems we have considered, these latter terms are only appreciable for Co/Pt/Co trilayers.

\begin{table*}
\centering
\setlength{\tabcolsep}{4.45pt}{
\caption{Expansion coefficients $C$, $A_1$, $B_1$ for the torquances, $\boldsymbol{\tau}_a\cdot\mhat_\alpha=C+A_1\cos\theta+B_1\sin\theta+A_2\cos2\theta+B_2\sin2\theta$, where $\theta$ is the angle between the rotating magnetization and the $\zhat$ or (if rotating in the $xy$ plane) the $\xhat$ axis. Second-harmonic coefficients $A_2$, $B_2$ are non-zero only in Co/Pt/Co trilayers; they are listed separately in Table \ref{tab:2nd}. Layer $a$ is at the top, $b$ at the top of the trilayer, and $\mhat_\times=\mhat_\one\times\mhat_\two$. Dashes indicate terms forbidden by symmetry. Units of \SI{e5}{\per\ohm\per\meter}. Typical error bars are on the order of 0.1 \SI{e5}{\per\ohm\per\meter}. A prime indicates that spin-orbit coupling is turned off in layer $a$ and in half of the adjacent spacer layer. Entry in square brackets: predicted by additivity (see Sec. \ref{sec:SOC}).}
\label{tab:trq}
\begin{tabular}{|c|c||c|c|c||c|c|c||c|c|c||c|c|c|}
\hline
\multirow{2}{*}{System} & \multirow{2}{*}{\shortstack[c]{Fixed\\layer}} & \multicolumn{3}{c||}{$\boldsymbol\tau_\two\cdot\mathbf{m}_\one$} & \multicolumn{3}{c||}{$\boldsymbol\tau_\two\cdot\mathbf{m}_\times$} & \multicolumn{3}{c||}{$\boldsymbol\tau_\one\cdot\mathbf{m}_\two$} & \multicolumn{3}{c|}{$\boldsymbol\tau_\one\cdot\mathbf{m}_\times$}\\
\cline{3-14}
& & $C$ & $A_1$ & $B_1$ & $C$ & $A_1$ & $B_1$ & $C$ & $A_1$ & $B_1$ & $C$ & $A_1$ & $B_1$ \\
\hline
\multirow{3}{*}{\shortstack[c]{Co$^\prime_4$Cu$_4$Co$_4$\\40 mRy}}
   &  $\mathbf{m}_\one\parallel -\hat x$ & \text{---} & 0.11 & 0.50 & \text{---} & -0.66 & 0.14 & \text{---} & 1.08 & 0.08 & \text{---} & 0.69 & -1.21 \\
   &  $\mathbf{m}_\one\parallel -\hat y$  & 0.06 & \text{---} & \text{---} & 0.12 & \text{---} & \text{---} & 0.45 & \text{---} & \text{---} & 0.04 & \text{---} & \text{---} \\
   &  $\mathbf{m}_\one\parallel -\hat z$  & \text{---} & -0.10 & 0.45 & \text{---} & 0.73 & 0.20 & \text{---} & -1.07 & 0.04 & \text{---} & -0.74 & -1.26 \\
\cline{2-14}
   & $\mathbf{m}_\two\parallel -\hat x$ & \text{---} & -0.10 & 0.06 & \text{---} & 0.78 & 0.05 & \text{---} & -1.07 & 0.50 & \text{---} & -0.78 & 0.05 \\
   & $\mathbf{m}_\two\parallel -\hat y$  & 0.48 & \text{---} & \text{---} & 0.13 & \text{---} & \text{---} & 0.04 & \text{---} & \text{---} & -1.13 & \text{---} & \text{---} \\
   & $\mathbf{m}_\two\parallel -\hat z$  & \text{---} & 0.11 & 0.05 & \text{---} & -0.73 & 0.08 & \text{---} & 1.04 & 0.49 & \text{---} & 0.79 & -0.00 \\
\hline
\emph{id.}, 80 mRy
   & $\mathbf{m}_\one\parallel -\hat x$  & \text{---} & 0.16 & 0.34 & \text{---} & -0.55 & -0.14 & \text{---} & 0.43 & 0.05 & \text{---} & 0.56 & -0.34 \\
   & $\mathbf{m}_\one\parallel -\hat y$  & -0.02 & \text{---} & \text{---} & 0.07 & \text{---} & \text{---} & 0.29 & \text{---} & \text{---} & 0.13 & \text{---} & \text{---} \\
\hline
\multirow{3}{*}{\shortstack[c]{[Co$_4$Cu$_4$Co$_4$]\\40 mRy\\ Predicted}}
   &  $\mathbf{m}_\one\parallel -\hat x$ & \text{---} & -0.97 & 0.95 & \text{---} & -1.35 & 0.18 & \text{---} & 0.97 & 0.14 & \text{---} & 1.35 & -1.09\\
   &  $\mathbf{m}_\one\parallel -\hat y$  & 0.12 & \text{---} & \text{---} & -1.115 & \text{---} & \text{---} & 0.925 & \text{---} & \text{---} & 0.21 & \text{---} & \text{---} \\
   &  $\mathbf{m}_\one\parallel -\hat z$  & \text{---} & 0.97 & 0.9 & \text{---} & 1.47 & 0.24 & \text{---} & -0.97 & 0.1 & \text{---} & -1.47 & -1.14 \\
\hline
\multirow{3}{*}{\shortstack[c]{Co$_4$Cu$_4$Co$_4$\\40 mRy}}
    & $\mathbf{m}_\one\parallel -\hat x$  & \text{---} & -0.94 & 0.97 & \text{---} & -1.33 & 0.15 & \text{---} & 0.92 & 0.16 & \text{---} & 1.36 & -1.06 \\
   & $\mathbf{m}_\one\parallel -\hat y$  & 0.16 & \text{---} & \text{---} & -1.02 & \text{---} & \text{---} & 0.89 & \text{---} & \text{---} & 0.27 & \text{---} & \text{---} \\
   &  $\mathbf{m}_\one\parallel -\hat z$ & \text{---} & 0.89 & 0.91 & \text{---} & 1.50 & 0.15 & \text{---} & -0.91 & 0.12 & \text{---} & -1.44 & -1.10 \\
\hline
\multirow{3}{*}{\shortstack[c]{Co$^\prime_4$Cu$_{12}$Co$_4$\\40 mRy}}
   & $\mathbf{m}_\one\parallel -\hat x$  & \text{---} & 0.13 & 0.90 & \text{---} & -1.07 & 0.23 & \text{---} & 2.46 & 0.17 & \text{---} & 1.30 & -2.73 \\
   & $\mathbf{m}_\one\parallel -\hat y$  & 0.04 & \text{---} & \text{---} & 0.06 & \text{---} & \text{---} & 0.72 & \text{---} & \text{---} & -0.05 & \text{---} & \text{---} \\
   & $\mathbf{m}_\one\parallel -\hat z$  & \text{---} & -0.10 & 0.82 & \text{---} & 1.02 & 0.22 & \text{---} & -2.46 & 0.13 & \text{---} & -1.18 & -2.63 \\
\hline
\emph{id.}, 80 mRy
   & $\mathbf{m}_\one\parallel -\hat x$  & \text{---} & 0.13 & 0.46 & \text{---} & -0.55 & -0.19 & \text{---} & 0.10 & 0.13 & \text{---} & 0.69 & -0.79 \\
   & $\mathbf{m}_\one\parallel -\hat y$  & -0.01 & \text{---} & \text{---} & 0.02 & \text{---} & \text{---} & 0.41 & \text{---} & \text{---} & 0.08 & \text{---} & \text{---} \\
\hline
\multirow{3}{*}{\shortstack[c]{Py$^\prime_4$Cu$_{4}$Py$_4$\\40 mRy}}
   &  $\mathbf{m}_\one\parallel -\hat x$ & \text{---} & 1.18 & 0.81 & \text{---} & -1.23 & -1.04 & \text{---} & 1.89 & 0.25 & \text{---} & 1.43 & -0.88 \\
   &  $\mathbf{m}_\one\parallel -\hat y$ & 0.06 & \text{---} & \text{---} & 0.42 & \text{---} & \text{---} & 0.96 & \text{---} & \text{---} & 0.87 & \text{---} & \text{---} \\
   &  $\mathbf{m}_\one\parallel -\hat z$ & \text{---} & -1.20 & 0.85 & \text{---} & 1.00 & -0.92 & \text{---} & -1.88 & 0.15 & \text{---} & -1.17 & -0.97 \\
\multirow{3}{*}{\shortstack[c]{Py$^\prime_4$Cu$_{4}$Py$_4$\\80 mRy}} 
   &  $\mathbf{m}_\one\parallel -\hat x$  & \text{---} & 0.25 & 0.14 & \text{---} & -0.57 & -0.44 & \text{---} & 0.66 & 0.01 & \text{---} & 0.69 & -0.19 \\
   & $\mathbf{m}_\one\parallel -\hat y$  & 0.08 & \text{---} & \text{---} & 0.08 & \text{---} & \text{---} & 0.08 & \text{---} & \text{---} & 0.24 & \text{---} & \text{---} \\
   & $\mathbf{m}_\one\parallel -\hat z$  & \text{---} & -0.23 & -0.01 & \text{---} & 0.35 & -0.37 & \text{---} & -0.64 & 0.19 & \text{---} & -0.39 & -0.38 \\
   \hline
\multirow{3}{*}{\shortstack[c]{Py$^\prime_4$Cu$_{4}$Py$_4$\\ (001) \\40 mRy}}
   &  $\mathbf{m}_\one\parallel -\hat x$ & \text{---} & 0.93 & 0.73 & \text{---} & -1.00 & -1.01 & \text{---} & 0.89 & 0.17 & \text{---} & 1.17 & 0.08 \\
   &  $\mathbf{m}_\one\parallel -\hat y$ & 0.06 & \text{---} & \text{---} & 0.02 & \text{---} & \text{---} & 0.72 & \text{---} & \text{---} & 0.98 & \text{---} & \text{---} \\
   &  $\mathbf{m}_\one\parallel -\hat z$ & \text{---} & -0.90 & 0.69 & \text{---} & 0.84 & -0.93 & \text{---} & -0.88 & 0.01 & \text{---} & -0.90 & -0.02 \\
   \hline
\multirow{3}{*}{\shortstack[c]{[Py$_4$Cu$_{4}$Py$_4$]\\40 mRy\\Predicted}}
   &  $\mathbf{m}_\one\parallel -\hat x$ & \text{---} & -0.71 & 1.77 & \text{---} & -2.66 & -0.17 & \text{---} & 0.71 & 0.31 & \text{---} & 2.66 & -0.46\\
   &  $\mathbf{m}_\one\parallel -\hat y$  & 0.26 & \text{---} & \text{---} & -0.51 & \text{---} & \text{---} & 1.79 & \text{---} & \text{---} & -0.11 & \text{---} & \text{---} \\
   &  $\mathbf{m}_\one\parallel -\hat z$  & \text{---} & 0.68 & 1.81 & \text{---} & 2.17 & -0.05 & \text{---} & -0.68 & 0.21 & \text{---} & -2.17 & -0.55 \\
   \hline
\multirow{3}{*}{\shortstack[c]{Py$_4$Cu$_{4}$Py$_4$\\40 mRy}}
   &   $\mathbf{m}_\one\parallel -\hat x$  & \text{---}  & -0.53 & 1.45 & \text{---} & -2.42 & -0.05 & \text{---} & 0.64 & 0.47 & \text{---} & 2.44 & -0.61 \\
   &   $\mathbf{m}_\one\parallel -\hat y$ & 0.44 & \text{---} & \text{---} & -0.51 & \text{---} & \text{---} & 1.48 & \text{---} & \text{---} & -0.15 & \text{---} & \text{---} \\
   &   $\mathbf{m}_\one\parallel -\hat z$ & \text{---} & 0.57 & 1.61 & \text{---} & 2.42 & 0.08 & \text{---} & -0.55 & 0.41 & \text{---} & -2.35 & -0.69 \\
   \hline
\multirow{3}{*}{\shortstack[c]{Py$_{12}^\prime$Cu$_{4}$Py$_{12}$\\40 mRy}}
   &  $\mathbf{m}_\one\parallel -\hat x$  & \text{---} & 0.42 & 0.38 & \text{---} & -0.59 & -0.54 & \text{---} & 1.54 & 0.08 & \text{---} & 0.71 & -0.88 \\
   &  $\mathbf{m}_\one\parallel -\hat y$  & 0.05 & \text{---} & \text{---} & 0.14 & \text{---} & \text{---} & 0.52 & \text{---} & \text{---} & 0.57 & \text{---} & \text{---} \\
   & $\mathbf{m}_\one\parallel -\hat z$  & \text{---} & -0.38 & 0.31 & \text{---} & 0.63 & -0.23 & \text{---} & -1.57 & 0.14 & \text{---} & -0.61 & -1.23 \\
\hline
\multirow{3}{*}{\shortstack[c]{Co$^\prime_4$Pt$^\prime$$_4$Co$_4$\\40 mRy}}
& $\mathbf{m}_\one\parallel -\hat x$ & \text{---} & 0.12 & -0.17 & \text{---} & 0.09 & -0.18 & \text{---} & -0.84 & -0.05 & \text{---} & -0.11 & 0.99 \\
&  $\mathbf{m}_\one\parallel -\hat y$ & 0.04 & \text{---} & \text{---} & 0.27 & \text{---} & \text{---} & -0.28 & \text{---} & \text{---} & -0.08 & \text{---} & \text{---} \\
& $\mathbf{m}_\one\parallel -\hat z$ & \text{---} & -0.16 & -0.22 & \text{---} & -0.24 & -0.17 & \text{---} & 0.86 & -0.10 & \text{---} & 0.30 & 1.06 \\
\hline
\multirow{3}{*}{\shortstack[c]{[Co$_4$Pt$_4$Co$_4$]\\40 mRy\\Predicted}}
  &  $\mathbf{m}_\one\parallel -\hat x$ & \text{---} & 0.96 & -0.45 & \text{---} & 0.20 & -0.26 & \text{---} & -0.96 & -0.01 & \text{---} & -0.20 & 1.26 \\
   &  $\mathbf{m}_\one\parallel -\hat y$  & -0.04 & \text{---} & \text{---} & 1.30 & \text{---} & \text{---} & -0.48 & \text{---} & \text{---} & -0.26 & \text{---} & \text{---} \\
   &  $\mathbf{m}_\one\parallel -\hat z$  & \text{---} & -1.02 & -0.50 & \text{---} & -0.54 & -0.25 & \text{---} & 1.02 & -0.06 & \text{---} & 0.54 & 1.33 \\
\hline
\multirow{3}{*}{\shortstack[c]{Co$_4$Pt$_4$Co$_4$\\40 mRy}}   
&  $\mathbf{m}_\one\parallel -\hat x$ & \text{---} & -1.02 & -0.33 & \text{---} & -2.20 & -0.19 & \text{---} & 1.60 & 1.03 & \text{---} & 2.38 & -1.50 \\
&  $\mathbf{m}_\one\parallel -\hat y$ & 1.01 & \text{---} & \text{---} & -1.32 & \text{---} & \text{---} & -0.28 & \text{---} & \text{---} & 0.31 & \text{---} & \text{---} \\
&  $\mathbf{m}_\one\parallel -\hat z$ & \text{---} & 1.70 & -0.22 & \text{---} & 2.27 & 0.42 & \text{---} & -1.14 & 0.96 & \text{---} & -2.17 & -1.28 \\
\hline
\multirow{3}{*}{\shortstack[c]{Co$_4$Pt$_{12}$Co$_4$\\ 40 mRy}}
&  $\mathbf{m}_\one\parallel -\hat x$ & \text{---} & -1.59 & 0.22 & \text{---} & -2.83 & -0.43 & \text{---} & 2.09 & 1.6 & \text{---} & 2.72 & -1.61 \\
&  $\mathbf{m}_\one\parallel -\hat y$ & 1.64 & \text{---} & \text{---} & -1.34 & \text{---} & \text{---} & 0.19 & \text{---} & \text{---} & -0.20 & \text{---} & \text{---} \\
&  $\mathbf{m}_\one\parallel -\hat z$ & \text{---} & 2.13 & 0.17 & \text{---} & 2.7 & -0.00 & \text{---} & -1.59 & 1.78 & \text{---} & -2.93 & -1.27 \\
\hline
\multirow{3}{*}{\shortstack[c]{Co$^\prime_4$Pt$_{12}$Co$_4$\\40 mRy}}
   &  $\mathbf{m}_\one\parallel -\hat x$  & \text{---} & -0.86 & 0.06 & \text{---} & -1.95 & -0.26 & \text{---} & 1.51 & 1.95 & \text{---} & 2.62 & -1.26 \\
   &  $\mathbf{m}_\one\parallel -\hat y$  & 1.01 & \text{---} & \text{---} & -0.50 & \text{---} & \text{---} & 0.01 & \text{---} & \text{---} & -0.51 & \text{---} & \text{---} \\
   & $\mathbf{m}_\one\parallel -\hat z$   & \text{---} & 1.30 & 0.25 & \text{---} & 1.54 & 0.65 & \text{---} & -1.05 & 1.70 & \text{---} & -2.40 & -1.19 \\
\hline
\end{tabular}}
\end{table*}

\begin{table*}[htb]
\caption{Second-order coefficients $A_2$, $B_2$ for the torquances in Co$_4$/Pt$_N$/Co$_4$ trilayers with $V_m=40$ mRy and $\mhat_\one\parallel-\hat y$. The definition is given in Table \ref{tab:trq}.}
\label{tab:2nd}
\centering
\setlength{\tabcolsep}{8.0pt}
\begin{tabular}{|c|c||c|c||c|c||c|c||c|c|}
\hline
\multirow{2}{*}{\( N \)} & \multirow{2}{*}{\shortstack[c]{Fixed\\layer}} & \multicolumn{2}{c||}{\( \boldsymbol\tau_\two\cdot\mathbf{m}_\one \)} & \multicolumn{2}{c||}{\( \boldsymbol\tau_\two\cdot\mathbf{m}_\times \)} & \multicolumn{2}{c||}{\( \boldsymbol\tau_\one\cdot\mathbf{m}_\two \)} & \multicolumn{2}{c|}{\( \boldsymbol\tau_\one\cdot\mathbf{m}_\times \)}\\
\cline{3-10}
& &  $A_2$ & $B_2$  & $A_2$ & $B_2$ & $A_2$ & $B_2$ & $A_2$ & $B_2$ \\
\hline  
   4 & $\mathbf{m}_\one\parallel -\hat y$ & 0.04 & -0.07 & -0.14 & 0.11 & 0.06 & -0.27 & -0.17 & -0.16 \\
\hline
  12 & $\mathbf{m}_\one\parallel -\hat y$ & 0.07 & -0.08 & 0.08 & -0.06 & -0.02 &  -0.21 & -0.17 & 0.01 \\
\cline{2-10}
\hline
\end{tabular}
\end{table*}

\begin{table*}[htb]
\caption{Correspondence of the expansion terms listed in Tables \ref{tab:trq} and \ref{tab:2nd} with the analytical expressions in Eqs. (\ref{FPdl})-(\ref{FPfl}). Coefficients $P_1$, $P_2$ give contributions from scaled $l=2$ harmonics, $\sqrt{5/3}\,\mathbf{Z}^{(1)}_{2,1}$ and $\sqrt{5/3}\,\mathbf{Z}^{(2)}_{2,1}$, respectively. The form for $\boldsymbol\tau_\one\cdot\mathbf{m}_\two$ and $\boldsymbol\tau_\one\cdot\mathbf{m}_\times$ components acting on layer 1 can be obtained by replacing $1 \leftrightarrow 2$ (which also changes the sign of $\mhat_\times$) and then reversing all signs. The coefficients for layers 1 and 2 are generally different, but they are related by symmetry for a symmetric trilayer.}
\begin{tabular}{|c||c|c|c|c|c||c|c|c|c|c|}
\hline
\multirow{2}{*}{\shortstack[c]{Fixed\\layer}} & \multicolumn{5}{c||}{$\boldsymbol\tau_\two\cdot\mathbf{m}_\one$} & \multicolumn{5}{c|}{$\boldsymbol\tau_\two\cdot\mathbf{m}_\times$} \\
\cline{2-11}
   & $C$ & $A_1$ & $B_1$ & $A_2$ & $B_2$ & $C$ & $A_1$ & $B_1$ & $A_2$ & $B_2$\\
\hline
   $\mathbf{m}_\one\parallel -\hat x$  & \text{---} & $-\fs-P_1$ & $-\fx-\fD$ & & &\text{---} & $-\ds+P_2$ & $-\dx-\dD$ & & \\
   $\mathbf{m}_\one\parallel -\hat y$  & $-\dm$ & \text{---} & \text{---} & $P_2$ & & $\fm$ & \text{---} & \text{---} & $P_1$ & \\
   $\mathbf{m}_\one\parallel -\hat z$  & \text{---} & $\fs-P_1+4f_2$ & $-\fx+\fD$ & & & \text{---} & $\ds+P_2+4d_2$ & $-\dx+\dD$ & & \\
   \cline{2-11}
   $\mathbf{m}_\two\parallel -\hat x$ & \text{---} & \scriptsize{$\fs-P_1+3f_2-f_z$} & \scriptsize{$-\dm-P_2-\dtwo-\dz$} & & & \text{---} & $\ds+P_2+3\dtwo-\dz$ & $\fm-P_1+f_2+\fz$ & & \\
   $\mathbf{m}_\two\parallel -\hat y$  & $-\fx$ & \text{---} & \text{---} & $\fD$ & $\dD$ & $-\dx$ & \text{---} & \text{---} & $\dD$ & $-\fD$\\
   $\mathbf{m}_\two\parallel -\hat z$  & \text{---} & \scriptsize{$-\fs-P_1$} & \scriptsize{$-\dm+P_2$} & & & \text{---} & $-\ds+P_2$ & $\fm+P_1$ & & \\
\hline
\end{tabular}
\label{tab:corr}
\end{table*}

\clearpage


\begin{thebibliography}{69}%
\makeatletter
\providecommand \@ifxundefined [1]{%
 \@ifx{#1\undefined}
}%
\providecommand \@ifnum [1]{%
 \ifnum #1\expandafter \@firstoftwo
 \else \expandafter \@secondoftwo
 \fi
}%
\providecommand \@ifx [1]{%
 \ifx #1\expandafter \@firstoftwo
 \else \expandafter \@secondoftwo
 \fi
}%
\providecommand \natexlab [1]{#1}%
\providecommand \enquote  [1]{``#1''}%
\providecommand \bibnamefont  [1]{#1}%
\providecommand \bibfnamefont [1]{#1}%
\providecommand \citenamefont [1]{#1}%
\providecommand \href@noop [0]{\@secondoftwo}%
\providecommand \href [0]{\begingroup \@sanitize@url \@href}%
\providecommand \@href[1]{\@@startlink{#1}\@@href}%
\providecommand \@@href[1]{\endgroup#1\@@endlink}%
\providecommand \@sanitize@url [0]{\catcode `\\12\catcode `\$12\catcode `\&12\catcode `\#12\catcode `\^12\catcode `\_12\catcode `\%12\relax}%
\providecommand \@@startlink[1]{}%
\providecommand \@@endlink[0]{}%
\providecommand \url  [0]{\begingroup\@sanitize@url \@url }%
\providecommand \@url [1]{\endgroup\@href {#1}{\urlprefix }}%
\providecommand \urlprefix  [0]{URL }%
\providecommand \Eprint [0]{\href }%
\providecommand \doibase [0]{https://doi.org/}%
\providecommand \selectlanguage [0]{\@gobble}%
\providecommand \bibinfo  [0]{\@secondoftwo}%
\providecommand \bibfield  [0]{\@secondoftwo}%
\providecommand \translation [1]{[#1]}%
\providecommand \BibitemOpen [0]{}%
\providecommand \bibitemStop [0]{}%
\providecommand \bibitemNoStop [0]{.\EOS\space}%
\providecommand \EOS [0]{\spacefactor3000\relax}%
\providecommand \BibitemShut  [1]{\csname bibitem#1\endcsname}%
\let\auto@bib@innerbib\@empty
\bibitem [{\citenamefont {Binasch}\ \emph {et~al.}(1989)\citenamefont {Binasch}, \citenamefont {Gr\"unberg}, \citenamefont {Saurenbach},\ and\ \citenamefont {Zinn}}]{PhysRevB.39.4828}%
  \BibitemOpen
  \bibfield  {author} {\bibinfo {author} {\bibfnamefont {G.}~\bibnamefont {Binasch}}, \bibinfo {author} {\bibfnamefont {P.}~\bibnamefont {Gr\"unberg}}, \bibinfo {author} {\bibfnamefont {F.}~\bibnamefont {Saurenbach}},\ and\ \bibinfo {author} {\bibfnamefont {W.}~\bibnamefont {Zinn}},\ }\href {https://doi.org/10.1103/PhysRevB.39.4828} {\bibfield  {journal} {\bibinfo  {journal} {Phys. Rev. B}\ }\textbf {\bibinfo {volume} {39}},\ \bibinfo {pages} {4828} (\bibinfo {year} {1989})}\BibitemShut {NoStop}%
\bibitem [{\citenamefont {Baibich}\ \emph {et~al.}(1988)\citenamefont {Baibich}, \citenamefont {Broto}, \citenamefont {Fert}, \citenamefont {Van~Dau}, \citenamefont {Petroff}, \citenamefont {Etienne}, \citenamefont {Creuzet}, \citenamefont {Friederich},\ and\ \citenamefont {Chazelas}}]{PhysRevLett.61.2472}%
  \BibitemOpen
  \bibfield  {author} {\bibinfo {author} {\bibfnamefont {M.~N.}\ \bibnamefont {Baibich}}, \bibinfo {author} {\bibfnamefont {J.~M.}\ \bibnamefont {Broto}}, \bibinfo {author} {\bibfnamefont {A.}~\bibnamefont {Fert}}, \bibinfo {author} {\bibfnamefont {F.~N.}\ \bibnamefont {Van~Dau}}, \bibinfo {author} {\bibfnamefont {F.}~\bibnamefont {Petroff}}, \bibinfo {author} {\bibfnamefont {P.}~\bibnamefont {Etienne}}, \bibinfo {author} {\bibfnamefont {G.}~\bibnamefont {Creuzet}}, \bibinfo {author} {\bibfnamefont {A.}~\bibnamefont {Friederich}},\ and\ \bibinfo {author} {\bibfnamefont {J.}~\bibnamefont {Chazelas}},\ }\href {https://doi.org/10.1103/PhysRevLett.61.2472} {\bibfield  {journal} {\bibinfo  {journal} {Phys. Rev. Lett.}\ }\textbf {\bibinfo {volume} {61}},\ \bibinfo {pages} {2472} (\bibinfo {year} {1988})}\BibitemShut {NoStop}%
\bibitem [{\citenamefont {Maekawa}\ and\ \citenamefont {Gafvert}(1982)}]{Maekawa1982}%
  \BibitemOpen
  \bibfield  {author} {\bibinfo {author} {\bibfnamefont {S.}~\bibnamefont {Maekawa}}\ and\ \bibinfo {author} {\bibfnamefont {U.}~\bibnamefont {Gafvert}},\ }\href {https://doi.org/10.1109/tmag.1982.1061834} {\bibfield  {journal} {\bibinfo  {journal} {{IEEE} Trans. Magn.}\ }\textbf {\bibinfo {volume} {18}},\ \bibinfo {pages} {707} (\bibinfo {year} {1982})}\BibitemShut {NoStop}%
\bibitem [{\citenamefont {Moodera}\ \emph {et~al.}(1995)\citenamefont {Moodera}, \citenamefont {Kinder}, \citenamefont {Wong},\ and\ \citenamefont {Meservey}}]{PhysRevLett.74.3273}%
  \BibitemOpen
  \bibfield  {author} {\bibinfo {author} {\bibfnamefont {J.~S.}\ \bibnamefont {Moodera}}, \bibinfo {author} {\bibfnamefont {L.~R.}\ \bibnamefont {Kinder}}, \bibinfo {author} {\bibfnamefont {T.~M.}\ \bibnamefont {Wong}},\ and\ \bibinfo {author} {\bibfnamefont {R.}~\bibnamefont {Meservey}},\ }\href {https://doi.org/10.1103/PhysRevLett.74.3273} {\bibfield  {journal} {\bibinfo  {journal} {Phys. Rev. Lett.}\ }\textbf {\bibinfo {volume} {74}},\ \bibinfo {pages} {3273} (\bibinfo {year} {1995})}\BibitemShut {NoStop}%
\bibitem [{\citenamefont {Butler}\ \emph {et~al.}(2001)\citenamefont {Butler}, \citenamefont {Zhang}, \citenamefont {Schulthess},\ and\ \citenamefont {MacLaren}}]{PhysRevB.63.054416}%
  \BibitemOpen
  \bibfield  {author} {\bibinfo {author} {\bibfnamefont {W.~H.}\ \bibnamefont {Butler}}, \bibinfo {author} {\bibfnamefont {X.-G.}\ \bibnamefont {Zhang}}, \bibinfo {author} {\bibfnamefont {T.~C.}\ \bibnamefont {Schulthess}},\ and\ \bibinfo {author} {\bibfnamefont {J.~M.}\ \bibnamefont {MacLaren}},\ }\href {https://doi.org/10.1103/PhysRevB.63.054416} {\bibfield  {journal} {\bibinfo  {journal} {Phys. Rev. B}\ }\textbf {\bibinfo {volume} {63}},\ \bibinfo {pages} {054416} (\bibinfo {year} {2001})}\BibitemShut {NoStop}%
\bibitem [{\citenamefont {Yuasa}\ \emph {et~al.}(2004)\citenamefont {Yuasa}, \citenamefont {Nagahama}, \citenamefont {Fukushima}, \citenamefont {Suzuki},\ and\ \citenamefont {Ando}}]{Yuasa2004}%
  \BibitemOpen
  \bibfield  {author} {\bibinfo {author} {\bibfnamefont {S.}~\bibnamefont {Yuasa}}, \bibinfo {author} {\bibfnamefont {T.}~\bibnamefont {Nagahama}}, \bibinfo {author} {\bibfnamefont {A.}~\bibnamefont {Fukushima}}, \bibinfo {author} {\bibfnamefont {Y.}~\bibnamefont {Suzuki}},\ and\ \bibinfo {author} {\bibfnamefont {K.}~\bibnamefont {Ando}},\ }\href {https://doi.org/10.1038/nmat1257} {\bibfield  {journal} {\bibinfo  {journal} {Nat. Mater.}\ }\textbf {\bibinfo {volume} {3}},\ \bibinfo {pages} {868} (\bibinfo {year} {2004})}\BibitemShut {NoStop}%
\bibitem [{\citenamefont {Parkin}\ \emph {et~al.}(2004)\citenamefont {Parkin}, \citenamefont {Kaiser}, \citenamefont {Panchula}, \citenamefont {Rice}, \citenamefont {Hughes}, \citenamefont {Samant},\ and\ \citenamefont {Yang}}]{Parkin2004}%
  \BibitemOpen
  \bibfield  {author} {\bibinfo {author} {\bibfnamefont {S.~S.~P.}\ \bibnamefont {Parkin}}, \bibinfo {author} {\bibfnamefont {C.}~\bibnamefont {Kaiser}}, \bibinfo {author} {\bibfnamefont {A.}~\bibnamefont {Panchula}}, \bibinfo {author} {\bibfnamefont {P.~M.}\ \bibnamefont {Rice}}, \bibinfo {author} {\bibfnamefont {B.}~\bibnamefont {Hughes}}, \bibinfo {author} {\bibfnamefont {M.}~\bibnamefont {Samant}},\ and\ \bibinfo {author} {\bibfnamefont {S.-H.}\ \bibnamefont {Yang}},\ }\href {https://doi.org/10.1038/nmat1256} {\bibfield  {journal} {\bibinfo  {journal} {Nat. Mater.}\ }\textbf {\bibinfo {volume} {3}},\ \bibinfo {pages} {862} (\bibinfo {year} {2004})}\BibitemShut {NoStop}%
\bibitem [{\citenamefont {Kent}\ and\ \citenamefont {Worledge}(2015)}]{Kent2015}%
  \BibitemOpen
  \bibfield  {author} {\bibinfo {author} {\bibfnamefont {A.~D.}\ \bibnamefont {Kent}}\ and\ \bibinfo {author} {\bibfnamefont {D.~C.}\ \bibnamefont {Worledge}},\ }\href {https://doi.org/10.1038/nnano.2015.24} {\bibfield  {journal} {\bibinfo  {journal} {Nat. Nanotechnol.}\ }\textbf {\bibinfo {volume} {10}},\ \bibinfo {pages} {187} (\bibinfo {year} {2015})}\BibitemShut {NoStop}%
\bibitem [{\citenamefont {Slonczewski}(1996)}]{Slonczewski1996}%
  \BibitemOpen
  \bibfield  {author} {\bibinfo {author} {\bibfnamefont {J.}~\bibnamefont {Slonczewski}},\ }\href {https://doi.org/10.1016/0304-8853(96)00062-5} {\bibfield  {journal} {\bibinfo  {journal} {J. Magn. Magn. Mater.}\ }\textbf {\bibinfo {volume} {159}},\ \bibinfo {pages} {L1} (\bibinfo {year} {1996})}\BibitemShut {NoStop}%
\bibitem [{\citenamefont {Berger}(1996)}]{PhysRevB.54.9353}%
  \BibitemOpen
  \bibfield  {author} {\bibinfo {author} {\bibfnamefont {L.}~\bibnamefont {Berger}},\ }\href {https://doi.org/10.1103/PhysRevB.54.9353} {\bibfield  {journal} {\bibinfo  {journal} {Phys. Rev. B}\ }\textbf {\bibinfo {volume} {54}},\ \bibinfo {pages} {9353} (\bibinfo {year} {1996})}\BibitemShut {NoStop}%
\bibitem [{\citenamefont {Manchon}\ \emph {et~al.}(2019)\citenamefont {Manchon}, \citenamefont {\ifmmode~\check{Z}\else \v{Z}\fi{}elezn\'y}, \citenamefont {Miron}, \citenamefont {Jungwirth}, \citenamefont {Sinova}, \citenamefont {Thiaville}, \citenamefont {Garello},\ and\ \citenamefont {Gambardella}}]{ManchonRMP2019}%
  \BibitemOpen
  \bibfield  {author} {\bibinfo {author} {\bibfnamefont {A.}~\bibnamefont {Manchon}}, \bibinfo {author} {\bibfnamefont {J.}~\bibnamefont {\ifmmode~\check{Z}\else \v{Z}\fi{}elezn\'y}}, \bibinfo {author} {\bibfnamefont {I.~M.}\ \bibnamefont {Miron}}, \bibinfo {author} {\bibfnamefont {T.}~\bibnamefont {Jungwirth}}, \bibinfo {author} {\bibfnamefont {J.}~\bibnamefont {Sinova}}, \bibinfo {author} {\bibfnamefont {A.}~\bibnamefont {Thiaville}}, \bibinfo {author} {\bibfnamefont {K.}~\bibnamefont {Garello}},\ and\ \bibinfo {author} {\bibfnamefont {P.}~\bibnamefont {Gambardella}},\ }\href {https://doi.org/10.1103/RevModPhys.91.035004} {\bibfield  {journal} {\bibinfo  {journal} {Rev. Mod. Phys.}\ }\textbf {\bibinfo {volume} {91}},\ \bibinfo {pages} {035004} (\bibinfo {year} {2019})}\BibitemShut {NoStop}%
\bibitem [{\citenamefont {Ramaswamy}\ \emph {et~al.}(2018)\citenamefont {Ramaswamy}, \citenamefont {Lee}, \citenamefont {Cai},\ and\ \citenamefont {Yang}}]{Ramaswamy2018}%
  \BibitemOpen
  \bibfield  {author} {\bibinfo {author} {\bibfnamefont {R.}~\bibnamefont {Ramaswamy}}, \bibinfo {author} {\bibfnamefont {J.~M.}\ \bibnamefont {Lee}}, \bibinfo {author} {\bibfnamefont {K.}~\bibnamefont {Cai}},\ and\ \bibinfo {author} {\bibfnamefont {H.}~\bibnamefont {Yang}},\ }\href {https://doi.org/10.1063/1.5041793} {\bibfield  {journal} {\bibinfo  {journal} {Appl. Phys. Rev.}\ }\textbf {\bibinfo {volume} {5}},\ \bibinfo {pages} {031107} (\bibinfo {year} {2018})}\BibitemShut {NoStop}%
\bibitem [{\citenamefont {Krizakova}\ \emph {et~al.}(2022)\citenamefont {Krizakova}, \citenamefont {Perumkunnil}, \citenamefont {Couet}, \citenamefont {Gambardella},\ and\ \citenamefont {Garello}}]{Krizakova2022}%
  \BibitemOpen
  \bibfield  {author} {\bibinfo {author} {\bibfnamefont {V.}~\bibnamefont {Krizakova}}, \bibinfo {author} {\bibfnamefont {M.}~\bibnamefont {Perumkunnil}}, \bibinfo {author} {\bibfnamefont {S.}~\bibnamefont {Couet}}, \bibinfo {author} {\bibfnamefont {P.}~\bibnamefont {Gambardella}},\ and\ \bibinfo {author} {\bibfnamefont {K.}~\bibnamefont {Garello}},\ }\href {https://doi.org/https://doi.org/10.1016/j.jmmm.2022.169692} {\bibfield  {journal} {\bibinfo  {journal} {J. Magn. Magn. Mater.}\ }\textbf {\bibinfo {volume} {562}},\ \bibinfo {pages} {169692} (\bibinfo {year} {2022})}\BibitemShut {NoStop}%
\bibitem [{\citenamefont {Sinova}\ \emph {et~al.}(2015)\citenamefont {Sinova}, \citenamefont {Valenzuela}, \citenamefont {Wunderlich}, \citenamefont {Back},\ and\ \citenamefont {Jungwirth}}]{RevModPhys.87.1213}%
  \BibitemOpen
  \bibfield  {author} {\bibinfo {author} {\bibfnamefont {J.}~\bibnamefont {Sinova}}, \bibinfo {author} {\bibfnamefont {S.~O.}\ \bibnamefont {Valenzuela}}, \bibinfo {author} {\bibfnamefont {J.}~\bibnamefont {Wunderlich}}, \bibinfo {author} {\bibfnamefont {C.~H.}\ \bibnamefont {Back}},\ and\ \bibinfo {author} {\bibfnamefont {T.}~\bibnamefont {Jungwirth}},\ }\href {https://doi.org/10.1103/RevModPhys.87.1213} {\bibfield  {journal} {\bibinfo  {journal} {Rev. Mod. Phys.}\ }\textbf {\bibinfo {volume} {87}},\ \bibinfo {pages} {1213} (\bibinfo {year} {2015})}\BibitemShut {NoStop}%
\bibitem [{\citenamefont {Liu}\ \emph {et~al.}(2011)\citenamefont {Liu}, \citenamefont {Moriyama}, \citenamefont {Ralph},\ and\ \citenamefont {Buhrman}}]{PhysRevLett.106.036601}%
  \BibitemOpen
  \bibfield  {author} {\bibinfo {author} {\bibfnamefont {L.}~\bibnamefont {Liu}}, \bibinfo {author} {\bibfnamefont {T.}~\bibnamefont {Moriyama}}, \bibinfo {author} {\bibfnamefont {D.~C.}\ \bibnamefont {Ralph}},\ and\ \bibinfo {author} {\bibfnamefont {R.~A.}\ \bibnamefont {Buhrman}},\ }\href {https://doi.org/10.1103/PhysRevLett.106.036601} {\bibfield  {journal} {\bibinfo  {journal} {Phys. Rev. Lett.}\ }\textbf {\bibinfo {volume} {106}},\ \bibinfo {pages} {036601} (\bibinfo {year} {2011})}\BibitemShut {NoStop}%
\bibitem [{\citenamefont {Liu}\ \emph {et~al.}(2012{\natexlab{a}})\citenamefont {Liu}, \citenamefont {Pai}, \citenamefont {Li}, \citenamefont {Tseng}, \citenamefont {Ralph},\ and\ \citenamefont {Buhrman}}]{Liu2012}%
  \BibitemOpen
  \bibfield  {author} {\bibinfo {author} {\bibfnamefont {L.}~\bibnamefont {Liu}}, \bibinfo {author} {\bibfnamefont {C.-F.}\ \bibnamefont {Pai}}, \bibinfo {author} {\bibfnamefont {Y.}~\bibnamefont {Li}}, \bibinfo {author} {\bibfnamefont {H.~W.}\ \bibnamefont {Tseng}}, \bibinfo {author} {\bibfnamefont {D.~C.}\ \bibnamefont {Ralph}},\ and\ \bibinfo {author} {\bibfnamefont {R.~A.}\ \bibnamefont {Buhrman}},\ }\href {https://doi.org/10.1126/science.1218197} {\bibfield  {journal} {\bibinfo  {journal} {Science}\ }\textbf {\bibinfo {volume} {336}},\ \bibinfo {pages} {555} (\bibinfo {year} {2012}{\natexlab{a}})}\BibitemShut {NoStop}%
\bibitem [{\citenamefont {Liu}\ \emph {et~al.}(2012{\natexlab{b}})\citenamefont {Liu}, \citenamefont {Lee}, \citenamefont {Gudmundsen}, \citenamefont {Ralph},\ and\ \citenamefont {Buhrman}}]{PhysRevLett.109.096602}%
  \BibitemOpen
  \bibfield  {author} {\bibinfo {author} {\bibfnamefont {L.}~\bibnamefont {Liu}}, \bibinfo {author} {\bibfnamefont {O.~J.}\ \bibnamefont {Lee}}, \bibinfo {author} {\bibfnamefont {T.~J.}\ \bibnamefont {Gudmundsen}}, \bibinfo {author} {\bibfnamefont {D.~C.}\ \bibnamefont {Ralph}},\ and\ \bibinfo {author} {\bibfnamefont {R.~A.}\ \bibnamefont {Buhrman}},\ }\href {https://doi.org/10.1103/PhysRevLett.109.096602} {\bibfield  {journal} {\bibinfo  {journal} {Phys. Rev. Lett.}\ }\textbf {\bibinfo {volume} {109}},\ \bibinfo {pages} {096602} (\bibinfo {year} {2012}{\natexlab{b}})}\BibitemShut {NoStop}%
\bibitem [{\citenamefont {Zhu}\ \emph {et~al.}(2021)\citenamefont {Zhu}, \citenamefont {Ralph},\ and\ \citenamefont {Buhrman}}]{Zhu2021}%
  \BibitemOpen
  \bibfield  {author} {\bibinfo {author} {\bibfnamefont {L.}~\bibnamefont {Zhu}}, \bibinfo {author} {\bibfnamefont {D.~C.}\ \bibnamefont {Ralph}},\ and\ \bibinfo {author} {\bibfnamefont {R.~A.}\ \bibnamefont {Buhrman}},\ }\href {https://doi.org/10.1063/5.0059171} {\bibfield  {journal} {\bibinfo  {journal} {Appl. Phys. Rev.}\ }\textbf {\bibinfo {volume} {8}},\ \bibinfo {pages} {031308} (\bibinfo {year} {2021})}\BibitemShut {NoStop}%
\bibitem [{\citenamefont {Chernyshov}\ \emph {et~al.}(2009)\citenamefont {Chernyshov}, \citenamefont {Overby}, \citenamefont {Liu}, \citenamefont {Furdyna}, \citenamefont {Lyanda-Geller},\ and\ \citenamefont {Rokhinson}}]{Chernyshov2009}%
  \BibitemOpen
  \bibfield  {author} {\bibinfo {author} {\bibfnamefont {A.}~\bibnamefont {Chernyshov}}, \bibinfo {author} {\bibfnamefont {M.}~\bibnamefont {Overby}}, \bibinfo {author} {\bibfnamefont {X.}~\bibnamefont {Liu}}, \bibinfo {author} {\bibfnamefont {J.~K.}\ \bibnamefont {Furdyna}}, \bibinfo {author} {\bibfnamefont {Y.}~\bibnamefont {Lyanda-Geller}},\ and\ \bibinfo {author} {\bibfnamefont {L.~P.}\ \bibnamefont {Rokhinson}},\ }\href {https://doi.org/10.1038/nphys1362} {\bibfield  {journal} {\bibinfo  {journal} {Nat. Phys.}\ }\textbf {\bibinfo {volume} {5}},\ \bibinfo {pages} {656} (\bibinfo {year} {2009})}\BibitemShut {NoStop}%
\bibitem [{\citenamefont {Miron}\ \emph {et~al.}(2010)\citenamefont {Miron}, \citenamefont {Gaudin}, \citenamefont {Auffret}, \citenamefont {Rodmacq}, \citenamefont {Schuhl}, \citenamefont {Pizzini}, \citenamefont {Vogel},\ and\ \citenamefont {Gambardella}}]{MihaiMiron2010}%
  \BibitemOpen
  \bibfield  {author} {\bibinfo {author} {\bibfnamefont {I.~M.}\ \bibnamefont {Miron}}, \bibinfo {author} {\bibfnamefont {G.}~\bibnamefont {Gaudin}}, \bibinfo {author} {\bibfnamefont {S.}~\bibnamefont {Auffret}}, \bibinfo {author} {\bibfnamefont {B.}~\bibnamefont {Rodmacq}}, \bibinfo {author} {\bibfnamefont {A.}~\bibnamefont {Schuhl}}, \bibinfo {author} {\bibfnamefont {S.}~\bibnamefont {Pizzini}}, \bibinfo {author} {\bibfnamefont {J.}~\bibnamefont {Vogel}},\ and\ \bibinfo {author} {\bibfnamefont {P.}~\bibnamefont {Gambardella}},\ }\href {https://doi.org/10.1038/nmat2613} {\bibfield  {journal} {\bibinfo  {journal} {Nat. Mater.}\ }\textbf {\bibinfo {volume} {9}},\ \bibinfo {pages} {230} (\bibinfo {year} {2010})}\BibitemShut {NoStop}%
\bibitem [{\citenamefont {Miron}\ \emph {et~al.}(2011)\citenamefont {Miron}, \citenamefont {Garello}, \citenamefont {Gaudin}, \citenamefont {Zermatten}, \citenamefont {Costache}, \citenamefont {Auffret}, \citenamefont {Bandiera}, \citenamefont {Rodmacq}, \citenamefont {Schuhl},\ and\ \citenamefont {Gambardella}}]{Miron2011}%
  \BibitemOpen
  \bibfield  {author} {\bibinfo {author} {\bibfnamefont {I.~M.}\ \bibnamefont {Miron}}, \bibinfo {author} {\bibfnamefont {K.}~\bibnamefont {Garello}}, \bibinfo {author} {\bibfnamefont {G.}~\bibnamefont {Gaudin}}, \bibinfo {author} {\bibfnamefont {P.-J.}\ \bibnamefont {Zermatten}}, \bibinfo {author} {\bibfnamefont {M.~V.}\ \bibnamefont {Costache}}, \bibinfo {author} {\bibfnamefont {S.}~\bibnamefont {Auffret}}, \bibinfo {author} {\bibfnamefont {S.}~\bibnamefont {Bandiera}}, \bibinfo {author} {\bibfnamefont {B.}~\bibnamefont {Rodmacq}}, \bibinfo {author} {\bibfnamefont {A.}~\bibnamefont {Schuhl}},\ and\ \bibinfo {author} {\bibfnamefont {P.}~\bibnamefont {Gambardella}},\ }\href {https://doi.org/10.1038/nature10309} {\bibfield  {journal} {\bibinfo  {journal} {Nature}\ }\textbf {\bibinfo {volume} {476}},\ \bibinfo {pages} {189} (\bibinfo {year} {2011})}\BibitemShut {NoStop}%
\bibitem [{\citenamefont {Manchon}\ \emph {et~al.}(2015)\citenamefont {Manchon}, \citenamefont {Koo}, \citenamefont {Nitta}, \citenamefont {Frolov},\ and\ \citenamefont {Duine}}]{Manchon2015}%
  \BibitemOpen
  \bibfield  {author} {\bibinfo {author} {\bibfnamefont {A.}~\bibnamefont {Manchon}}, \bibinfo {author} {\bibfnamefont {H.~C.}\ \bibnamefont {Koo}}, \bibinfo {author} {\bibfnamefont {J.}~\bibnamefont {Nitta}}, \bibinfo {author} {\bibfnamefont {S.~M.}\ \bibnamefont {Frolov}},\ and\ \bibinfo {author} {\bibfnamefont {R.~A.}\ \bibnamefont {Duine}},\ }\href {https://doi.org/10.1038/nmat4360} {\bibfield  {journal} {\bibinfo  {journal} {Nat. Mater.}\ }\textbf {\bibinfo {volume} {14}},\ \bibinfo {pages} {871} (\bibinfo {year} {2015})}\BibitemShut {NoStop}%
\bibitem [{\citenamefont {Go}\ \emph {et~al.}(2020)\citenamefont {Go}, \citenamefont {Freimuth}, \citenamefont {Hanke}, \citenamefont {Xue}, \citenamefont {Gomonay}, \citenamefont {Lee}, \citenamefont {Bl\"ugel}, \citenamefont {Haney}, \citenamefont {Lee},\ and\ \citenamefont {Mokrousov}}]{PhysRevResearch.2.033401}%
  \BibitemOpen
  \bibfield  {author} {\bibinfo {author} {\bibfnamefont {D.}~\bibnamefont {Go}}, \bibinfo {author} {\bibfnamefont {F.}~\bibnamefont {Freimuth}}, \bibinfo {author} {\bibfnamefont {J.-P.}\ \bibnamefont {Hanke}}, \bibinfo {author} {\bibfnamefont {F.}~\bibnamefont {Xue}}, \bibinfo {author} {\bibfnamefont {O.}~\bibnamefont {Gomonay}}, \bibinfo {author} {\bibfnamefont {K.-J.}\ \bibnamefont {Lee}}, \bibinfo {author} {\bibfnamefont {S.}~\bibnamefont {Bl\"ugel}}, \bibinfo {author} {\bibfnamefont {P.~M.}\ \bibnamefont {Haney}}, \bibinfo {author} {\bibfnamefont {H.-W.}\ \bibnamefont {Lee}},\ and\ \bibinfo {author} {\bibfnamefont {Y.}~\bibnamefont {Mokrousov}},\ }\href {https://doi.org/10.1103/PhysRevResearch.2.033401} {\bibfield  {journal} {\bibinfo  {journal} {Phys. Rev. Research}\ }\textbf {\bibinfo {volume} {2}},\ \bibinfo {pages} {033401} (\bibinfo {year} {2020})}\BibitemShut {NoStop}%
\bibitem [{\citenamefont {Go}\ and\ \citenamefont {Lee}(2020)}]{PhysRevResearch.2.013177}%
  \BibitemOpen
  \bibfield  {author} {\bibinfo {author} {\bibfnamefont {D.}~\bibnamefont {Go}}\ and\ \bibinfo {author} {\bibfnamefont {H.-W.}\ \bibnamefont {Lee}},\ }\href {https://doi.org/10.1103/PhysRevResearch.2.013177} {\bibfield  {journal} {\bibinfo  {journal} {Phys. Rev. Research}\ }\textbf {\bibinfo {volume} {2}},\ \bibinfo {pages} {013177} (\bibinfo {year} {2020})}\BibitemShut {NoStop}%
\bibitem [{\citenamefont {Safranski}\ \emph {et~al.}(2018)\citenamefont {Safranski}, \citenamefont {Montoya},\ and\ \citenamefont {Krivorotov}}]{Safranski2018}%
  \BibitemOpen
  \bibfield  {author} {\bibinfo {author} {\bibfnamefont {C.}~\bibnamefont {Safranski}}, \bibinfo {author} {\bibfnamefont {E.~A.}\ \bibnamefont {Montoya}},\ and\ \bibinfo {author} {\bibfnamefont {I.~N.}\ \bibnamefont {Krivorotov}},\ }\href {https://doi.org/10.1038/s41565-018-0282-0} {\bibfield  {journal} {\bibinfo  {journal} {Nat. Nanotechnol.}\ }\textbf {\bibinfo {volume} {14}},\ \bibinfo {pages} {27} (\bibinfo {year} {2018})}\BibitemShut {NoStop}%
\bibitem [{\citenamefont {Safranski}\ \emph {et~al.}(2020)\citenamefont {Safranski}, \citenamefont {Sun}, \citenamefont {Xu},\ and\ \citenamefont {Kent}}]{PhysRevLett.124.197204}%
  \BibitemOpen
  \bibfield  {author} {\bibinfo {author} {\bibfnamefont {C.}~\bibnamefont {Safranski}}, \bibinfo {author} {\bibfnamefont {J.~Z.}\ \bibnamefont {Sun}}, \bibinfo {author} {\bibfnamefont {J.-W.}\ \bibnamefont {Xu}},\ and\ \bibinfo {author} {\bibfnamefont {A.~D.}\ \bibnamefont {Kent}},\ }\href {https://doi.org/10.1103/PhysRevLett.124.197204} {\bibfield  {journal} {\bibinfo  {journal} {Phys. Rev. Lett.}\ }\textbf {\bibinfo {volume} {124}},\ \bibinfo {pages} {197204} (\bibinfo {year} {2020})}\BibitemShut {NoStop}%
\bibitem [{\citenamefont {Amin}\ \emph {et~al.}(2019)\citenamefont {Amin}, \citenamefont {Li}, \citenamefont {Stiles},\ and\ \citenamefont {Haney}}]{SAHE}%
  \BibitemOpen
  \bibfield  {author} {\bibinfo {author} {\bibfnamefont {V.~P.}\ \bibnamefont {Amin}}, \bibinfo {author} {\bibfnamefont {J.}~\bibnamefont {Li}}, \bibinfo {author} {\bibfnamefont {M.~D.}\ \bibnamefont {Stiles}},\ and\ \bibinfo {author} {\bibfnamefont {P.~M.}\ \bibnamefont {Haney}},\ }\href {https://doi.org/10.1103/PhysRevB.99.220405} {\bibfield  {journal} {\bibinfo  {journal} {Phys. Rev. B}\ }\textbf {\bibinfo {volume} {99}},\ \bibinfo {pages} {220405} (\bibinfo {year} {2019})}\BibitemShut {NoStop}%
\bibitem [{\citenamefont {Mook}\ \emph {et~al.}(2020)\citenamefont {Mook}, \citenamefont {Neumann}, \citenamefont {Johansson}, \citenamefont {Henk},\ and\ \citenamefont {Mertig}}]{PhysRevResearch.2.023065}%
  \BibitemOpen
  \bibfield  {author} {\bibinfo {author} {\bibfnamefont {A.}~\bibnamefont {Mook}}, \bibinfo {author} {\bibfnamefont {R.~R.}\ \bibnamefont {Neumann}}, \bibinfo {author} {\bibfnamefont {A.}~\bibnamefont {Johansson}}, \bibinfo {author} {\bibfnamefont {J.}~\bibnamefont {Henk}},\ and\ \bibinfo {author} {\bibfnamefont {I.}~\bibnamefont {Mertig}},\ }\href {https://doi.org/10.1103/PhysRevResearch.2.023065} {\bibfield  {journal} {\bibinfo  {journal} {Phys. Rev. Research}\ }\textbf {\bibinfo {volume} {2}},\ \bibinfo {pages} {023065} (\bibinfo {year} {2020})}\BibitemShut {NoStop}%
\bibitem [{\citenamefont {Amin}\ \emph {et~al.}(2020)\citenamefont {Amin}, \citenamefont {Haney},\ and\ \citenamefont {Stiles}}]{Amin2020}%
  \BibitemOpen
  \bibfield  {author} {\bibinfo {author} {\bibfnamefont {V.~P.}\ \bibnamefont {Amin}}, \bibinfo {author} {\bibfnamefont {P.~M.}\ \bibnamefont {Haney}},\ and\ \bibinfo {author} {\bibfnamefont {M.~D.}\ \bibnamefont {Stiles}},\ }\href {https://doi.org/10.1063/5.0024019} {\bibfield  {journal} {\bibinfo  {journal} {J. Appl. Phys.}\ }\textbf {\bibinfo {volume} {128}},\ \bibinfo {pages} {151101} (\bibinfo {year} {2020})}\BibitemShut {NoStop}%
\bibitem [{\citenamefont {Belashchenko}\ \emph {et~al.}(2020{\natexlab{a}})\citenamefont {Belashchenko}, \citenamefont {Kovalev},\ and\ \citenamefont {van Schilfgaarde}}]{PhysRevB.101.020407}%
  \BibitemOpen
  \bibfield  {author} {\bibinfo {author} {\bibfnamefont {K.~D.}\ \bibnamefont {Belashchenko}}, \bibinfo {author} {\bibfnamefont {A.~A.}\ \bibnamefont {Kovalev}},\ and\ \bibinfo {author} {\bibfnamefont {M.}~\bibnamefont {van Schilfgaarde}},\ }\href {https://doi.org/10.1103/PhysRevB.101.020407} {\bibfield  {journal} {\bibinfo  {journal} {Phys. Rev. B}\ }\textbf {\bibinfo {volume} {101}},\ \bibinfo {pages} {020407(R)} (\bibinfo {year} {2020}{\natexlab{a}})}\BibitemShut {NoStop}%
\bibitem [{\citenamefont {Yu}\ \emph {et~al.}(2014)\citenamefont {Yu}, \citenamefont {Upadhyaya}, \citenamefont {Fan}, \citenamefont {Alzate}, \citenamefont {Jiang}, \citenamefont {Wong}, \citenamefont {Takei}, \citenamefont {Bender}, \citenamefont {Chang}, \citenamefont {Jiang}, \citenamefont {Lang}, \citenamefont {Tang}, \citenamefont {Wang}, \citenamefont {Tserkovnyak}, \citenamefont {Amiri},\ and\ \citenamefont {Wang}}]{Yu2014}%
  \BibitemOpen
  \bibfield  {author} {\bibinfo {author} {\bibfnamefont {G.}~\bibnamefont {Yu}}, \bibinfo {author} {\bibfnamefont {P.}~\bibnamefont {Upadhyaya}}, \bibinfo {author} {\bibfnamefont {Y.}~\bibnamefont {Fan}}, \bibinfo {author} {\bibfnamefont {J.~G.}\ \bibnamefont {Alzate}}, \bibinfo {author} {\bibfnamefont {W.}~\bibnamefont {Jiang}}, \bibinfo {author} {\bibfnamefont {K.~L.}\ \bibnamefont {Wong}}, \bibinfo {author} {\bibfnamefont {S.}~\bibnamefont {Takei}}, \bibinfo {author} {\bibfnamefont {S.~A.}\ \bibnamefont {Bender}}, \bibinfo {author} {\bibfnamefont {L.-T.}\ \bibnamefont {Chang}}, \bibinfo {author} {\bibfnamefont {Y.}~\bibnamefont {Jiang}}, \bibinfo {author} {\bibfnamefont {M.}~\bibnamefont {Lang}}, \bibinfo {author} {\bibfnamefont {J.}~\bibnamefont {Tang}}, \bibinfo {author} {\bibfnamefont {Y.}~\bibnamefont {Wang}}, \bibinfo {author} {\bibfnamefont {Y.}~\bibnamefont {Tserkovnyak}}, \bibinfo {author} {\bibfnamefont {P.~K.}\ \bibnamefont {Amiri}},\ and\ \bibinfo {author} {\bibfnamefont {K.~L.}\ \bibnamefont
  {Wang}},\ }\href {https://doi.org/10.1038/nnano.2014.94} {\bibfield  {journal} {\bibinfo  {journal} {Nat. Nanotechnol.}\ }\textbf {\bibinfo {volume} {9}},\ \bibinfo {pages} {548} (\bibinfo {year} {2014})}\BibitemShut {NoStop}%
\bibitem [{\citenamefont {MacNeill}\ \emph {et~al.}(2016)\citenamefont {MacNeill}, \citenamefont {Stiehl}, \citenamefont {Guimaraes}, \citenamefont {Buhrman}, \citenamefont {Park},\ and\ \citenamefont {Ralph}}]{MacNeill2016}%
  \BibitemOpen
  \bibfield  {author} {\bibinfo {author} {\bibfnamefont {D.}~\bibnamefont {MacNeill}}, \bibinfo {author} {\bibfnamefont {G.~M.}\ \bibnamefont {Stiehl}}, \bibinfo {author} {\bibfnamefont {M.~H.~D.}\ \bibnamefont {Guimaraes}}, \bibinfo {author} {\bibfnamefont {R.~A.}\ \bibnamefont {Buhrman}}, \bibinfo {author} {\bibfnamefont {J.}~\bibnamefont {Park}},\ and\ \bibinfo {author} {\bibfnamefont {D.~C.}\ \bibnamefont {Ralph}},\ }\href {https://doi.org/10.1038/nphys3933} {\bibfield  {journal} {\bibinfo  {journal} {Nat. Phys.}\ }\textbf {\bibinfo {volume} {13}},\ \bibinfo {pages} {300} (\bibinfo {year} {2016})}\BibitemShut {NoStop}%
\bibitem [{\citenamefont {Liu}\ \emph {et~al.}(2021)\citenamefont {Liu}, \citenamefont {Zhou}, \citenamefont {Shu}, \citenamefont {Li}, \citenamefont {Zhao}, \citenamefont {Lin}, \citenamefont {Deng}, \citenamefont {Xie}, \citenamefont {Chen}, \citenamefont {Zhou}, \citenamefont {Guo}, \citenamefont {Wang}, \citenamefont {Yu}, \citenamefont {Shi}, \citenamefont {Yang}, \citenamefont {Pennycook}, \citenamefont {Manchon},\ and\ \citenamefont {Chen}}]{Liu2021}%
  \BibitemOpen
  \bibfield  {author} {\bibinfo {author} {\bibfnamefont {L.}~\bibnamefont {Liu}}, \bibinfo {author} {\bibfnamefont {C.}~\bibnamefont {Zhou}}, \bibinfo {author} {\bibfnamefont {X.}~\bibnamefont {Shu}}, \bibinfo {author} {\bibfnamefont {C.}~\bibnamefont {Li}}, \bibinfo {author} {\bibfnamefont {T.}~\bibnamefont {Zhao}}, \bibinfo {author} {\bibfnamefont {W.}~\bibnamefont {Lin}}, \bibinfo {author} {\bibfnamefont {J.}~\bibnamefont {Deng}}, \bibinfo {author} {\bibfnamefont {Q.}~\bibnamefont {Xie}}, \bibinfo {author} {\bibfnamefont {S.}~\bibnamefont {Chen}}, \bibinfo {author} {\bibfnamefont {J.}~\bibnamefont {Zhou}}, \bibinfo {author} {\bibfnamefont {R.}~\bibnamefont {Guo}}, \bibinfo {author} {\bibfnamefont {H.}~\bibnamefont {Wang}}, \bibinfo {author} {\bibfnamefont {J.}~\bibnamefont {Yu}}, \bibinfo {author} {\bibfnamefont {S.}~\bibnamefont {Shi}}, \bibinfo {author} {\bibfnamefont {P.}~\bibnamefont {Yang}}, \bibinfo {author} {\bibfnamefont {S.}~\bibnamefont {Pennycook}}, \bibinfo {author} {\bibfnamefont
  {A.}~\bibnamefont {Manchon}},\ and\ \bibinfo {author} {\bibfnamefont {J.}~\bibnamefont {Chen}},\ }\href {https://doi.org/10.1038/s41565-020-00826-8} {\bibfield  {journal} {\bibinfo  {journal} {Nat. Nanotechnol.}\ }\textbf {\bibinfo {volume} {16}},\ \bibinfo {pages} {277} (\bibinfo {year} {2021})}\BibitemShut {NoStop}%
\bibitem [{\citenamefont {Humphries}\ \emph {et~al.}(2017)\citenamefont {Humphries}, \citenamefont {Wang}, \citenamefont {Edwards}, \citenamefont {Allen}, \citenamefont {Shaw}, \citenamefont {Nembach}, \citenamefont {Xiao}, \citenamefont {Silva},\ and\ \citenamefont {Fan}}]{Humphries2017}%
  \BibitemOpen
  \bibfield  {author} {\bibinfo {author} {\bibfnamefont {A.~M.}\ \bibnamefont {Humphries}}, \bibinfo {author} {\bibfnamefont {T.}~\bibnamefont {Wang}}, \bibinfo {author} {\bibfnamefont {E.~R.~J.}\ \bibnamefont {Edwards}}, \bibinfo {author} {\bibfnamefont {S.~R.}\ \bibnamefont {Allen}}, \bibinfo {author} {\bibfnamefont {J.~M.}\ \bibnamefont {Shaw}}, \bibinfo {author} {\bibfnamefont {H.~T.}\ \bibnamefont {Nembach}}, \bibinfo {author} {\bibfnamefont {J.~Q.}\ \bibnamefont {Xiao}}, \bibinfo {author} {\bibfnamefont {T.~J.}\ \bibnamefont {Silva}},\ and\ \bibinfo {author} {\bibfnamefont {X.}~\bibnamefont {Fan}},\ }\href {https://doi.org/10.1038/s41467-017-00967-w} {\bibfield  {journal} {\bibinfo  {journal} {Nat. Commun.}\ }\textbf {\bibinfo {volume} {8}},\ \bibinfo {pages} {911} (\bibinfo {year} {2017})}\BibitemShut {NoStop}%
\bibitem [{\citenamefont {Baek}\ \emph {et~al.}(2018)\citenamefont {Baek}, \citenamefont {Amin}, \citenamefont {Oh}, \citenamefont {Go}, \citenamefont {Lee}, \citenamefont {Lee}, \citenamefont {Kim}, \citenamefont {Stiles}, \citenamefont {Park},\ and\ \citenamefont {Lee}}]{Baek2018}%
  \BibitemOpen
  \bibfield  {author} {\bibinfo {author} {\bibfnamefont {S.~C.}\ \bibnamefont {Baek}}, \bibinfo {author} {\bibfnamefont {V.~P.}\ \bibnamefont {Amin}}, \bibinfo {author} {\bibfnamefont {Y.-W.}\ \bibnamefont {Oh}}, \bibinfo {author} {\bibfnamefont {G.}~\bibnamefont {Go}}, \bibinfo {author} {\bibfnamefont {S.-J.}\ \bibnamefont {Lee}}, \bibinfo {author} {\bibfnamefont {G.-H.}\ \bibnamefont {Lee}}, \bibinfo {author} {\bibfnamefont {K.-J.}\ \bibnamefont {Kim}}, \bibinfo {author} {\bibfnamefont {M.~D.}\ \bibnamefont {Stiles}}, \bibinfo {author} {\bibfnamefont {B.-G.}\ \bibnamefont {Park}},\ and\ \bibinfo {author} {\bibfnamefont {K.-J.}\ \bibnamefont {Lee}},\ }\href {https://doi.org/10.1038/s41563-018-0041-5} {\bibfield  {journal} {\bibinfo  {journal} {Nat. Mater.}\ }\textbf {\bibinfo {volume} {17}},\ \bibinfo {pages} {509} (\bibinfo {year} {2018})}\BibitemShut {NoStop}%
\bibitem [{\citenamefont {Fert}\ and\ \citenamefont {Lee}(1996)}]{PhysRevB.53.6554}%
  \BibitemOpen
  \bibfield  {author} {\bibinfo {author} {\bibfnamefont {A.}~\bibnamefont {Fert}}\ and\ \bibinfo {author} {\bibfnamefont {S.-F.}\ \bibnamefont {Lee}},\ }\href {https://doi.org/10.1103/PhysRevB.53.6554} {\bibfield  {journal} {\bibinfo  {journal} {Phys. Rev. B}\ }\textbf {\bibinfo {volume} {53}},\ \bibinfo {pages} {6554} (\bibinfo {year} {1996})}\BibitemShut {NoStop}%
\bibitem [{\citenamefont {Kovalev}\ \emph {et~al.}(2002)\citenamefont {Kovalev}, \citenamefont {Brataas},\ and\ \citenamefont {Bauer}}]{PhysRevB.66.224424}%
  \BibitemOpen
  \bibfield  {author} {\bibinfo {author} {\bibfnamefont {A.~A.}\ \bibnamefont {Kovalev}}, \bibinfo {author} {\bibfnamefont {A.}~\bibnamefont {Brataas}},\ and\ \bibinfo {author} {\bibfnamefont {G.~E.~W.}\ \bibnamefont {Bauer}},\ }\href {https://doi.org/10.1103/PhysRevB.66.224424} {\bibfield  {journal} {\bibinfo  {journal} {Phys. Rev. B}\ }\textbf {\bibinfo {volume} {66}},\ \bibinfo {pages} {224424} (\bibinfo {year} {2002})}\BibitemShut {NoStop}%
\bibitem [{\citenamefont {Barna{\'s}}\ \emph {et~al.}(2005)\citenamefont {Barna{\'s}}, \citenamefont {Fert}, \citenamefont {Gmitra}, \citenamefont {Weymann},\ and\ \citenamefont {Dugaev}}]{Barnas.Fert.eaPRB2005}%
  \BibitemOpen
  \bibfield  {author} {\bibinfo {author} {\bibfnamefont {J.}~\bibnamefont {Barna{\'s}}}, \bibinfo {author} {\bibfnamefont {A.}~\bibnamefont {Fert}}, \bibinfo {author} {\bibfnamefont {M.}~\bibnamefont {Gmitra}}, \bibinfo {author} {\bibfnamefont {I.}~\bibnamefont {Weymann}},\ and\ \bibinfo {author} {\bibfnamefont {V.~K.}\ \bibnamefont {Dugaev}},\ }\href {https://doi.org/10.1103/PhysRevB.72.024426} {\bibfield  {journal} {\bibinfo  {journal} {Phys. Rev. B}\ }\textbf {\bibinfo {volume} {72}},\ \bibinfo {eid} {024426} (\bibinfo {year} {2005})}\BibitemShut {NoStop}%
\bibitem [{\citenamefont {Amin}\ \emph {et~al.}(2018)\citenamefont {Amin}, \citenamefont {Zemen},\ and\ \citenamefont {Stiles}}]{PhysRevLett.121.136805}%
  \BibitemOpen
  \bibfield  {author} {\bibinfo {author} {\bibfnamefont {V.~P.}\ \bibnamefont {Amin}}, \bibinfo {author} {\bibfnamefont {J.}~\bibnamefont {Zemen}},\ and\ \bibinfo {author} {\bibfnamefont {M.~D.}\ \bibnamefont {Stiles}},\ }\href {https://doi.org/10.1103/PhysRevLett.121.136805} {\bibfield  {journal} {\bibinfo  {journal} {Phys. Rev. Lett.}\ }\textbf {\bibinfo {volume} {121}},\ \bibinfo {pages} {136805} (\bibinfo {year} {2018})}\BibitemShut {NoStop}%
\bibitem [{\citenamefont {Amin}\ and\ \citenamefont {Stiles}(2016{\natexlab{a}})}]{PhysRevB.94.104419}%
  \BibitemOpen
  \bibfield  {author} {\bibinfo {author} {\bibfnamefont {V.~P.}\ \bibnamefont {Amin}}\ and\ \bibinfo {author} {\bibfnamefont {M.~D.}\ \bibnamefont {Stiles}},\ }\href {https://doi.org/10.1103/PhysRevB.94.104419} {\bibfield  {journal} {\bibinfo  {journal} {Phys. Rev. B}\ }\textbf {\bibinfo {volume} {94}},\ \bibinfo {pages} {104419} (\bibinfo {year} {2016}{\natexlab{a}})}\BibitemShut {NoStop}%
\bibitem [{\citenamefont {Amin}\ and\ \citenamefont {Stiles}(2016{\natexlab{b}})}]{PhysRevB.94.104420}%
  \BibitemOpen
  \bibfield  {author} {\bibinfo {author} {\bibfnamefont {V.~P.}\ \bibnamefont {Amin}}\ and\ \bibinfo {author} {\bibfnamefont {M.~D.}\ \bibnamefont {Stiles}},\ }\href {https://doi.org/10.1103/PhysRevB.94.104420} {\bibfield  {journal} {\bibinfo  {journal} {Phys. Rev. B}\ }\textbf {\bibinfo {volume} {94}},\ \bibinfo {pages} {104420} (\bibinfo {year} {2016}{\natexlab{b}})}\BibitemShut {NoStop}%
\bibitem [{\citenamefont {Freimuth}\ \emph {et~al.}(2018)\citenamefont {Freimuth}, \citenamefont {Bl\"ugel},\ and\ \citenamefont {Mokrousov}}]{Freimuth2018}%
  \BibitemOpen
  \bibfield  {author} {\bibinfo {author} {\bibfnamefont {F.}~\bibnamefont {Freimuth}}, \bibinfo {author} {\bibfnamefont {S.}~\bibnamefont {Bl\"ugel}},\ and\ \bibinfo {author} {\bibfnamefont {Y.}~\bibnamefont {Mokrousov}},\ }\href {https://doi.org/10.1103/PhysRevB.98.024419} {\bibfield  {journal} {\bibinfo  {journal} {Phys. Rev. B}\ }\textbf {\bibinfo {volume} {98}},\ \bibinfo {pages} {024419} (\bibinfo {year} {2018})}\BibitemShut {NoStop}%
\bibitem [{\citenamefont {Tsymbal}\ and\ \citenamefont {Pettifor}(2001)}]{Tsymbal2001}%
  \BibitemOpen
  \bibfield  {author} {\bibinfo {author} {\bibfnamefont {E.~Y.}\ \bibnamefont {Tsymbal}}\ and\ \bibinfo {author} {\bibfnamefont {D.~G.}\ \bibnamefont {Pettifor}},\ }in\ \href {https://doi.org/10.1016/S0081-1947(01)80019-9} {\emph {\bibinfo {booktitle} {Solid State Physics}}},\ Vol.~\bibinfo {volume} {56},\ \bibinfo {editor} {edited by\ \bibinfo {editor} {\bibfnamefont {H.}~\bibnamefont {Ehrenreich}}\ and\ \bibinfo {editor} {\bibfnamefont {F.}~\bibnamefont {Spaepen}}}\ (\bibinfo  {publisher} {Academic Press},\ \bibinfo {year} {2001})\ pp.\ \bibinfo {pages} {113--237}\BibitemShut {NoStop}%
\bibitem [{\citenamefont {Aronov}\ and\ \citenamefont {Lyanda-Geller}(1989)}]{ALG1989}%
  \BibitemOpen
  \bibfield  {author} {\bibinfo {author} {\bibfnamefont {A.}~\bibnamefont {Aronov}}\ and\ \bibinfo {author} {\bibfnamefont {Y.}~\bibnamefont {Lyanda-Geller}},\ }\href@noop {} {\bibfield  {journal} {\bibinfo  {journal} {JETP Lett.}\ }\textbf {\bibinfo {volume} {50}},\ \bibinfo {pages} {431} (\bibinfo {year} {1989})}\BibitemShut {NoStop}%
\bibitem [{\citenamefont {Edelstein}(1990)}]{Edelstein1990}%
  \BibitemOpen
  \bibfield  {author} {\bibinfo {author} {\bibfnamefont {V.~M.}\ \bibnamefont {Edelstein}},\ }\href {https://doi.org/https://doi.org/10.1016/0038-1098(90)90963-C} {\bibfield  {journal} {\bibinfo  {journal} {Solid State Commun.}\ }\textbf {\bibinfo {volume} {73}},\ \bibinfo {pages} {233} (\bibinfo {year} {1990})}\BibitemShut {NoStop}%
\bibitem [{\citenamefont {Ganichev}\ \emph {et~al.}(2019)\citenamefont {Ganichev}, \citenamefont {Trushin},\ and\ \citenamefont {Schliemann}}]{Ganichev-handbook}%
  \BibitemOpen
  \bibfield  {author} {\bibinfo {author} {\bibfnamefont {S.~D.}\ \bibnamefont {Ganichev}}, \bibinfo {author} {\bibfnamefont {M.}~\bibnamefont {Trushin}},\ and\ \bibinfo {author} {\bibfnamefont {J.}~\bibnamefont {Schliemann}},\ }in\ \href@noop {} {\emph {\bibinfo {booktitle} {Spintronics Handbook: Spin Transport and Magnetism, Vol. 2: Semiconductor Spintronics, 2nd Edition}}},\ \bibinfo {editor} {edited by\ \bibinfo {editor} {\bibfnamefont {E.}~\bibnamefont {Tsymbal}}\ and\ \bibinfo {editor} {\bibfnamefont {I.}~\bibnamefont {Zutic}}}\ (\bibinfo  {publisher} {CRC Press},\ \bibinfo {year} {2019})\ Chap.~\bibinfo {chapter} {7}, pp.\ \bibinfo {pages} {317--338}\BibitemShut {NoStop}%
\bibitem [{\citenamefont {Belashchenko}\ \emph {et~al.}(2019{\natexlab{a}})\citenamefont {Belashchenko}, \citenamefont {Kovalev},\ and\ \citenamefont {van Schilfgaarde}}]{PhysRevMaterials.3.011401}%
  \BibitemOpen
  \bibfield  {author} {\bibinfo {author} {\bibfnamefont {K.~D.}\ \bibnamefont {Belashchenko}}, \bibinfo {author} {\bibfnamefont {A.~A.}\ \bibnamefont {Kovalev}},\ and\ \bibinfo {author} {\bibfnamefont {M.}~\bibnamefont {van Schilfgaarde}},\ }\href {https://doi.org/10.1103/PhysRevMaterials.3.011401} {\bibfield  {journal} {\bibinfo  {journal} {Phys. Rev. Materials}\ }\textbf {\bibinfo {volume} {3}},\ \bibinfo {pages} {011401(R)} (\bibinfo {year} {2019}{\natexlab{a}})}\BibitemShut {NoStop}%
\bibitem [{\citenamefont {Taniguchi}\ \emph {et~al.}(2015)\citenamefont {Taniguchi}, \citenamefont {Grollier},\ and\ \citenamefont {Stiles}}]{Taniguchi2015}%
  \BibitemOpen
  \bibfield  {author} {\bibinfo {author} {\bibfnamefont {T.}~\bibnamefont {Taniguchi}}, \bibinfo {author} {\bibfnamefont {J.}~\bibnamefont {Grollier}},\ and\ \bibinfo {author} {\bibfnamefont {M.~D.}\ \bibnamefont {Stiles}},\ }\href {https://doi.org/10.1103/PhysRevApplied.3.044001} {\bibfield  {journal} {\bibinfo  {journal} {Phys. Rev. Appl.}\ }\textbf {\bibinfo {volume} {3}},\ \bibinfo {pages} {044001} (\bibinfo {year} {2015})}\BibitemShut {NoStop}%
\bibitem [{\citenamefont {Brataas}\ \emph {et~al.}(2000)\citenamefont {Brataas}, \citenamefont {Nazarov},\ and\ \citenamefont {Bauer}}]{Brataas2000}%
  \BibitemOpen
  \bibfield  {author} {\bibinfo {author} {\bibfnamefont {A.}~\bibnamefont {Brataas}}, \bibinfo {author} {\bibfnamefont {Y.~V.}\ \bibnamefont {Nazarov}},\ and\ \bibinfo {author} {\bibfnamefont {G.~E.~W.}\ \bibnamefont {Bauer}},\ }\href {https://doi.org/10.1103/PhysRevLett.84.2481} {\bibfield  {journal} {\bibinfo  {journal} {Phys. Rev. Lett.}\ }\textbf {\bibinfo {volume} {84}},\ \bibinfo {pages} {2481} (\bibinfo {year} {2000})}\BibitemShut {NoStop}%
\bibitem [{\citenamefont {Brataas}\ \emph {et~al.}(2006)\citenamefont {Brataas}, \citenamefont {Bauer},\ and\ \citenamefont {Kelly}}]{Brataas2006}%
  \BibitemOpen
  \bibfield  {author} {\bibinfo {author} {\bibfnamefont {A.}~\bibnamefont {Brataas}}, \bibinfo {author} {\bibfnamefont {G.~E.~W.}\ \bibnamefont {Bauer}},\ and\ \bibinfo {author} {\bibfnamefont {P.}~\bibnamefont {Kelly}},\ }\href {https://doi.org/10.1016/j.physrep.2006.01.001} {\bibfield  {journal} {\bibinfo  {journal} {Phys. Rep.}\ }\textbf {\bibinfo {volume} {427}},\ \bibinfo {pages} {157} (\bibinfo {year} {2006})}\BibitemShut {NoStop}%
\bibitem [{\citenamefont {Lifshits}\ and\ \citenamefont {Dyakonov}(2009)}]{PhysRevLett.103.186601}%
  \BibitemOpen
  \bibfield  {author} {\bibinfo {author} {\bibfnamefont {M.~B.}\ \bibnamefont {Lifshits}}\ and\ \bibinfo {author} {\bibfnamefont {M.~I.}\ \bibnamefont {Dyakonov}},\ }\href {https://doi.org/10.1103/PhysRevLett.103.186601} {\bibfield  {journal} {\bibinfo  {journal} {Phys. Rev. Lett.}\ }\textbf {\bibinfo {volume} {103}},\ \bibinfo {pages} {186601} (\bibinfo {year} {2009})}\BibitemShut {NoStop}%
\bibitem [{\citenamefont {Pauyac}\ \emph {et~al.}(2018)\citenamefont {Pauyac}, \citenamefont {Chshiev}, \citenamefont {Manchon},\ and\ \citenamefont {Nikolaev}}]{PhysRevLett.120.176802}%
  \BibitemOpen
  \bibfield  {author} {\bibinfo {author} {\bibfnamefont {C.~O.}\ \bibnamefont {Pauyac}}, \bibinfo {author} {\bibfnamefont {M.}~\bibnamefont {Chshiev}}, \bibinfo {author} {\bibfnamefont {A.}~\bibnamefont {Manchon}},\ and\ \bibinfo {author} {\bibfnamefont {S.~A.}\ \bibnamefont {Nikolaev}},\ }\href {https://doi.org/10.1103/PhysRevLett.120.176802} {\bibfield  {journal} {\bibinfo  {journal} {Phys. Rev. Lett.}\ }\textbf {\bibinfo {volume} {120}},\ \bibinfo {pages} {176802} (\bibinfo {year} {2018})}\BibitemShut {NoStop}%
\bibitem [{\citenamefont {Baez~Flores}\ \emph {et~al.}(2020)\citenamefont {Baez~Flores}, \citenamefont {Kovalev}, \citenamefont {van Schilfgaarde},\ and\ \citenamefont {Belashchenko}}]{Flores2020}%
  \BibitemOpen
  \bibfield  {author} {\bibinfo {author} {\bibfnamefont {G.~G.}\ \bibnamefont {Baez~Flores}}, \bibinfo {author} {\bibfnamefont {A.~A.}\ \bibnamefont {Kovalev}}, \bibinfo {author} {\bibfnamefont {M.}~\bibnamefont {van Schilfgaarde}},\ and\ \bibinfo {author} {\bibfnamefont {K.~D.}\ \bibnamefont {Belashchenko}},\ }\href {https://doi.org/10.1103/PhysRevB.101.224405} {\bibfield  {journal} {\bibinfo  {journal} {Phys. Rev. B}\ }\textbf {\bibinfo {volume} {101}},\ \bibinfo {pages} {224405} (\bibinfo {year} {2020})}\BibitemShut {NoStop}%
\bibitem [{\citenamefont {G\'eranton}\ \emph {et~al.}(2017)\citenamefont {G\'eranton}, \citenamefont {Zimmermann}, \citenamefont {Long}, \citenamefont {Mavropoulos}, \citenamefont {Bl\"ugel}, \citenamefont {Freimuth},\ and\ \citenamefont {Mokrousov}}]{Geranton2017}%
  \BibitemOpen
  \bibfield  {author} {\bibinfo {author} {\bibfnamefont {G.}~\bibnamefont {G\'eranton}}, \bibinfo {author} {\bibfnamefont {B.}~\bibnamefont {Zimmermann}}, \bibinfo {author} {\bibfnamefont {N.~H.}\ \bibnamefont {Long}}, \bibinfo {author} {\bibfnamefont {P.}~\bibnamefont {Mavropoulos}}, \bibinfo {author} {\bibfnamefont {S.}~\bibnamefont {Bl\"ugel}}, \bibinfo {author} {\bibfnamefont {F.}~\bibnamefont {Freimuth}},\ and\ \bibinfo {author} {\bibfnamefont {Y.}~\bibnamefont {Mokrousov}},\ }\href {https://doi.org/10.1103/PhysRevB.95.134449} {\bibfield  {journal} {\bibinfo  {journal} {Phys. Rev. B}\ }\textbf {\bibinfo {volume} {95}},\ \bibinfo {pages} {134449} (\bibinfo {year} {2017})}\BibitemShut {NoStop}%
\bibitem [{\citenamefont {Kimata}\ \emph {et~al.}(2019)\citenamefont {Kimata}, \citenamefont {Chen}, \citenamefont {Kondou}, \citenamefont {Sugimoto}, \citenamefont {Muduli}, \citenamefont {Ikhlas}, \citenamefont {Omori}, \citenamefont {Tomita}, \citenamefont {MacDonald}, \citenamefont {Nakatsuji},\ and\ \citenamefont {Otani}}]{Kimata2019}%
  \BibitemOpen
  \bibfield  {author} {\bibinfo {author} {\bibfnamefont {M.}~\bibnamefont {Kimata}}, \bibinfo {author} {\bibfnamefont {H.}~\bibnamefont {Chen}}, \bibinfo {author} {\bibfnamefont {K.}~\bibnamefont {Kondou}}, \bibinfo {author} {\bibfnamefont {S.}~\bibnamefont {Sugimoto}}, \bibinfo {author} {\bibfnamefont {P.~K.}\ \bibnamefont {Muduli}}, \bibinfo {author} {\bibfnamefont {M.}~\bibnamefont {Ikhlas}}, \bibinfo {author} {\bibfnamefont {Y.}~\bibnamefont {Omori}}, \bibinfo {author} {\bibfnamefont {T.}~\bibnamefont {Tomita}}, \bibinfo {author} {\bibfnamefont {A.~H.}\ \bibnamefont {MacDonald}}, \bibinfo {author} {\bibfnamefont {S.}~\bibnamefont {Nakatsuji}},\ and\ \bibinfo {author} {\bibfnamefont {Y.}~\bibnamefont {Otani}},\ }\href {https://doi.org/10.1038/s41586-018-0853-0} {\bibfield  {journal} {\bibinfo  {journal} {Nature}\ }\textbf {\bibinfo {volume} {565}},\ \bibinfo {pages} {627} (\bibinfo {year} {2019})}\BibitemShut {NoStop}%
\bibitem [{\citenamefont {Kondou}\ \emph {et~al.}(2021)\citenamefont {Kondou}, \citenamefont {Chen}, \citenamefont {Tomita}, \citenamefont {Ikhlas}, \citenamefont {Higo}, \citenamefont {MacDonald}, \citenamefont {Nakatsuji},\ and\ \citenamefont {Otani}}]{Kondou2021}%
  \BibitemOpen
  \bibfield  {author} {\bibinfo {author} {\bibfnamefont {K.}~\bibnamefont {Kondou}}, \bibinfo {author} {\bibfnamefont {H.}~\bibnamefont {Chen}}, \bibinfo {author} {\bibfnamefont {T.}~\bibnamefont {Tomita}}, \bibinfo {author} {\bibfnamefont {M.}~\bibnamefont {Ikhlas}}, \bibinfo {author} {\bibfnamefont {T.}~\bibnamefont {Higo}}, \bibinfo {author} {\bibfnamefont {A.~H.}\ \bibnamefont {MacDonald}}, \bibinfo {author} {\bibfnamefont {S.}~\bibnamefont {Nakatsuji}},\ and\ \bibinfo {author} {\bibfnamefont {Y.}~\bibnamefont {Otani}},\ }\href {https://doi.org/10.1038/s41467-021-26453-y} {\bibfield  {journal} {\bibinfo  {journal} {Nat. Commun.}\ }\textbf {\bibinfo {volume} {12}},\ \bibinfo {pages} {6491} (\bibinfo {year} {2021})}\BibitemShut {NoStop}%
\bibitem [{\citenamefont {Pashov}\ \emph {et~al.}(2020)\citenamefont {Pashov}, \citenamefont {Acharya}, \citenamefont {Lambrecht}, \citenamefont {Jackson}, \citenamefont {Belashchenko}, \citenamefont {Chantis}, \citenamefont {Jamet},\ and\ \citenamefont {{van Schilfgaarde}}}]{Questaal}%
  \BibitemOpen
  \bibfield  {author} {\bibinfo {author} {\bibfnamefont {D.}~\bibnamefont {Pashov}}, \bibinfo {author} {\bibfnamefont {S.}~\bibnamefont {Acharya}}, \bibinfo {author} {\bibfnamefont {W.~R.}\ \bibnamefont {Lambrecht}}, \bibinfo {author} {\bibfnamefont {J.}~\bibnamefont {Jackson}}, \bibinfo {author} {\bibfnamefont {K.~D.}\ \bibnamefont {Belashchenko}}, \bibinfo {author} {\bibfnamefont {A.}~\bibnamefont {Chantis}}, \bibinfo {author} {\bibfnamefont {F.}~\bibnamefont {Jamet}},\ and\ \bibinfo {author} {\bibfnamefont {M.}~\bibnamefont {{van Schilfgaarde}}},\ }\href {https://doi.org/https://doi.org/10.1016/j.cpc.2019.107065} {\bibfield  {journal} {\bibinfo  {journal} {Comput. Phys. Commun.}\ }\textbf {\bibinfo {volume} {249}},\ \bibinfo {pages} {107065} (\bibinfo {year} {2020})}\BibitemShut {NoStop}%
\bibitem [{\citenamefont {Faleev}\ \emph {et~al.}(2005)\citenamefont {Faleev}, \citenamefont {L\'eonard}, \citenamefont {Stewart},\ and\ \citenamefont {van Schilfgaarde}}]{Faleev2005}%
  \BibitemOpen
  \bibfield  {author} {\bibinfo {author} {\bibfnamefont {S.~V.}\ \bibnamefont {Faleev}}, \bibinfo {author} {\bibfnamefont {F.}~\bibnamefont {L\'eonard}}, \bibinfo {author} {\bibfnamefont {D.~A.}\ \bibnamefont {Stewart}},\ and\ \bibinfo {author} {\bibfnamefont {M.}~\bibnamefont {van Schilfgaarde}},\ }\href {https://doi.org/10.1103/PhysRevB.71.195422} {\bibfield  {journal} {\bibinfo  {journal} {Phys. Rev. B}\ }\textbf {\bibinfo {volume} {71}},\ \bibinfo {pages} {195422} (\bibinfo {year} {2005})}\BibitemShut {NoStop}%
\bibitem [{\citenamefont {Belashchenko}\ \emph {et~al.}(2019{\natexlab{b}})\citenamefont {Belashchenko}, \citenamefont {Kovalev},\ and\ \citenamefont {van Schilfgaarde}}]{Belashchenko2019}%
  \BibitemOpen
  \bibfield  {author} {\bibinfo {author} {\bibfnamefont {K.~D.}\ \bibnamefont {Belashchenko}}, \bibinfo {author} {\bibfnamefont {A.~A.}\ \bibnamefont {Kovalev}},\ and\ \bibinfo {author} {\bibfnamefont {M.}~\bibnamefont {van Schilfgaarde}},\ }\href {https://doi.org/10.1103/PhysRevMaterials.3.011401} {\bibfield  {journal} {\bibinfo  {journal} {Phys. Rev. Mater.}\ }\textbf {\bibinfo {volume} {3}},\ \bibinfo {pages} {011401(R)} (\bibinfo {year} {2019}{\natexlab{b}})}\BibitemShut {NoStop}%
\bibitem [{\citenamefont {Belashchenko}\ \emph {et~al.}(2020{\natexlab{b}})\citenamefont {Belashchenko}, \citenamefont {Kovalev},\ and\ \citenamefont {van Schilfgaarde}}]{Belashchenko2020}%
  \BibitemOpen
  \bibfield  {author} {\bibinfo {author} {\bibfnamefont {K.~D.}\ \bibnamefont {Belashchenko}}, \bibinfo {author} {\bibfnamefont {A.~A.}\ \bibnamefont {Kovalev}},\ and\ \bibinfo {author} {\bibfnamefont {M.}~\bibnamefont {van Schilfgaarde}},\ }\href {https://doi.org/10.1103/PhysRevB.101.020407} {\bibfield  {journal} {\bibinfo  {journal} {Phys. Rev. B}\ }\textbf {\bibinfo {volume} {101}},\ \bibinfo {pages} {020407} (\bibinfo {year} {2020}{\natexlab{b}})}\BibitemShut {NoStop}%
\bibitem [{\citenamefont {Belashchenko}\ \emph {et~al.}(2023)\citenamefont {Belashchenko}, \citenamefont {Baez~Flores}, \citenamefont {Fang}, \citenamefont {Kovalev}, \citenamefont {van Schilfgaarde}, \citenamefont {Haney},\ and\ \citenamefont {Stiles}}]{belashchenko2023breakdown}%
  \BibitemOpen
  \bibfield  {author} {\bibinfo {author} {\bibfnamefont {K.~D.}\ \bibnamefont {Belashchenko}}, \bibinfo {author} {\bibfnamefont {G.~G.}\ \bibnamefont {Baez~Flores}}, \bibinfo {author} {\bibfnamefont {W.}~\bibnamefont {Fang}}, \bibinfo {author} {\bibfnamefont {A.~A.}\ \bibnamefont {Kovalev}}, \bibinfo {author} {\bibfnamefont {M.}~\bibnamefont {van Schilfgaarde}}, \bibinfo {author} {\bibfnamefont {P.~M.}\ \bibnamefont {Haney}},\ and\ \bibinfo {author} {\bibfnamefont {M.~D.}\ \bibnamefont {Stiles}},\ }\href {https://doi.org/10.1103/PhysRevB.108.144433} {\bibfield  {journal} {\bibinfo  {journal} {Phys. Rev. B}\ }\textbf {\bibinfo {volume} {108}},\ \bibinfo {pages} {144433} (\bibinfo {year} {2023})}\BibitemShut {NoStop}%
\bibitem [{\citenamefont {Starikov}\ \emph {et~al.}(2018)\citenamefont {Starikov}, \citenamefont {Liu}, \citenamefont {Yuan},\ and\ \citenamefont {Kelly}}]{Starikov2018}%
  \BibitemOpen
  \bibfield  {author} {\bibinfo {author} {\bibfnamefont {A.~A.}\ \bibnamefont {Starikov}}, \bibinfo {author} {\bibfnamefont {Y.}~\bibnamefont {Liu}}, \bibinfo {author} {\bibfnamefont {Z.}~\bibnamefont {Yuan}},\ and\ \bibinfo {author} {\bibfnamefont {P.~J.}\ \bibnamefont {Kelly}},\ }\href {https://doi.org/10.1103/PhysRevB.97.214415} {\bibfield  {journal} {\bibinfo  {journal} {Phys. Rev. B}\ }\textbf {\bibinfo {volume} {97}},\ \bibinfo {pages} {214415} (\bibinfo {year} {2018})}\BibitemShut {NoStop}%
\bibitem [{\citenamefont {Ho}\ \emph {et~al.}(1983)\citenamefont {Ho}, \citenamefont {Ackerman}, \citenamefont {Wu}, \citenamefont {Havill}, \citenamefont {Bogaard}, \citenamefont {Matula}, \citenamefont {Oh},\ and\ \citenamefont {James}}]{Py-expt}%
  \BibitemOpen
  \bibfield  {author} {\bibinfo {author} {\bibfnamefont {C.~Y.}\ \bibnamefont {Ho}}, \bibinfo {author} {\bibfnamefont {M.~W.}\ \bibnamefont {Ackerman}}, \bibinfo {author} {\bibfnamefont {K.~Y.}\ \bibnamefont {Wu}}, \bibinfo {author} {\bibfnamefont {T.~N.}\ \bibnamefont {Havill}}, \bibinfo {author} {\bibfnamefont {R.~H.}\ \bibnamefont {Bogaard}}, \bibinfo {author} {\bibfnamefont {R.~A.}\ \bibnamefont {Matula}}, \bibinfo {author} {\bibfnamefont {S.~G.}\ \bibnamefont {Oh}},\ and\ \bibinfo {author} {\bibfnamefont {H.~M.}\ \bibnamefont {James}},\ }\href {https://doi.org/10.1063/1.555684} {\bibfield  {journal} {\bibinfo  {journal} {J. Phys. Chem. Ref. Data}\ }\textbf {\bibinfo {volume} {12}},\ \bibinfo {pages} {183} (\bibinfo {year} {1983})}\BibitemShut {NoStop}%
\bibitem [{\citenamefont {Wang}\ \emph {et~al.}(2019)\citenamefont {Wang}, \citenamefont {Wang}, \citenamefont {Amin}, \citenamefont {Wang}, \citenamefont {Radhakrishnan}, \citenamefont {Davidsons}, \citenamefont {Allen}, \citenamefont {Silva}, \citenamefont {Ohldag}, \citenamefont {Balzar}, \citenamefont {Zink}, \citenamefont {Haney}, \citenamefont {Xiao}, \citenamefont {Cahill}, \citenamefont {Lorenz},\ and\ \citenamefont {Fan}}]{Wang2019}%
  \BibitemOpen
  \bibfield  {author} {\bibinfo {author} {\bibfnamefont {W.}~\bibnamefont {Wang}}, \bibinfo {author} {\bibfnamefont {T.}~\bibnamefont {Wang}}, \bibinfo {author} {\bibfnamefont {V.~P.}\ \bibnamefont {Amin}}, \bibinfo {author} {\bibfnamefont {Y.}~\bibnamefont {Wang}}, \bibinfo {author} {\bibfnamefont {A.}~\bibnamefont {Radhakrishnan}}, \bibinfo {author} {\bibfnamefont {A.}~\bibnamefont {Davidsons}}, \bibinfo {author} {\bibfnamefont {S.~R.}\ \bibnamefont {Allen}}, \bibinfo {author} {\bibfnamefont {T.~J.}\ \bibnamefont {Silva}}, \bibinfo {author} {\bibfnamefont {H.}~\bibnamefont {Ohldag}}, \bibinfo {author} {\bibfnamefont {D.}~\bibnamefont {Balzar}}, \bibinfo {author} {\bibfnamefont {B.~L.}\ \bibnamefont {Zink}}, \bibinfo {author} {\bibfnamefont {P.~M.}\ \bibnamefont {Haney}}, \bibinfo {author} {\bibfnamefont {J.~Q.}\ \bibnamefont {Xiao}}, \bibinfo {author} {\bibfnamefont {D.~G.}\ \bibnamefont {Cahill}}, \bibinfo {author} {\bibfnamefont {V.~O.}\ \bibnamefont {Lorenz}},\ and\ \bibinfo {author} {\bibfnamefont
  {X.}~\bibnamefont {Fan}},\ }\href {https://doi.org/10.1038/s41565-019-0504-0} {\bibfield  {journal} {\bibinfo  {journal} {Nature Nanotech.}\ }\textbf {\bibinfo {volume} {14}},\ \bibinfo {pages} {819} (\bibinfo {year} {2019})}\BibitemShut {NoStop}%
\bibitem [{\citenamefont {Baez~Flores}\ and\ \citenamefont {Belashchenko}(2022)}]{PhysRevB.105.054405}%
  \BibitemOpen
  \bibfield  {author} {\bibinfo {author} {\bibfnamefont {G.~G.}\ \bibnamefont {Baez~Flores}}\ and\ \bibinfo {author} {\bibfnamefont {K.~D.}\ \bibnamefont {Belashchenko}},\ }\href {https://doi.org/10.1103/PhysRevB.105.054405} {\bibfield  {journal} {\bibinfo  {journal} {Phys. Rev. B}\ }\textbf {\bibinfo {volume} {105}},\ \bibinfo {pages} {054405} (\bibinfo {year} {2022})}\BibitemShut {NoStop}%
\bibitem [{\citenamefont {Carva}\ and\ \citenamefont {Turek}(2007)}]{CarvaSMC}%
  \BibitemOpen
  \bibfield  {author} {\bibinfo {author} {\bibfnamefont {K.}~\bibnamefont {Carva}}\ and\ \bibinfo {author} {\bibfnamefont {I.}~\bibnamefont {Turek}},\ }\href {https://doi.org/10.1103/PhysRevB.76.104409} {\bibfield  {journal} {\bibinfo  {journal} {Phys. Rev. B}\ }\textbf {\bibinfo {volume} {76}},\ \bibinfo {pages} {104409} (\bibinfo {year} {2007})}\BibitemShut {NoStop}%
\bibitem [{\citenamefont {Xiao}\ \emph {et~al.}(2007)\citenamefont {Xiao}, \citenamefont {Zangwill},\ and\ \citenamefont {Stiles}}]{BoltzmannStiles2}%
  \BibitemOpen
  \bibfield  {author} {\bibinfo {author} {\bibfnamefont {J.}~\bibnamefont {Xiao}}, \bibinfo {author} {\bibfnamefont {A.}~\bibnamefont {Zangwill}},\ and\ \bibinfo {author} {\bibfnamefont {M.~D.}\ \bibnamefont {Stiles}},\ }\href {https://doi.org/DOI: 10.1140/epjb/e2007-00004-0} {\bibfield  {journal} {\bibinfo  {journal} {Eur. Phys. J. B}\ }\textbf {\bibinfo {volume} {59}},\ \bibinfo {pages} {415–427} (\bibinfo {year} {2007})}\BibitemShut {NoStop}%
\bibitem [{\citenamefont {Penn}\ and\ \citenamefont {Stiles}(1999)}]{BoltzmannStiles1}%
  \BibitemOpen
  \bibfield  {author} {\bibinfo {author} {\bibfnamefont {D.~R.}\ \bibnamefont {Penn}}\ and\ \bibinfo {author} {\bibfnamefont {M.~D.}\ \bibnamefont {Stiles}},\ }\href {https://doi.org/10.1103/PhysRevB.59.13338} {\bibfield  {journal} {\bibinfo  {journal} {Phys. Rev. B}\ }\textbf {\bibinfo {volume} {59}},\ \bibinfo {pages} {13338} (\bibinfo {year} {1999})}\BibitemShut {NoStop}%
\bibitem [{\citenamefont {Haney}\ \emph {et~al.}(2013)\citenamefont {Haney}, \citenamefont {Lee}, \citenamefont {Lee}, \citenamefont {Manchon},\ and\ \citenamefont {Stiles}}]{PhysRevB.87.174411}%
  \BibitemOpen
  \bibfield  {author} {\bibinfo {author} {\bibfnamefont {P.~M.}\ \bibnamefont {Haney}}, \bibinfo {author} {\bibfnamefont {H.-W.}\ \bibnamefont {Lee}}, \bibinfo {author} {\bibfnamefont {K.-J.}\ \bibnamefont {Lee}}, \bibinfo {author} {\bibfnamefont {A.}~\bibnamefont {Manchon}},\ and\ \bibinfo {author} {\bibfnamefont {M.~D.}\ \bibnamefont {Stiles}},\ }\href {https://doi.org/10.1103/PhysRevB.87.174411} {\bibfield  {journal} {\bibinfo  {journal} {Phys. Rev. B}\ }\textbf {\bibinfo {volume} {87}},\ \bibinfo {pages} {174411} (\bibinfo {year} {2013})}\BibitemShut {NoStop}%
\end{thebibliography}
\end{document}